# Retrieval of Water Vapor Column Abundance and Aerosol Properties from ChemCam Passive Sky Spectroscopy

Timothy H. McConnochie[a,b]
Michael D. Smith[b]
Michael J. Wolff[c]
Steve Bender[d]
Mark Lemmon[e]
Roger C. Wiens[f]
Sylvestre Maurice[g]
Olivier Gasnault[g]
Jeremie Lasue[g]
Pierre-Yves Meslin[g]
Ari-Matti Harri[h]
Maria Genzer[h]
Osku Kemppinen[h,i]
Germán M. Martínez[j]
Lauren DeFlores[k]
Diana Blaney[k]
Jeffrey R. Johnson[l]
James F. Bell III[m]

[a] Department of Astronomy, University of Maryland, College Park, MD 20742, United States
[b] NASA Goddard Space Flight Center, Greenbelt, MD 20771, United States
[c] Space Science Institute, Boulder, CO 80301, United States
[d] Planetary Science Institute, Tucson, AZ 85719, United States
[e] Texas A&M University, College Station, TX 77843, United States
[f] Los Alamos National Laboratory, Los Alamos, NM 87544, USA
[g] Institut de Recherche en Astrophysique et Planetologie, Toulouse, France
[h] Finnish Meteorological Institute, Helsinki, Finland
[i] Kansas State University, Manhattan, KS 66506, United States
[j] Department of Climate and Space Sciences and Engineering, University of Michigan, Ann Arbor, MI 48109, USA
[k] Jet Propulsion Laboratory, Pasadena, CA 91109, United States
[l] Johns Hopkins University Applied Physics Laboratory, Laurel, MD 20723-6005, United States
[m] Arizona State University, Tempe, AZ 85287-0002, United States

Abstract:

We derive water vapor column abundances and aerosol properties from Mars Science Laboratory (MSL) ChemCam passive mode observations of scattered sky light. This paper covers the methodology and initial results for water vapor and also provides preliminary results for aerosols. The data set presented here includes the results of 113 observations spanning from Mars Year 31 $L_s$ = 291° (March 30, 2013) to Mars Year 33 $L_s$=127° (March 24, 2016).

Each ChemCam passive sky observation acquires spectra at two different elevation angles. We fit these spectra with a discrete-ordinates multiple scattering radiative transfer model, using the correlated-k approximation for gas absorption bands. The retrieval proceeds by first fitting the continuum of the ratio of the two elevation angles to solve for aerosol properties, and then fitting the continuum-removed ratio to solve for gas abundances. The final step of the retrieval makes use of the observed $CO_2$ absorptions and the known $CO_2$ abundance to correct the retrieved water vapor abundance for the effects of the vertical distribution of scattering aerosols and to derive an aerosol scale height parameter.

Our water vapor results give water vapor column abundance with a precision of +/- 0.6 precipitable microns and systematic errors no larger than +/- 0.3 precipitable microns, assuming uniform vertical mixing. The ChemCam-retrieved water abundances show, with only a few exceptions, the same seasonal behavior and the same timing of seasonal minima and maxima as the TES, CRISM, and REMS-H data sets that we compare them to. However ChemCam-retrieved water abundances are generally lower than zonal and regional scale from-orbit water vapor data, while at the same time being significantly larger than pre-dawn REMS-H abundances. Pending further analysis of REMS-H volume mixing ratio uncertainties, the differences between ChemCam and REMS-H pre-dawn mixing ratios appear to be much too large to be explained by large scale circulations and thus they tend to support the hypothesis of substantial diurnal interactions of water vapor with the surface. Our preliminary aerosol results, meanwhile, show the expected seasonal pattern in dust particle size but also indicate a surprising inter-annual increase in water-ice cloud opacities.

Keywords: Mars, atmosphere; Spectroscopy; Radiative transfer

Short title: ChemCam Passive Sky Spectroscopy

Highlights:

- We measure water vapor abundances and aerosol properties at Gale Crater, Mars.
- Precipitable water column is measured with a precision of +/- 0.6 μm.
- Measured quantities include dust & ice fractions, particle sizes, and scale heights.
- Results suggest substantial diurnal interactions of water vapor with the surface.
- A large interannual change in water ice cloud / haze opacity is observed.

Corresponding author / contact information: Timothy McConnochie; email tmcconno@umd.edu, phone 607-351-8345; mailing address 4 Wickersham Ln, Pittsford NY, 14534

# 1 Introduction

The Mars Science Laboratory's (MSL) ChemCam spectrometer (Wiens et al., 2012, Maurice et al. 2012) measures atmospheric aerosol properties and gas abundances by operating in passive mode and observing scattered sky light at two different elevation angles. ChemCam was designed primarily for laser induced breakdown spectroscopy (LIBS) of Martian surface materials (Wiens et al., 2015; Maurice et al. 2016), but it has been used extensively for both imaging (Le Mouélic et al., 2015) and passive spectroscopy (Johnson et al., 2015). This paper covers the methodology and initial results of ChemCam passive sky spectroscopy with ChemCam's VNIR (visible and near infrared) spectrometer, focusing on water vapor abundances and providing preliminary results for aerosols. We also retrieve molecular oxygen, but further data analysis refinements will be required before we are ready to report detailed $O_2$ results.

Other than ChemCam passive sky spectroscopy, our most direct information about water vapor at MSL's work site in Gale Crater comes from MSL's Rover Environmental Monitoring Station (REMS) humidity sensor (Harri et al., 2014b), which provides routine in-situ monitoring of *relative* humidity at a height of 1.6 m above the surface and yields estimates of absolute mixing ratio when relative humidity values are sufficiently high. Water vapor on Mars has of course been measured extensively from orbit and from Earth, and the most densely sampled and high-resolution orbital measurements can provide a useful amount of Gale-Crater-specific information. In this paper we compare ChemCam results with the Gale-specific information that can be provided by the Mars Global Surveyor (MGS) Thermal Emission Spectrometer (TES) data set (Smith et al., 2002) and by Mars Reconnaissance Orbiter's (MRO) Compact Reconnaissance Imaging Spectrometer for Mars (CRISM) (Smith et al., 2009; Toigo et al., 2013).

Interest in Gale Crater water vapor has focused on possible exchanges of water vapor with the surface. Savijärvi et al. (2016) argue that the REMS humidity sensor (REMS-H) data is best explained by diurnal adsorption of water on soil grains, and Martín-Torres et al. (2015) have argued that temperature and humidity conditions at Gale are compatible with the formation of liquid brines on the surface via deliquescence. Meanwhile ChemCam LIBS observations show elevated hydrogen in soils but no evidence for diurnal change in the hydrogen (Meslin et al., 2013, Schröder et al., 2015). Elsewhere on Mars, evidence for a near-surface diurnal cycle of water vapor has been found at both Viking Lander sites (Jakosky et al.,1997) and at the Phoenix lander site (Tampari et al., 2010; Savijärvi and Määttänen, 2010).

The possibility of exchange of water vapor with the surface, by adsorption in particular, remains a controversial but potentially significant factor in the global water cycle. Such exchanges are one way to avoid modeled water columns substantially larger than observed (Böttger et al., 2004), but Montmessin et al. (2004) show that a detailed accounting of the effects of clouds can accomplish the same thing. The radiative effects of those clouds are another aspect of the Martian water cycle that has attracted a lot of recent interest. Models with radiatively active water ice clouds have tended to produce a water cycle that is too dry. This situation has been improved to a significant extent with work by Navarro et al. (2014) that includes detailed cloud microphysics, but that model is still too dry relative to TES water columns at low latitudes, making alternative and additional equatorial water vapor measurements such as those provided by ChemCam particularly valuable.

Clearly aerosols have an important influence on the water cycle and vice versa, so obtaining aerosol and water vapor information in tandem is particular valuable. It has also become clear that water and dust aerosol dynamics are strongly coupled to each other, with vertical distributions playing an important role in that coupling (Kahre et al., 2015). Particle size, too, plays an important role in the influence of dust on dynamics (e.g. Kahre et al., 2008). ChemCam passive sky observations provide a valuable opportunity to capture many of these aerosol properties at the same time, although it should be noted that ChemCam is only one of several MSL instruments that play a significant role in monitoring aerosols. Others include REMS UV photodiodes (Smith et al. 2016) Navcam (e.g. Kloos et al., 2016; Moore et al., 2016), and Mastcam (Lemmon, 2014).

We begin this paper with an overview of the ChemCam passive sky observations and our measurement procedures (section 2.1) and then describe our methods in detail (remainder of section 2). We analyze the sensitivity of our results to various input assumptions in Section 3, then present and discuss our results in Section 4 and summarize our conclusions in section 5. The appendices (A – C) provide additional information which, although necessary for any replication or extension of the results in this paper, is not essential for understanding them: Appendix A addresses external data sets, Appendix B addresses methodological details, and Appendix C shows how to identify ChemCam passive sky data in the Planetary Data System. Data tables covering the complete water vapor and aerosol data set presented here can be found in the supplemental materials. This data set includes the results of 113 observations spanning from Mars Year 31 $L_s$ = 291° (March 30, 2013) to Mars Year 33 $L_s$=127° (March 24, 2016).

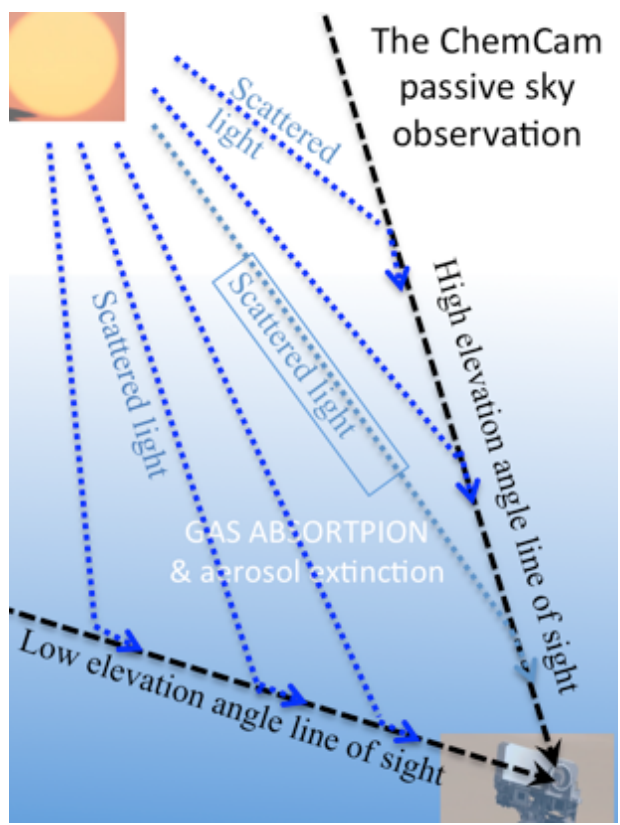

**Figure 1:** Schematic of the ChemCam passive sky observation geometry.

## 2  Methods

### *2.1  Overview*

Figure 1 illustrates the ChemCam passive sky observing strategy. We observe at two different elevation angles, collecting scattered skylight that has traced two significantly different path lengths through the atmosphere, and then ratio the low elevation signal to the high elevation signal to eliminate solar spectral features and instrument response uncertainties. Figures 2 and 3 show an example of the raw signal levels as well as the resulting ratio and continuum-removed ratio, with and without a correction for spatially-variable detector background. Using a discrete-ordinates multiple-scattering radiative transfer model, we fit the continuum of the ratio to solve for aerosol properties (Fig. 4) and then the continuum-removed ratio to solve for gas abundances (Fig. 5). Lastly, we use the known $CO_2$ abundance to correct the gas abundances and derive an aerosol vertical profile parameter (Fig. 6).

The following subsections present the four major steps of the ChemCam passive sky measurement process in order, together with a brief overview of their technical details.

### 2.1.1   Fit the continuum ratio to solve for aerosol properties

We fit the continuum at 15 evenly spaced wavelengths ranging from 550 to 830 nm. The aerosol property parameters used to fit the continuum ratio are dust particle effective radius, ice particle effective radius, and the fraction of 880 nm opacity contributed by dust. We use 880 nm as the reference wavelength for expressing all opacities even though we don't use it for fitting ChemCam data due to ChemCam's relatively low optical response (c.f. Wiens et al., 2012) and signal levels (Fig. 2) at that wavelength. We chose the 880 nm reference wavelength because Mastcam measures opacity in a narrow-band filter centered at 880 nm. The total 880 nm opacity is constrained by Mastcam direct-sun imaging and the water ice 880 nm opacity is thus set equal to the difference between the total 880 nm opacity and the dust 880 nm opacity.

### 2.1.2   Fit the continuum-removed ratio to solve for gas abundances

The gas abundances used to fit the continuum-removed ratio are $O_2$ volume mixing ratio based on the $O_2$ "A" band near 762 nm, $CO_2$ volume mixing ratio based on bands near 783 nm and 789 nm, and water vapor column in precipitable microns based on line groups in the 719 – 730 nm and 810 – 835 nm ranges. We treat all gasses as uniformly vertically mixed in our standard retrievals.

### 2.1.3   Use the known $CO_2$ abundance to correct gas abundances and derive an aerosol vertical profile parameter

The aerosol profiles for both dust and ice aerosols are initialized from TES limb-sounding climatology (Guzewich et al. 2013). In the final stages of our retrieval we iteratively adjust an aerosol extinction scale height parameter (expressed as gas scale height over dust or ice extinction scale height) and repeat the gas abundance fit to find the unique parametric relationship between $CO_2$ abundance and the other gas abundances. We use this to correct the $O_2$ and water vapor abundances, and also estimate the aerosol extinction scale height, based on the known (Mahaffy et al. 2013) correct value for $CO_2$ (96.0%).

### 2.1.4   Estimate uncertainties

After completing all retrievals we use the fit-residual covariance matrices calculated from the complete set of retrievals to make a Monte-Carlo estimate of the gas abundance uncertainties of each retrieval. The $CO_2$ fit residuals determine the uncertainty estimate for the aerosol scale height parameter. In addition they contribute to the final uncertainty estimate for both $O_2$ and water vapor because of the correction procedure described in the previous paragraph. For the remaining aerosol parameters, the statistical instrumental uncertainties are negligible compared to the effects of systematic errors and uncertain inputs, and so a discussion of those errors will be postponed to Section 3, which deals with sensitivity testing.

## 2.2   Sky Observations

The passive sky observation is designed to meet instrument safety requirements while maximizing the atmospheric path length contrast and achieving a signal-to-noise-ratio (SNR) value of ~3000 for individual detector pixels in the 710 – 840 nm portion of the VNIR spectrometer (which has a spectral range of 473-906 nm). It's also designed to accommodate the calibration requirements of the violet (VIO) and UV spectrometers (382-469, 241-341 nm, respectively), which are not used in this paper but may be significant for future work. Other considerations include minimizing rover time and data volume resources, the significant variations in lighting geometry during the observation, and accommodating the plane-parallel geometry of the radiative transfer model. Note that the SNR target of 3000 refers to the SNR of ratio spectra after accounting for systematic detector background errors and the imperfect correction thereof. 3000 is the objective and is typical of the best quality observations but is not a minimum requirement; we obtain useful results with SNR as low as 500.

The SNR objective drives some secondary objectives related to the detector background contribution to noise. Minimizing the sky brightness contrast is extremely valuable for achieving the desired SNR with ChemCam because it minimizes the effect of detector background uncertainty on the ratio spectra. Minimizing the average time elapsed between the low elevation and high elevation spectra being ratioed helps as well because of the ChemCam noise characteristics. So does avoiding high detector temperatures.

To avoid the risk of sun damage, all sky observations are performed with the ChemCam telescope in the "sun safe" focus position of 2.0 meters focal distance, resulting in an approximate 54 milliradian (3.1°) geometric field of view. (The ChemCam telescope aperture is 108.4 mm.) As an additional precaution, pointings within 22.5° of the sun's path on the given sol are avoided.

Most passive sky observations are performed near noon local solar time because ChemCam detector temperatures are generally too warm later in the day and because early morning local solar times are more demanding of rover resources. We have however acquired a small number of early morning passive sky observations.

For our two pointing positions we use a low elevation angle of 20° and a high elevation angle of 65°, 69°, or 72°. We typically use an azimuth of 0° or 180° (relative to true north) for the low pointing, generally choosing the side of the sky opposite the noon position of the sun. Between sol 540 and 650 we also used azimuth of 90° and 270° for the low pointing. The only other deviations from the 0° or 180° low pointing strategy are six occasions with azimuths ranging from 340° to 30°.

For the high pointing we initially selected the same azimuth as the low pointing (except for 6 occasions between sols 295 and 330 where it was pointed in the opposite direction), but after sol 790 we began adjusting the azimuth of the high pointing to minimize the difference between the predicted sky brightness of the high pointing and the predicted sky brightness of the low pointing. In other words, if the high elevation angle pointing would have had higher signal levels than the lower elevation angle pointing its azimuth was adjusted to move it further from the sun and bring its predicted signal levels down to match those of the low elevation angle pointing. If it would have had lower signal levels, we adjusted the azimuth to move it closer to the sun and raise its predicted signal levels to match those of the low elevation angle pointing.

Each observation sequence moves back and forth between the two pointing positions multiple times to allow the effects of lighting geometry to be separated from those of aerosol properties and to diminish the effects of temporal changes in detector background. At each visit to each pointing position we acquire multiple "collects", with each "collect" being an individual command to ChemCam to acquire 75 spectral exposures and return an average. The integration time of the individual exposures is chosen to produce VNIR peak DN values approximately 20% less than the detector saturation level, based on the predicted sky brightness. (DN or "data number" is the unit for digitized raw detector output.) Every integration time in an observation sequence is the same, with typical integration times being 70 – 350 milliseconds. All three ChemCam spectrometers integrate simultaneously.

The ChemCam CCD detectors are two-dimensional arrays, but except for a special diagnostic mode detector rows are always summed in the horizontal register during the readout process. Standard ChemCam observations, LIBS and passive alike, use detector rows 100 – 300, 75 – 275, and 100 – 300 for the UV, VIO, and VNIR spectrometers, respectively, because these rows cover the illuminated portion of the detector. The choice of rows is commandable, however, and the passive sky observation devotes some of its collects to the non-illuminated ("dark") rows in order to help estimate the detector background. When non-illuminated rows are collected for a particular spectrometer, the detectors rows used are 310 – 510. The non-illuminated rows are not in fact completely dark because the detector readout process sweeps all detector pixels through the illuminated portion of the chip, but all rows experience the same readout contribution which allows the difference between illuminated and non-illuminated rows

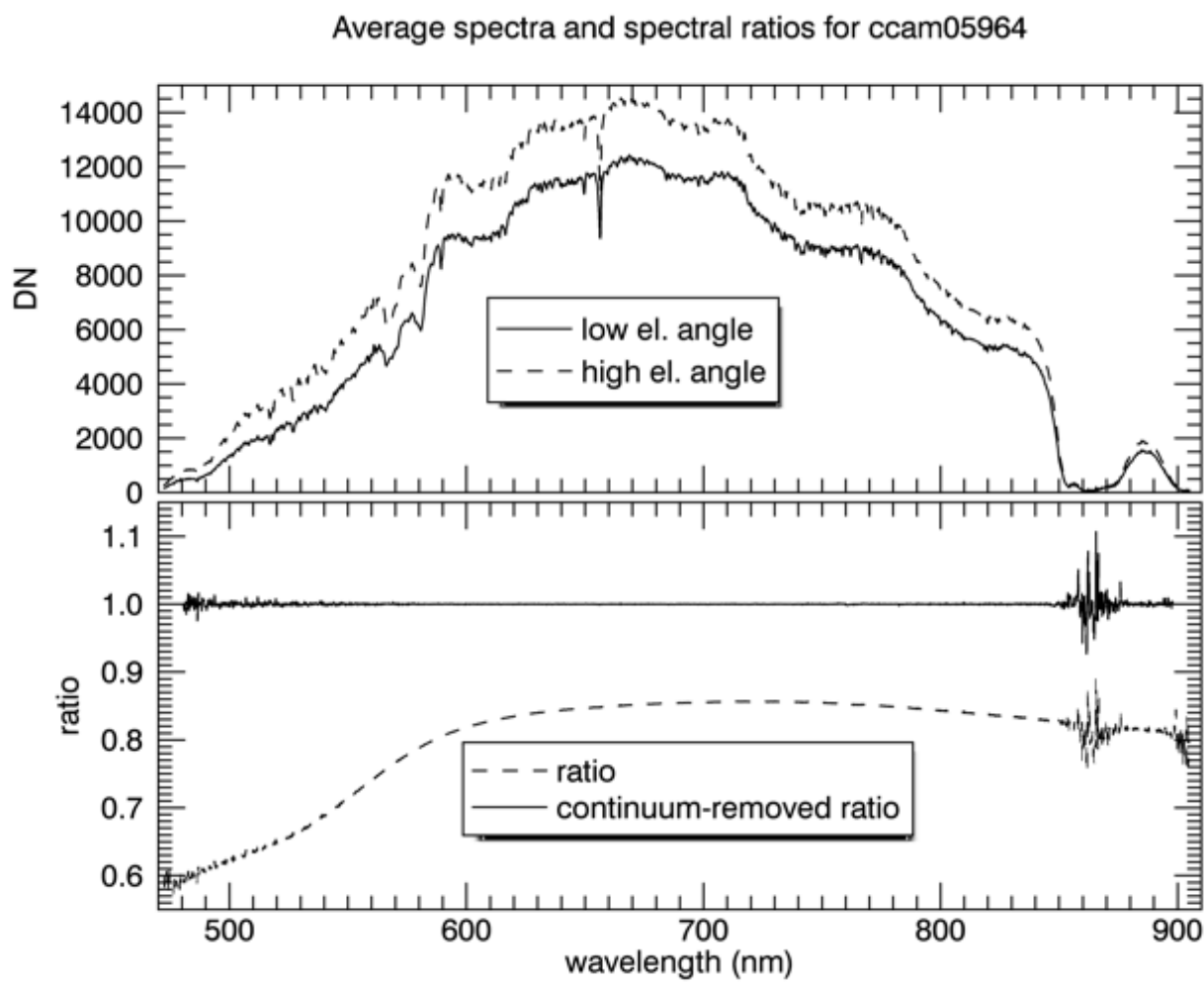

**Figure 2:** Top panel: Raw data number "DN" values after subtracting the spatially-uniform component of detector background. The solid and dashed lines show the average of all low elevation angle collects and all high elevation angle collects, respectively. Bottom panel, dashed line: The ratio of the low elevation angle average to the high elevation angle average. Bottom panel, solid line: the same ratio, after continuum removal (most features in this line are not visible at this scale.)

to remove a substantial portion of the detector background.

All passive sky observation sequences include 32 collects of illuminated VNIR rows – 16 at the high pointing position and 16 at the low pointing position – and during those 32 illuminated VNIR collects the UV and VIO spectrometers are alternated between illuminated and non-illuminated rows. In addition to those 32 collects all passive sky observation sequences include one collect at the beginning and one collect at the end that acquires non-illuminated rows for all three detectors. Over the course of the mission, we have tried different approaches to the timing of slews between the high and low pointing positions, and some passive sky sequences have used additional collects of non-illuminated rows for all detectors. In total, we have employed three different passive sky observing sequences, which we will label passive sky types 1, 2, and 3, and which are detailed in Table 1. Type 1 was used earlier in the mission, and type 2 is the current default. Note that Table 1 also gives the label typically given to these passive sky activities in the MSL "SOWG Documentarian" reports, which are available in NASA's Planetary Data System (PDS).

**Table 1**: ChemCam passive sky observing sequence types

| Type | Documented as | Sequence details |
|---|---|---|
| 1 | "CCAM_passive_sky" | * d LlLl LlLl * HhHh HhHh * LlLl LlLl * HhHh HhHh d |
| 2 | "CCAM_passive_sky_ 2X" | * d LlLl * HhHh * LlLl * HhHh * LlLl * HhHh * LlLl * HhHh d |
| 3 | "CCAM_passive_sky_ 2X_improved" | * d LlLl * HhHh d HhHh * LlLl d LlLl * HhHh d HhHh * LlLl d |

| Definitions |
|---|
| * = Rover remote sensing mast (RSM) motion |
| d = collects of non-illuminated detector rows for all 3 detectors |
| L = Low elevation angle pointing collects with illuminated rows for all three detectors |
| l = Low elevation angle pointing with VNIR illuminated rows, UV and VIO non-illuminated rows |
| H = High elevation angle pointing collects with illuminated rows for all three detectors |
| h = High elevation angle pointing with VNIR illuminated rows, UV and VIO non-illuminated rows |

### 2.3    Calibration observations

We have used two different types of calibration observations, both of which are designed specifically for use with passive sky measurements. Other non-sky-specific passive-mode calibration observations (e.g., Johnson et al., 2015) are generally not useful because of the extremely high S/N required by the passive sky gas retrievals. We use these calibration observations to measure the spatial variations of the electronic background across the detector, which are primarily due to CCD dark current and which cannot be completely removed using the non-illuminated row observations alone because those non-illuminated rows are located on a different physical region of the detector than the illuminated rows.

Our primary calibration observation sequence uses the ChemCam calibration targets (c.f. Wiens et al. 2012; Johnson et al. 2015) to generate a ratio of a dark surface spectrum to a bright surface spectrum. Since the calibration targets have no narrow-band spectral features and since they are all illuminated by the same combination of direct solar and diffuse sky light, the continuum-removed dark-to-bright calibration target spectral ratio contains only the narrow-band features contributed by pixel-to-pixel variations in detector background. The effectiveness of a calibration observation decreases with time – see Appendix B section B.7 and Figure B.1 for further discussion of this. Unfortunately, we did not develop and implement this calibration approach until well into the MSL mission – these calibration-target-for-passive-sky observations were first used on sol 964.

The ChemCam calibration target for passive sky observing sequence is identical to the first half of a type 3 passive sky observation, i.e. * d LlLl * HhHh d HhHh * LlLl d (using the letter code defined in Table 1) but with the "L" or "l" and "H" or "h" collects pointing at bright and dark calibration targets, respectively, and with Remote Micro Imager (Maurice et al., 2012) documentation. In MSL SOWG Documentarian reports (available in the PDS) these activities are labeled as "CCAM_CCCT_for_passive_sky". We have used calibration target #5 (graphite) for the dark target and either #9 (high-sulfur smectite) or #10 (titanium) for the bright target. (The chosen targets can be identified from the PDS header – see Appendix C)

Prior to developing the calibration-target-for-passive-sky observation, we experimented with a different detector calibration approach that we eventually realized was nearly as effective, although it was more difficult to fit into rover operations. This was a sequence of ChemCam nighttime observations designed to acquire a S/N level and pattern of illuminated vs. non-illuminated collects similar to that of an actual passive sky observation, while the ChemCam detector temperatures were similar to those typical of local solar noon. This calibration sequence was performed on sol 611 (SEQUENCE_ID = "ccam04610") – it alternated between collects of illuminated rows for all detectors and collects of non-illuminated rows for all detectors, and it included two different integration times. The detector background pattern derived from this nighttime calibration is used whenever it is a better fit than that from the calibration target calibration, which only occurs for observations acquired closer in time to sol 611 than to the earliest calibration target calibration on sol 964.

### 2.4    Other inputs

The external inputs required by the ChemCam passive sky aerosol and gas retrievals are surface pressure, total column opacity, surface albedo, aerosol optical properties, a vertical temperature profile, and an initial guess vertical profile of aerosols.

#### 2.4.1    Pressure, opacity, surface albedo

The surface pressure comes from MSL's Rover Environmental Monitoring Station (REMS) pressure sensor (Harri et al., 2014a). Total column opacity is measured by MSL's Mastcam using direct solar imaging in narrow-band solar filters center at 440 and 880nm. This data set is described by Lemmon (2014). For surface albedo we adopt a lambert albedo that is a weighted sum of endmember spectra from Mustard and Bell (1994), using a CRISM atmospherically corrected lambert albedo data cube (Arvidson et al., 2005) to guide our choice of weights. We found that the appropriate dark region weights in the vicinity of the MSL operating area are were 10%-20% with 10% appearing (subjectively) to be most typical – thus we adopt the 10% dark region weight but we consider the possibility of a 20% weight in our sensitivity testing (section 3). See Appendix A for further details on our surface pressure, column opacity, and surface albedo inputs.

#### 2.4.2    Vertical profiles

We obtain both a vertical temperature profile and an initial aerosol extinction profile from the binned Mars Global Survey Thermal Emission Spectrometer limb-sounding data set (Guzewich et al. 2013). This data set is binned at 10° $L_s$ intervals (0° – 10°, 10° – 20°, …) in time and 30° and 10° intervals in longitude and latitude. The bin that includes MSL covers 120 °– 150° east, 0° – 10° south. We select a time bin corresponding to the seasonal (i.e. $L_s$) interval that includes the passive sky observation being modeled, and if that seasonal interval is covered in multiple Martian years in the TES data set, we select the earliest Martian year. The TES temperature profile is used as is – it is simply interpolated onto the radiative transfer model grid. The TES ice and dust extinction mixing ratio profiles are modified and then smoothed, and in the bottom two scale heights the dust profile is replaced with an idealized profile based on a parameterization. The profile modifications are described in detail in Appendix A (A.4). The parameterization of the bottom scale heights is described in the following section.

### 2.5    Parameterization of the aerosol vertical profile

The idealized, parameterized profile at two scale heights and below is introduced to allow us to adjust a single parameter to help match our observations. It is defined by the vertical extent above the local surface of a near surface mixed layer, $z_m$, and by a relative scale height parameter, $H'$, defined by

$$H' = H_{gas} / H_{aerosol} \qquad (1)$$

where $H_{gas}$ is the gas scale height and $H_{aerosol}$ is the scale height of the extinction coefficient and defined analogously to the gas scale height. Specifically, for extinction coefficient at some reference level $\beta_{ref}$ the extinction coefficient $\beta$ at some other level $z$ by our definitions is:

$$\beta = \beta_{ref} \ \exp(- (z - z_{ref}) / H_{aerosol}) \qquad (2)$$

Unlike $H_{gas}$, $H_{aerosol}$ can be infinite or negative as would be the case if the number density and hence the resulting extinction coefficient of an aerosol species was constant or increasing with height. A well-mixed aerosol would have $H'$=1. We use $z_m = 0.25 \ H_{gas}$ in all cases, and we initialize the value of $H'$ by determining the $H_{aerosol}$ value that causes an exponential function to pass through the pre-existing TES aerosol profile points at $z = 1 H_{gas}$ and $z = 2 H_{gas}$. The initial guess dust aerosol profile is then set equal to the pre-existing TES derived dust aerosol profile at $z = 2 \ H_{gas}$ and above, equal to the exponential defined by $H'$ from $z = 2 H_{gas}$ down to $z = z_m$, and set to a constant from $z_m$ to the local surface. We use $z_{ref} = 2 H_{gas}$ so that after any adjustments to $H'$ the adjusted profile will still match the background profile at 2 scale heights above the local surface.

### 2.6    Radiative transfer model

We use a discrete ordinates (c.f. Thomas and Stamnes, 1999) radiative transfer model with a correlated-k approximation (Lacis and Oinas, 1991) for gas absorption. Our discrete ordinates code is nearly identical to that employed by Smith et al. (2009) except for the aerosol size and shape assumptions and except for the detailed choices of model parameters. See Appendix B (B.2) for further discussion of these assumptions and details. The one significant change that we have made to the underlying radiative transfer code of Smith et al. (2009) is the incorporation of the delta-approximation (e.g. Goody and Yung, 1989) to allow for more efficient modeling of highly-forward scattering particle phase functions. In this approximation the mathematical representation of the scattering phase function is given by the sum of a Dirac delta-function that represents the forward scattering peak, and the usual series of Legendre polynomials that represents the phase function outside the forward peak.

### 2.7    Data processing

To prepare the ChemCam passive sky data for our retrievals, we first read all of the EDRs associated with a given passive sky observation, then calculate and subtract the constant-across-the-detector component of detector background, then calculate calibrated wavelengths for each spectral pixel, then generate the continuum ratio and continuum-removed ratio, and then lastly generate bootstrap standard errors for the continuum-removed ratio.

#### 2.7.1    Spatially uniform detector background

To measure the constant-across-the-detector component of detector background we select pixels between 859 nm and 869 nm nominal wavelength in non-illuminated row collects. Then we fit a parabola to those pixels and treat the minimum of the parabola as our estimate of the true local minimum. In passive sky observations the minimum VNIR signal levels always fall within this group of pixels because of a strong minimum in spectrometer system response (c.f. Wiens et al., 2012). By selecting the non-illuminated rows from the wavelength range with minimal system response we nearly eliminate the contribution of scene photons. There is still, however, a very small contribution of photon signal that accumulates in the non-illuminated row pixels during the readout process even in these minimum-signal wavelengths. To account for this we compare the fitted-parabola minimum in each non-illuminated row collect to the fitted-parabola minimum in the same nominal wavelength range of the nearest illuminated-pixels collect, and then use this relationship to extrapolate to the true background level. See appendix B (B.4) for further discussion of this procedure and its uncertainties, which turn out to be negligible because the non-illuminated-rows minimum is so close to the true background.

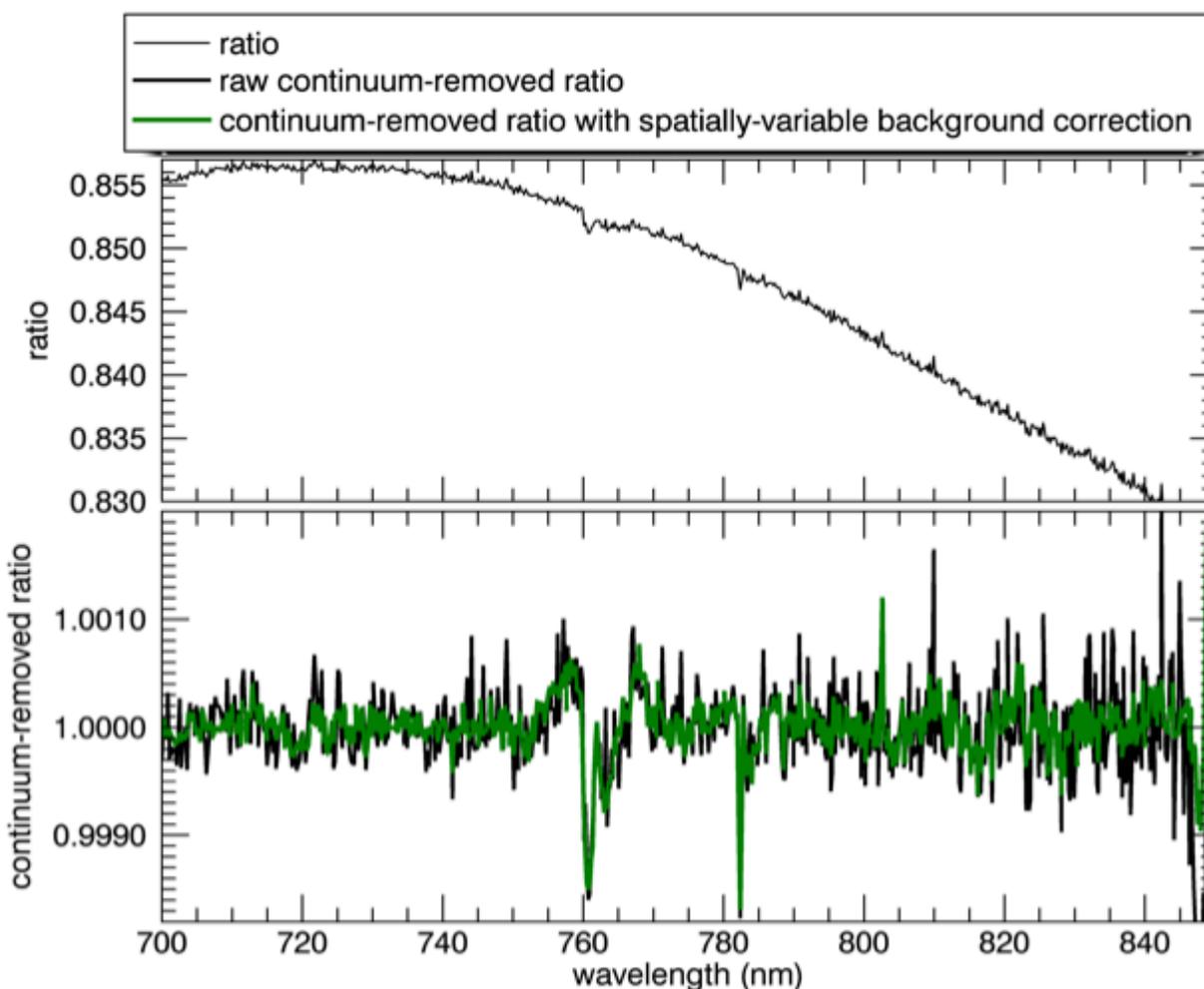

**Figure 3:** The same data as in Fig. 2, but zoomed-in to the wavelengths used in gas abundance fitting and with spatially-variable background correction included in the green continuum-removed spectrum. Note the different scales in the top and bottom panels – the top panel shows the ratio and bottom top panel shows the continuum-removed ratio with and without spatially-variable background correction.

### 2.7.2   Wavelength calibration

We use the wavelength calibration function from the ChemCam PDS archive to calculate the wavelength of each EDR pixel, using the average wavelength over the course of a passive sky observation sequence because variations during a single sequence are negligible. See Appendix B (B.1) for further details.

### 2.7.3   Continuum-removed ratio spectra

All passive sky observing sequences contain 16 low-elevation-angle VNIR illuminated-row collects and 16 high-elevation-angle VNIR illuminated-row collects. After subtraction of the spatially uniform component of detector background, all collects at each elevation angle are averaged (Fig. 2 top panel) and then the ratio spectrum is calculated as the low-elevation average divided by the high-elevation average (Fig. 2 bottom panel, Fig. 3 top panel). The continuum-removed ratio spectrum (Fig. 2 bottom panel, Fig. 3 bottom panel) is calculated from the ratio spectrum by performing two iterations of dividing the spectrum by a smoothed representation of itself. In the first iteration, the original ratio spectrum is convolved with a 75 pixel wide boxcar smoothing kernel and then the original ratio spectrum is divided by the result of that convolution to yield the spectrum carried forward to the second iteration. In the second iteration the spectrum from the first iteration is convolved with a 75 pixel wide boxcar smoothing kernel and then the spectrum from the first iteration is divided by the result of that convolution to yield what we call the continuum removed ratio spectrum. We develop the formal definition of this procedure in Appendix B section B.5.

### 2.7.4   Ratio spectrum for aerosol property fits

The ratio spectrum that we use for aerosol property fitting has a much simpler definition than the continuum removed ratio, but it starts with the same (Fig. 2 top) low-elevation average and high-elevation average spectra. To form this ratio spectrum (the points in Fig. 4 are examples) we smooth the low-elevation-angle average and the high-elevation-angle average with a 25 pixel FWHM Gaussian kernel, and then we divide the smoothed low-elevation-angle average by the smoothed high-elevation-angle average. We then interpolate this ratio to the wavelength samples chosen for the aerosol-continuum model fitting.

### 2.7.5   Bootstrap standard errors

To estimate standard errors, we repeatedly generate first a random sample of the 16 low elevation collects and then a random sample of the 16 high elevation collects, and then perform the procedure in the previous section to generate a new continuum removed ratio spectrum at each repetition. All random samples are drawn with replacement and from a uniform distribution. We perform 32,000 repetitions. We then use the standard deviations of this ensemble of spectra as the standard errors for each pixel in the continuum-removed ratio.

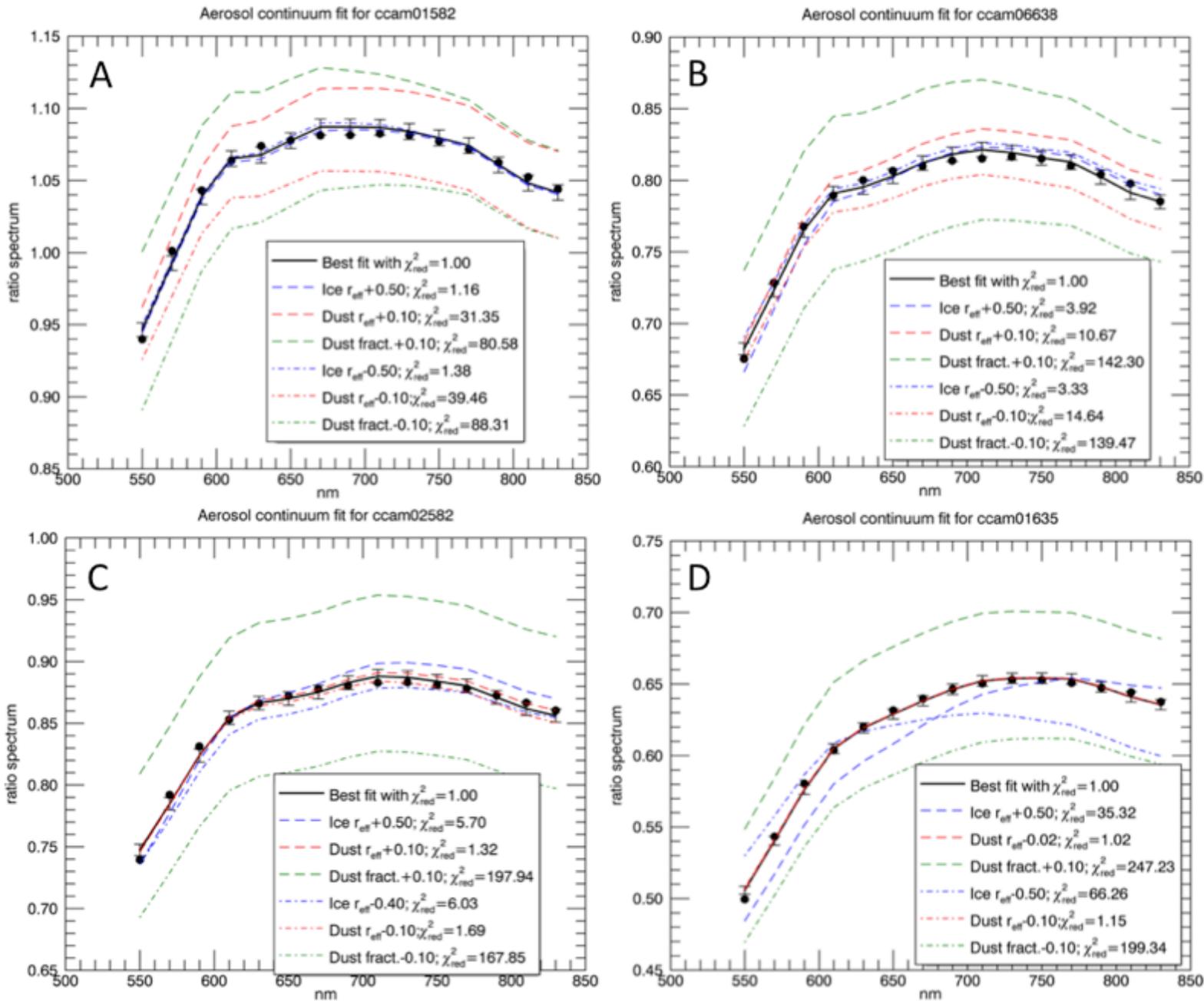

**Figure 4:** Best model fit (heavy black line) to the continuum ratio spectrum (black points) obtained by varying aerosol properties for four different ChemCam passive sky observation sequences, displayed in four different panels. The error bars are obtained by scaling an assumed constant fractional uncertainty in order to obtain a reduced-$\chi^2$ ("$\chi^2_{red}$") of 1. The dashed lines represent models perturbed by increasing one of three aerosol parameters and the dash-dot lines represent models perturbed by decreasing one of the aerosol parameters, as shown in the legend. The legend gives the reduced-$\chi^2$ values obtained by comparing each model to the observed spectrum. For each panel (A, B, C, or D), the sequence ID, season ($L_s$), local true solar time, and the azimuth (az.) and elevation (el.) angles for both pointings are as follows. A: ccam01582, $L_s$=108°, LTST=11:25, 180° az. 72° el. & 180° az. 20° el.; B: ccam06638, $L_s$=135°, LTST =11:38, 180° az. 72° el. & 180° az. 20° el.; C: ccam02582, $L_s$=108°, LTST=12:04, 270° az. 72° el. & 270° az. 20° el.; D: ccam01635, $L_s$=133°, LTST =11:36, 95° az. 72° el. & 95° az. 20° el..

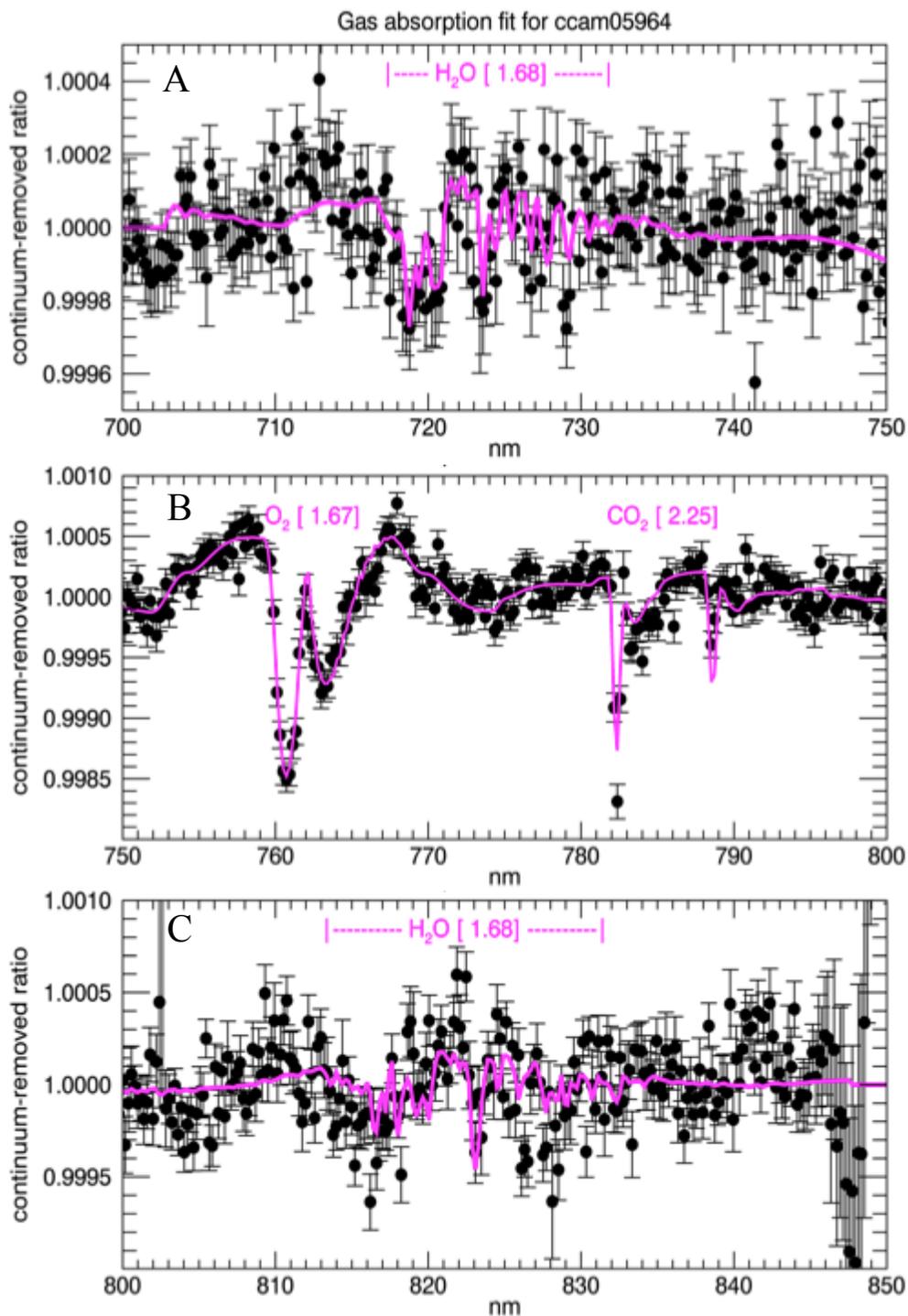

**Figure 5:** Example best fit model (magenta line) for the continuum-removed ratio obtained by varying $O_2$, $CO_2$, and $H_2O$ gas column abundances. Panels A, B, and C show different portions of the spectrum. The major absorption features are labeled with the molecule that forms it, and the reduced-$\chi^2$ values obtained by fitting the abundance of each molecule are given next to the label. The observed continuum-removed, spatially-variable-background-corrected ratio is shown with black points and error bars. It is identical to the green line in Fig. 3.

### 2.8    Aerosol fitting with the continuum ratio spectrum

To solve for aerosol properties we model the continuum ratio spectrum at 15 equally spaced wavelength samples ranging from 550 to 830 nm. Figure 4 shows examples. The data value at each of these points is generated by smoothing, then ratioing, and then interpolating as previously described (2.7.4). Since the random errors of the ratio spectra are extremely small, on the order of $10^{-4}$ to $10^{-3}$, they are a negligible source of uncertainty in aerosol continuum fitting, and we therefore assign each data point a fixed fractional error of 5% for purposes of calculating the $\chi^2$ of a model fit. This assigned uncertainty should be thought of as an initial guess at the combined uncertainties of the radiative transfer model and, especially, its inputs.

To minimize $\chi^2$ we use the downhill simplex method using the 'amoeba' routine from Numerical Recipes in C (Press et al., 1992) as implemented by IDL. Dust $r_{eff}$, water ice $r_{eff}$, and the fraction of 880 nm column opacity contributed by dust are varied to minimize $\chi^2$ with a 1% tolerance. Dust $r_{eff}$ between 0.5 and 2.5 microns, water ice $r_{eff}$ between 1.0 and 4.0 microns, and dust fractions between 0 and 1 are considered. We use initial values of 1.5, 2.5, and 0.52 for those quantities, respectively. We have not observed any sensitivity to that initial guess.

### 2.9    Gas Abundance Fitting

This section first describes the various key pieces of the gas abundance fitting procedure, including wavelength selections (2.9.1), molecular abundance fitting (2.9.2), model look-up tables (2.9.3), and spatially variable background scale factor fitting (2.9.4). It then concludes with showing how these pieces fit together by enumerating (in subsection 2.9.5) the complete set of steps required to solve for the abundance of all three gasses. Fig. 5 shows an example fit.

Some of the low-level details and justifications for this procedure are left for Appendix B. In particular, see Appendix B for a formal definition of model fitting with continuum-removed spectra (B.5) and spatially variable background correction (B.7).

#### 2.9.1    Wavelength ranges for fitting molecular absorptions

For each of our three target molecules we adopt a fitting wavelength range within which all data spectral samples contribute to the $\chi^2$ for that molecule; and we adopt a model calculation range within which we must recalculate the model spectrum at each iteration for that particular molecule because the abundance of that molecule makes a non-negligible contribution. These wavelength ranges are shown in Table 2.

**Table 2**: ChemCam passive sky VNIR wavelength ranges

| Molecule | $\chi^2$ fitting range (nm) | Model calculation range (nm) |
|---|---|---|
| $O_2$ | 758.5 – 768.5 | 749.0 – 776.5 |
| $CO_2$ | 780.5 – 794.0 | 776.5 – 795.0 |
| $H_2O$ | 716.0 – 734.0, 811.0 – 835.0 | 701.0 – 749.0, 795.0 – 845.0 |

We fit our three target molecules sequentially – first $O_2$, then $CO_2$, then $H_2O$. To avoid unnecessary model runs, we also build up our model spectrum sequentially. We start by calculating the model from 695 nm to 855 nm with no gas absorption, then we restore the gas absorption and update the model *only* in the $O_2$ model calculation range while fitting for $O_2$ in its $\chi^2$ fitting range, and then while fitting for $CO_2$ we update that model only in the $CO_2$ model calculation range, and finally while fitting for $H_2O$ we update the model only in the $H_2O$ model calculation range. Since the boxcar smoothing that we use in generating continuum-removed spectra consists of two iterations with a 75 pixel, 15.2 nm width, it is apparent from Table 2 that the $O_2$ fitting range could in principle be influenced by the $CO_2$ absorption strength which has not yet been solved for at the time of $O_2$ fitting, and similarly that the $CO_2$ fitting range could be influenced by the not-yet-solved-for $H_2O$ lines. However these effects turn out to be negligible as we discuss in Appendix B (B.6).

### 2.9.2   Molecular abundance grid search

To fit each molecule we perform a one-dimensional grid search for the minimum $\chi^2$ (appendix B.5 equation B23). The grid spacing interval is chosen to be less than the best observed uncertainties in the mixing ratio, and the grid is expanded if the lowest $\chi^2$ is at the edge of the range. The final result of each grid search is obtained by taking the minimum of a parabola fitted through the four smallest $\chi^2$ grid points.

### 2.9.3   Adaptive model look-up table

The grid-search spacing – 1% volume mixing ratio for $CO_2$ and 0.1 precipitable microns for water vapor – is too small to allow the radiative transfer model to be run at each grid point in a reasonable amount of time. For this reason prior to the grid search we generate an adaptive look-up table of model output for each gas and each observation sequence. The grid search queries the look-up table to get a model result at each grid point, and those queries usually take negligible time because the model look-up table is mostly pre-computed. The adaptive model look-up table also has the advantage that it can be reused as long as no model parameters are changed – we use this capability when we re-run the grid-search after solving for the spatially-variable detector background scale factor (section 2.9.4), and to enable our Monte-Carlo techniques for estimating uncertainties (section 2.11).

The model look-up table has its own set of gas abundance calculation points which can be more widely separated than the grid-search points due to the approximate linearity of the model with respect to small abundance changes. The model look-up table is adaptive in the sense that the spacing of the look-up table is adjusted until it meets a linearity test, and in the sense that the upper and lower limits of its range can be and are expanded if it is queried for a gas abundance that is out of its current limits. Appendix B (B.3) gives more details on how the model look-up table is generated and maintained.

### 2.9.4   Fitting for the scale factor of the spatially variable detector background

To complete the gas abundance fitting process, we must account for the influence of spatially variable detector background. We call the expression of the spatially variable detector background in the continuum-removed ratio spectra $e_{4sky}$ (see Appendix B.5). It's defined by Eq. (B16) and calculated according to Eq. (B22). This equation contains terms that come from the observation itself: integration time $t$, temperature $T$, DN values $A_i$ and $B_i$ for the low-elevation-angle and high-elevation-angle spectra respectively. It also contains terms that come from one of the calibration observations described in section 2.3: $t_{cal}$, $T_{cal}$, $A_{ical}$, $B_{ical}$, and $e_{4cal}$, all of which are analogous to their sky observation counterparts. Appendix B (B.7) provides more details on calculating these terms as well as an assessment of the overall effectiveness of the spatially variable background correction.

Lastly and most importantly for our fitting algorithms, $e_{4sky}$ contains a scale factor $\gamma$, the magnitude of which we must solve for. In fact we must find a simultaneous solution for both gas abundance and $\gamma$. We do so as follows:

<u>Gas & background fitting procedure</u>

i. We first solve for the abundance of a particular gas as described in 2.9.2, with $\gamma$ set equal to zero.

ii. We next solve for $\gamma$ by minimizing the $\chi^2$ differences between $e_{4sky}$ and the residuals of the first radiative transfer modeling pass, using a grid-search approach that follows the same grid resolution and domain expansion rules that were applied for the gas abundance. To avoid multiple iterations of this procedure, we must remove any covariances between the model and $e_{4sky}$ before calculating $\chi^2$. To do so we calculate the model derivatives of the first-pass best fit model, then calculate and subtract the vector projection of these model derivatives onto the first-pass residuals, then perform the same procedure on $e_{4sky}$.

iii. After solving for $\gamma$, we repeat the subsection 2.9.2 grid search fit for the mixing ratio. In principal we could continue to iteratively solve for $\gamma$ and then the mixing ratio, but in our testing we find that our projection-subtraction procedure for removing covariances between the model and $e_{4sky}$ is so effective that multiple iterations to solve for $\gamma$ produce no changes in the result.

### 2.9.5   Gas abundance solution

To generate the gas abundance solution for all three molecules, we perform the following steps:

1. To select the calibration observation that we will use, we perform the "gas & background fitting procedure" (section 2.9.4) for $O_2$ multiple times – once for each of the calibration observations in our data set (including both calibration target and nighttime background calibration observations). Then we select the one calibration observation that produces the smallest $\chi^2$.

2. We require the detector background fit parameterized by $\gamma$ to produce a substantial improvement in $\chi^2$. If the best $\chi^2$ after fitting for $\gamma$ is *more* than 0.64 times the $\chi^2$ with $\gamma = 0$, we choose $\gamma = 0$ as the preferred solution which means that no variable-detector-background subtraction is applied to that sky observation for any of the gases.

3. We fit for $CO_2$ and $H_2O$ (in that order) following the "gas & background fitting procedure" (section 2.9.4) and using the same calibration observation selected in step 1. That means that we use the same calibration observation for all three molecules, but each molecule has its own solution for $\gamma$. If we chose $\gamma = 0$ in step 2 then the $\gamma = 0$ solution is adopted for all three molecules.

We chose the factor of 0.64 for improvement in $\chi^2$ empirically, based on our subjective evaluation of whether the $e_{4sky}$ vector is matching real features in the initial fit residuals. When the improved $\chi^2$ is below this threshold there are always identifiable matching features and we are confident that we are not over-fitting. A failure to cross this $\chi^2$ threshold typically occurs when $A_{isky}$ is nearly equal to $B_{isky}$ rendering the detector background pattern undetectable. It also occurs in some cases when a sky observation does not have a calibration observation sufficiently nearby in time. In all cases where the improvement does cross the 0.64 threshold, the improvement in $\chi^2$ passes an F-test with greater than 99% confidence, meaning that our subjective threshold is conservative relative to the standard quantitative metric.

### 2.10   Matching the known CO₂ mixing ratio

The $CO_2$ mixing ratio value that we retrieve from the steps described in section 2.9 typically differs significantly from the well-established value of 96% (Mahaffy et al. 2013). These differences are expected, because the depth of all of the Mars-atmosphere absorption lines depends on the distribution of scattering sources along the lines of sight. In the low-elevation-angle-to-high-elevation-angle ratio spectra, moving the scattering source lower in the atmosphere decreases the absorption line strength because it decreases the absolute path length difference between the two pointings. Thus if our initial estimate of the distribution of scatterers tends to have those scatterers higher in the atmosphere than the true distribution, we will overestimate the absorption line strength for a given mixing ratio and therefore under-estimate the mixing ratio in our retrieval from an observed absorption line.

Recall that our estimate of the vertical distribution of dust aerosol is described by $H'_{dust}$, which is defined by (1), and it is initialized from TES climatology (Guzewich et al., 2013). The TES climatology is derived from a different Martian year, from measurements averaged over a region much larger than Gale Crater, and from TES measurements that don't resolve the bottom scale height of the atmosphere. We therefore expect our initial $H'_{dust}$ to be significantly in error and so we interpret deviations of the retrieved $CO_2$ mixing ratio from the expected 96% as information about the correct value of $H'_{dust}$. Note that aerosol properties other than the vertical distribution, for example particle size or ice fraction or total opacity, can change the effective vertical distribution of the scattering source function. We choose to vary $H'_{dust}$ simply because unlike those other aerosol properties it is not constrained by our continuum fits or by other measurements. Also note that the true value of the $CO_2$ mixing ratio does vary seasonally, but by less than 0.5 percentage points, which is negligible compared to our other measurement uncertainties.

Figure 6 shows the relationship between the retrieved water vapor mixing ratio and the retrieved $CO_2$ mixing ratio as $H'_{dust}$ is varied for a particular sky observation. Each point on this graph represents a repeat of the complete gas-mixing-ratio fitting procedure (including spatially variable background) described in section 2.9, but with different $H'_{dust}$. $CO_2$ and water vapor are seen to have a linear relationship, as do $CO_2$ and $O_2$, but the slope is different for different sky observations and different for different molecules. The slope is observed to be independent of the parameter being varied – varying particle size or ice fraction or opacity or surface albedo produces the same (to within measurement uncertainties) slope, and the correction to gas mixing ratios indicated by this slope is independent of the accuracy of these other variables and independent of our choice of $H'_{dust}$ as the parameter to be varied. The value of $H'_{dust}$ at which $CO_2$ reaches 96% does however depend on the other aerosol parameters and so our retrieved value for $H'_{dust}$ has some sensitivity to their accuracy.

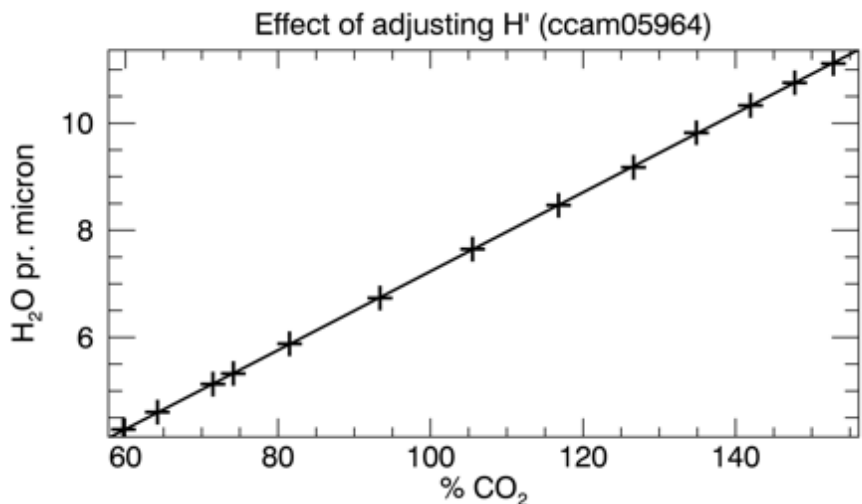

**Figure 6:** The effect of adjusting the dust relative scale height ($H'$) parameter on the solutions obtained for water vapor and $CO_2$. Each black cross gives the water vapor and $CO_2$ results for a particular value of $H'$, and the line shows the best linear fit.

The final step of every gas mixing ratio retrieval is to generate a graph as shown in Figure 6 and estimate: 1) the slope of the relationship between water vapor and $CO_2$ as well as $O_2$ and $CO_2$; 2) the retrieved $H'_{dust}$, which is determined by a quadratic interpolation to the point at which the $CO_2$ mixing ratio reaches 96%. In a few cases we cannot reach 96% $CO_2$ simply by varying $H'_{dust}$. This occurs when the dust opacity is low and the observed $CO_2$ absorption implies more low-altitude scattering than can be supplied by $H'_{dust}$. We resolve this by adjusting the ice aerosol vertical distribution, defining an $H'_{ice}$ parameter analogous to $H'_{dust}$. However once we introduce this additional parameter we can no longer obtain a unique solution for the aerosol vertical distribution with our current methodology. Thus we do not attempt to routinely explore the full $H'_{dust}$, $H'_{ice}$ parameter space and instead fix $H'_{dust} = 5$ and vary $H'_{ice}$ to reach 96% $CO_2$. In future work it may be possible to jointly constrain both $H'$ and $H'_{ice}$ by fitting both the continuum and the $CO_2$ absorption simultaneously, but for now we simply report these cases in our results as $H'_{dust} = 5$ which should be taken to mean a large contribution from near-surface scatterers whose detailed properties are indeterminate but likely different from higher altitude particles in terms of composition and/or size. Another reason that we do not attempt to fully explore the possible physical solutions or report a particular value when $H'_{ice}$ is needed is that our current definition of $H'_{ice}$ is flawed in terms of physical interpretation due to the fact that it uses $z_{ref} = 2H_{gas}$ just as does $H'_{dust}$ (see section 2.5). The ice extinction can be approaching zero at this level which forces $H'_{ice}$ to be very large (up to ~7) to contribute enough near-surface scattering.

The slope of the relationship between gas mixing ratios, $m_{H2O}$ (or $m_{O2}$) is used to produce a final estimate $(X_{H2O})_{final}$ by

$$(X_{H_2O})_{final} = (X_{H_2O})_{initial} + m_{H_2O}\left(0.96 - (X_{CO_2})_{initial}\right) \tag{3}$$

*or* by

$$(X_{H_2O})_{final} = (X_{H_2O})_{initial} + \langle m_{H_2O}\rangle\left(0.96 - \langle(X_{CO_2})_{initial}\rangle\right). \tag{4}$$

The $< >$ brackets indicate an average of the slopes and $CO_2$ mixing ratios from other individual sky observations within (in most cases) 10° of solar longitude. The average is restricted to other sky observations with similar low-elevation-angle pointing azimuths – those within 45°, and those in the same category of local solar time – either before 10:00 or after 10:00. For observations before 10:00 local solar time, other sky observations within 20° of solar longitude are used.

We employ the "smoothed" moving average version in Eq. (4) to diminish the effects of uncertainties and outliers in the $CO_2$ solution. Sky conditions could in principle change more rapidly than the averaging window we employ but in practice this effect doesn't seem to be significant – see Appendix B (B.8) for further discussion of this.

### 2.11  Gas mixing ratio uncertainties

We estimate that uncertainty of each gas mixing ratio fit for each sky observation by means of a Monte-Carlo model based on the covariances of the residuals of all the other fits for that gas in the data set. We start by assembling the residuals for the same set of pixels used in the current gas mixing ratio fit, scaling the residuals for each sky observation by a constant factor so as to make their root-mean-square equal to that of the observation being modeled. If the spatially-variable background correction was applied for the current gas mixing ratio fit we include only residuals of observations for which the spatially-variable background correction was also used, and if the spatially-variable background correction was not applied we include only residuals for which the spatially-variable background correction was also not applied. After assembling the set of scaled residuals, we calculate the covariance matrix of these scaled residuals, and then generate a random set of perturbations with the same covariance matrix as that of the scaled residuals. For each member of this ensemble of perturbations, we add it to the continuum-removed ratio spectrum for the observations being modeled and repeat the gas retrieval, including the $\gamma$ fit if the spatially variable background correction was used for the observation being modeled, as described in section 2.9. We adopt the standard deviation of the resulting ensemble of gas mixing ratios as the standard error for the gas mixing ratio.

In calculating the final uncertainty for water vapor, the uncertainties for the initial estimates of water vapor and $CO_2$ are propagated through Eq. (3) or Eq. (4). The uncertainty for $H'$ is determined from the Monte-Carlo uncertainty of $CO_2$ and the slope of observed $H'$ vs. $CO_2$ curve at 96% $CO_2$.

## 3   Sensitivity Analysis

We quantify the sensitivity of our results to uncertainties in our input assumptions and constraints by performing a range of perturbation experiments on those inputs. We have tested perturbations to the Mastcam-derived total column opacity, to the weight given to the dark endmember for the surface albedo spectrum, to the initial guess vertical profile of aerosols, to the width of the assumed ChemCam line-spread function, and to the molecular absorption parameters used by our radiative transfer model. We have also tested alternative aerosol and temperature vertical profiles drawn from the Mars Climate Sounder (MCS) data set (Kleinböhl et al., 2009) instead of the TES data set. The results of these perturbation experiments are shown in Table 3 for two different passive sky observations, one on sol 1275 that had low total opacity, high water ice opacity, and an extreme value of the initial guess of $H'$, and one on sol 975 with high dust opacity, minimal contribution from water ice, and a more typical initial guess for $H'$. The $H'_{ice}$ parameter was required for sol 1275 but not for sol 975. For each observation, the first row of entries shows the nominal case with no perturbation. For the values that depend on the gas absorption lines – $H'$ and water vapor abundance – these first rows give the uncertainties calculated as described in section 2.11. Both of these observations are a near-best case in terms of goodness-of-fit for the water vapor and carbon dioxide absorption lines which makes their uncertainties for water vapor and $H'$ or $H'_{ice}$ relatively small. Therefore perturbations that cause changes smaller than the "±" uncertainties provided in the first rows for each observation are of relatively minor concern, although they are still of interest if they suggest persistent biases.

### 3.1   Sensitivity to Mastcam-measured opacity

We use perturbations to the Mastcam-measured opacity of +5% and -5% because the 10 percentage point difference between these values exceeds the standard errors of the Mastcam measurement in all cases while being an approximate upper limit for the magnitude of interpolation errors when a Mastcam observation is not available on the same day as the ChemCam passive sky observation. The results of these experiments show that the Mastcam opacity uncertainties have no effect on the dust aerosol particle size retrieval. They also show a very small effect on the dust-versus-ice opacity fraction – a 3% change in response to a 10% change in the assumed input total opacity. Of course when the dust opacity fraction is large even these small errors lead to large relative errors in the magnitude of the ice contribution to opacity. In these cases where ice makes only a very small contribution to the observed spectrum, the effect of Mastcam opacity errors on ice aerosol fraction becomes significant (20% change in response to a 10% Mastcam error), but the effect becomes negligible when the ice contribution becomes substantial.

**Table 3**: Perturbation experiments for sensitivity analysis

| Perturbation experiments for sol 1275, LTST 12:46 | Inputs | | | Results | | | | | |
|---|---|---|---|---|---|---|---|---|---|
| | Initial $H'$ | Surface lambert albedo at 800 nm | Mastcam opacity ($\tau$) at 880 nm | Dust fract-ion | Dust $r_{eff}$ ($\mu$m) | Ice $r_{eff}$ ($\mu$m) | $\chi^2$ for Aero-sol | $H'$, $H'_{ice}$ [2] | Water Vapor pr. $\mu$m [1] |
| Nominal | -2.40 | 0.288 | 0.42 | 0.67 | 0.60 | 2.5 | 0.14 | 5.0, 5.0 ±0.5 | 9.3 ±1.1 |
| $\tau$ increased by 5% | -2.40 | 0.288 | 0.44 | 0.65 | 0.61 | 2.6 | 0.14 | 5.0, 5.0 | 9.3 |
| $\tau$ decreased by 5% | -2.40 | 0.288 | 0.40 | 0.67 | 0.60 | 2.5 | 0.14 | 5.0, 5.1 | 9.3 |
| 0% dark region albedo | -2.40 | 0.301 | 0.42 | 0.75 | 0.56 | 4.0 | 0.16 | 4.4, - | 9.5 |
| 20% dark region albedo | -2.40 | 0.275 | 0.42 | 0.61 | 0.71 | 4.0 | 0.09 | 5.0, 5.2 | 9.3 |
| $H'$ = 1.0 | 1.01 | 0.288 | 0.42 | 0.62 | 0.60 | 4.0 | 0.30 | 5.0, 5.2 | 9.4 |
| $H'$ = 0.0 | 0.01 | 0.288 | 0.42 | 0.70 | 0.58 | 4.0 | 0.14 | 5.0, 4.8 | 9.4 |
| MCS profile (MY 29) | 0.43 | 0.288 | 0.42 | 0.70 | 0.56 | 3.7 | 0.19 | 5.0, −0.7 | 9.3 |
| Line spread FWHM = 0.47 nm | -2.40 | 0.288 | 0.42 | 0.67 | 0.60 | 2.5 | 0.14 | 5.0, 4.9 | 9.6 |
| If -15% $CO_2$ line intensity error[4] | -2.40 | 0.288 | 0.42 | 0.67 | 0.60 | 2.5 | 0.14 | 5.0, 5.7 | 11.0 |
| If +15% $CO_2$ line intensity error[4] | -2.40 | 0.288 | 0.42 | 0.67 | 0.60 | 2.5 | 0.14 | 5.0, 0.0 | 8.0 |

| Perturbation experiments for sol 975, LTST 10:10 | Inputs | | | Results | | | | | |
|---|---|---|---|---|---|---|---|---|---|
| | Initial $H'$ | Surface lambert albedo at 800 nm | Mastcam opacity ($\tau$) at 880 nm | Dust fract-ion | Dust $r_{eff}$ | Ice $r_{eff}$ | $\chi^2$ for Aero-sol | $H'$ [2] | Water Vapor pr. $\mu$m [1] |
| Nominal | 0.37 | 0.288 | 1.21 | 0.96 | 1.6 | 2.1 | 0.19 | 1.4 ±0.1 | 4.2 ±0.4 |
| $\tau$ increased by 5% | 0.37 | 0.288 | 1.27 | 0.97 | 1.7 | 1.7 | 0.19 | 1.2 | 4.2 |
| $\tau$ decreased by 5% | 0.37 | 0.288 | 1.15 | 0.94 | 1.6 | 2.0 | 0.19 | 1.5 | 4.2 |
| 0% dark region albedo | 0.37 | 0.301 | 1.21 | 0.94 | 1.6 | 2.4 | 0.20 | 1.4 | 4.3 |
| 20% dark region albedo | 0.37 | 0.275 | 1.21 | 0.96 | 1.7 | 1.7 | 0.19 | 1.3 | 4.2 |
| $H'$ = 1.0 | 1.01 | 0.288 | 1.21 | 0.95 | 1.6 | 2.1 | 0.20 | 1.4 | 4.2 |
| MCS profile (MY 28) | 0.44 | 0.288 | 1.21 | 0.95 | 1.7 | 2.0 | 0.19 | 1.4 | 4.2 |
| $H'$ = 0.0 | 0.01 | 0.288 | 1.21 | 0.96 | 1.7 | 2.0 | 0.19 | 1.4 | 4.2 |
| Line spread FWHM = 0.47 nm | 0.37 | 0.288 | 1.21 | 0.96 | 1.6 | 2.1 | 0.19 | 1.2 | 4.5 |

| | | | | | | | | |
|---|---|---|---|---|---|---|---|---|
| If -15% CO$_2$ line intensity error[4] | 0.37 | 0.288 | 1.21 | 0.96 | 1.6 | 2.1 | 0.19 | 1.9 | 5.0 |
| If +15% CO$_2$ line intensity error[4] | 0.37 | 0.288 | 1.21 | 0.96 | 1.6 | 2.1 | 0.19 | 0.9 | 3.7 |

Notes: 1. The water vapor pr. micron values in this table *are not* scaled by surface pressure.
2. If only one value is given in the *H'*, *H'*$_{ice}$ or *H'* columns, the given values it refers to *H'* and the *H'*$_{ice}$ was not used in the retrieval because a successful solution was found with *H'* alone.
3. For the experiments were the water ice particle size was forced to a certain value, the "$\chi^2$ for Aerosol" parameter is omitted because no aerosol fitting was performed.
4. The values are the results *after correcting for* the given line intensity error.

Considering the Mastcam-opacity effect on the gas absorption line, we see the expected effect on the CO$_2$ absorption band depth in the form of a statistically significant change in the retrieved vertical profile parameter *H'*$_{dust}$. Also as expected, there is no effect on retrieved water vapor column abundance because we use the CO$_2$ absorption band to correct for aerosol effects on the water vapor absorption bands (section 2.10).

### 3.2    Sensitivity to assumed surface albedo spectrum

To represent the uncertainty in surface albedo spectrum due to the diversity of surface characteristics in the vicinity of MSL, we include an experiment with the dark region endmember contributing 20% to the surface spectrum instead of the nominal 10%. Note that as previously discussed (section 2.4.1) 10% – 20% dark region endmember contributions span the range of plausible surface albedos. We include a 0% dark region experiment for reference but this should not be taken as a likely scenario.

Using the 20% dark region endmember has very little effect when the atmospheric opacity is high (as in the sol 975 sensitivity test)– the only effect is a 20% change in the particle size retrieved for the very small ice aerosol contribution. When atmospheric opacity is low the surface albedo effect is significant – in our 1275 perturbation experiment it lowers the dust contribution to opacity by 9% which corresponds to an 18% increase in the ice contribution. However these changes represented changes to absolute opacity contributions of only ~0.03, and so even though they are likely persistent systematic effects that depend on overall opacity and the direction the passive sky observation was pointing they are small compared to the main seasonal trends. They do limit our ability to interpret diurnal variations. The change in dust *r*$_{eff}$ in our 20% dark region endmember experiment is similarly modest – it is an 18% increase but the 0.11 micron change is small compared to observed seasonal variations. The change in ice *r*$_{eff}$ is much more important – our experiment has it increasing from 2.5 to 4.0 microns which is similar in magnitude to the apparent variations in our data set. This means that our current aerosol retrieval procedures are only providing a limited amount of information about ice particle sizes.

Our surface albedo perturbation experiments show no significant effect on the vertical profile parameter and no effect on water vapor.

### 3.3    Sensitivity to the initial guess vertical profile

We test the *H'* = 1.0 case of dust extinction being uniformly mixed with atmospheric gasses by mass, and the *H'* = 0 case of dust being uniformly mixed by volume. When these tests represent modest changes from the nominal initial guess as for our sol 975 experiments, the perturbation has no effect on any results. When the change in initial guess is very large as for sol 1275, we see modest changes in aerosol fraction and large changes in ice aerosol particle size, similar to the case of perturbed surface albedo with low total opacity. In this case however we don't see much effect on dust aerosol particle size.

The significance of these vertical profile sensitivities is similar to that of the surface albedo sensitivities. It is a minor effect relative to seasonal changes and a cause for some concern in interpreting diurnal behavior. It does add an additional reason to not believe our current ice particle size results, but only when the initial guess for $H'$ differs greatly from what we ultimately retrieve.

We observe no significant effect of the initial guess vertical profile on the retrieved $H'$ and no effect on retrieved water vapor.

### 3.4    Sensitivity to the ChemCam VNIR line-spread function

As discussed in Appendix B section B.2 there is some modest uncertainty as to the width and shape of ChemCam's line-spread function in passive mode. Considering the estimated upper limit of 0.47 nm FWHM for the width of the ChemCam VNIR line spread function (Appendix B, B.2), we get (as shown in Table 3) a column water vapor abundance ~0.3 precipitable microns larger than for our nominal case of 0.42 nm FWHM. The size of the absolute change appears to be independent of the size of the nominal column abundance value. Additional experiments with the line-spread function (not shown) show that the absolute change in water vapor column abundance is proportional to the percentage change in line-spread function FWHM. This sensitivity results from the effects of the line-spread function on the peak absorption for a given column of gas. The effect appears to be slightly larger for water vapor than for $CO_2$ presumably because of differences in the intrinsic widths of the absorption lines we are using, and this difference leads to the sensitivity that we observe after our $CO_2$-based correction (section 2.10). Since the slight line-shape mismatch in modeled-vs.-observed line shape that we noted in section 2.5 also affects the peak-to-area ratio of the lines in a fashion similar to the FWHM, the 3.5% difference in area we noted represents an estimated potential effect equivalent to an additional 3.5 percentage-point increase in the FWHM (less than one-third as large as 12 percentage point increase already considered). So the true uncertainty in the water vapor column due to uncertainty about the line spread function details is slightly larger than 0.3 precipitable microns before scaling to a reference surface pressure. (It is slightly less than 0.3 precipitable microns after scaling to a reference surface pressure and less than 4 ppm in terms of column-averaged volume mixing ratio.) This effect is always smaller than the statistical uncertainty of individual water vapor measurements, but since it applies systematically to all passive sky observations it could in theory be of interest when comparing large-scale averages with other data sets.

The effect of the wider line spread function on the $H'_{dust}$ solution turns out to be statistically significant in the best-case retrieval of $H'_{dust}$ represented by our sol 975 observations. This is not surprising since the effect on peak line intensity in this case is not mitigated by reference to another absorption line as it is for water vapor. However since $H'$ is by nature a rough approximation of the vertical profile and since the other perturbations we consider have comparably large or larger effects, the line-spread-function-induced uncertainty in $H'$ is not an important issue.

### 3.5   Sensitivity to HITRAN molecular absorption parameter uncertainties

As discussed in Appendix B section B.2, the molecular absorption parameters that we take from the HITRAN 2008 database (Rothman et al., 2009) have significant uncertainties. Appendix B section B.2 also describes our approach to calculating the sensitivities to these uncertainties. Note that we are assuming a worst case 100% correlation in the errors of individual lines; but we will also assume that errors in $CO_2$ lines are independent of errors in $H_2O$ lines.

The impact of the $CO_2$ line parameter uncertainties is shown on Table 3 – the values we show are the result of hypothesizing that the $CO_2$ line intensities were in error by ± 15% and then correcting for that error. For example the "If -15% $CO_2$ line intensity error" row and "water vapor pr. μm" column of the sol 1275 portion of the table should be read as: "If the HITRAN 2008 $CO_2$ line intensities that we used for our nominal retrieval were all 15% too low then the correct value for water vapor on sol 1275 is actually 11 pr. μm instead of the 9.3 pr. μm from our nominal retrieval." $CO_2$ line parameter uncertainties affect both water vapor results and $H'$ results.

Table 3 shows that correcting water vapor results for hypothetical $CO_2$ line intensity parameter errors could make water vapor data points lower or higher by between −14% – +18%. The direct effect of the water vapor line intensity parameter uncertainties on water vapor is relatively simple and relatively small and therefore not shown on Table 3: by itself it contributes ± 7.5% but makes the combined uncertainty only slightly larger given our assumption that it is statistically independent of $CO_2$ line intensity errors. The combined line intensity parameter uncertainty means that corrected water vapor results could be globally higher or lower by between −16% – +20%. Whatever the error might be, every data point must be affected by the same percentage (which is what is meant by "globally higher or lower"). Thus for water vapor the uncertainty only affects comparisons with other data sets. This uncertainty has a minimal impact on our conclusions in this paper due to the existing uncertainties in other data sets, but it could become significant for future comparisons.

For $H'$ the magnitude of the $CO_2$ line parameter uncertainty effect is variable for a given assumed line intensity parameter error because the relationship between $H'$ and $CO_2$ is non-linear and depends on the details of the aerosols. We observe changes of +0.5 – +2 or −0.5 – −1 when correcting −15% or +15% line intensity parameter errors, respectively. (The example in Table 3 turns out to be a best-case scenario for the $H'$ uncertainty.) The sign of the effect is however the same for all observations and the variations in magnitude show no clear temporal pattern, so the impact of the line parameter uncertainty is almost entirely on comparisons with other data sets. (Note that when $H'_{ice}$ is used the numerical effect of $CO_2$ line parameter uncertainty on $H'_{ice}$ could be very large as seen in Table 3 but this is not physically meaningful because of the inherent flaws in $H'_{ice}$ that were described in section 2.10.)

### 3.6   Sensitivity to temperature and aerosol vertical profile inputs

To assess the sensitivity of our results to vertical profile inputs we experimented with substituting MCS-derived profiles for our usual TES-derived profiles. Using MCS-derived temperature profiles had no effect on any of our retrieved quantities, and there was still no effect even when we perturbed our temperature profiles by twice the difference between a TES profile and the corresponding MCS profile. Likewise, MCS-derived aerosol profiles had no effect on water vapor retrievals. MCS-derived aerosol profiles did have a small effect on retrieved aerosol parameters in the icy, low-opacity sol 1275 case, but that effect is mostly similar to the effect of using different vertical profile parameters. See Appendix A section A.4.2 for further discussion and examples of MCS vs. TES aerosol profiles, including a discussion of the small differences between our procedure for generating MCS-derived profiles and our procedure for generating TES-derived profiles.

One effect of the MCS profiles that might appear significant is the large change to the $H'_{ice}$ parameter in the low-opacity sol 1275 case, but this is in fact not physically meaningful due to the limitations of our $H'$, $H'_{ice}$ retrieval (see section 2.10). What is actually happening with the $H'_{ice}$ parameter in the MCS-derived sol 1275 case is that a modest amount of additional extinction in MCS relative to TES near 2 scale heights above the surface is coincidentally deactivating the flaw in our $H'_{ice}$ methodology (see section 2.10). (Recall that this flaw produced very high values of $H'_{ice}$ when the original ice profile was very small near 2 scale heights.) Even with this flaw deactivated we do not get a unique value for $H'_{ice}$ because of the degeneracy with $H'_{dust}$ solution, so the new $H'_{ice}$ is still not physically meaningful.

### 3.7   Vertical averaging kernel

We use a different type of sensitivity analysis to assess the sensitivity of the water vapor column abundance that we retrieve to the real vertical distribution of water vapor mixing ratio. The objective of this analysis is to derive the averaging kernel for water vapor, which describes, as a function of altitude, the scaling factor between a change in the quantity of water vapor at a given altitude and the change in the retrieved quantity of water vapor. In other words our observation process produces a weighted average of water vapor mixing ratio as a function of altitude, and the averaging kernel is the vector of weights. We can apply this averaging kernel to a proposed model distribution of water vapor to determine what we would observe, and we can apply it to a hypothesized change in water vapor at a certain altitude to determine whether it would be observable.

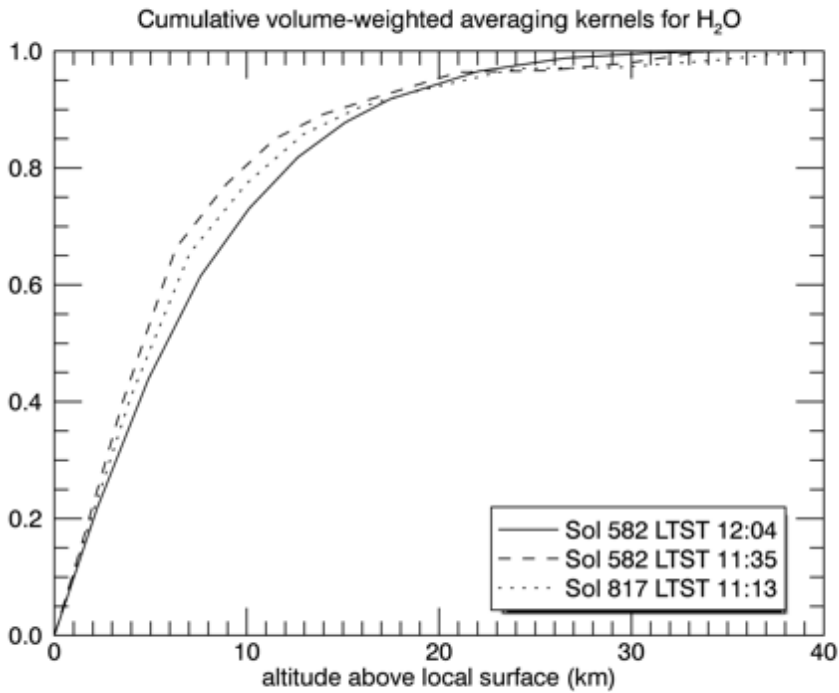

**Figure 7:** Cumulative volume-weighted vertical averaging kernels for water vapor for three different ChemCam passive sky observations. The y-axis is unitless cumulative response. These observations are a typical high-opacity ($\tau = 1.19$) case – sol 817 LTST; a typical low-opacity ($\tau = 0.46$) case – sol 582 @ 11:35 LTST; and a case with low opacity ($\tau = 0.46$) but also westward-looking geometry and unusually low near-surface scattering i.e. unusually low H'– sol 582 @ 12:04.

These vertical averaging kernels are constructed by generating a forward model solution for the continuum-removed ratio spectrum, performing the gas retrieval on the model spectrum, and then generating a perturbed forward model and repeating the retrieval. All of these retrievals use the same assumptions as the original retrieval. The perturbation is to the water vapor mixing ratio at a single model level, and we repeat

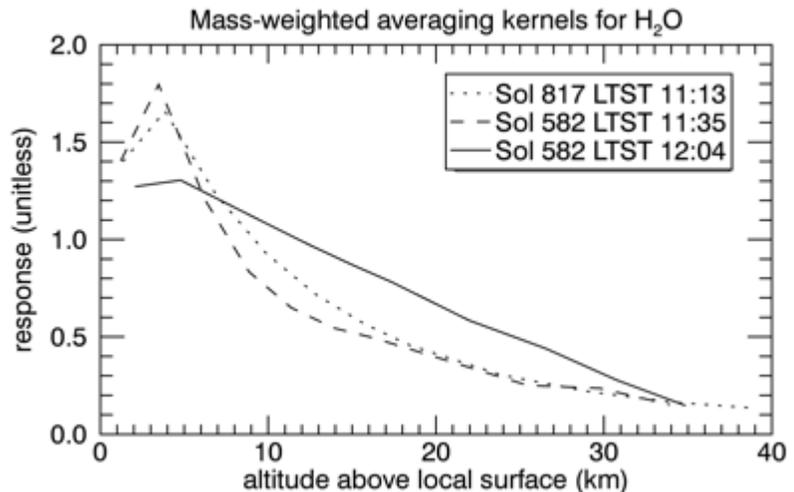

**Figure 8**: Mass-weighted vertical averaging kernels for water vapor for three different ChemCam passive sky observations. The three cases plotted are identical to those in Figure 7.

the process for a series of model levels to build up the complete kernel.

We present the averaging kernel in Figure 7 as a cumulative volume-weighted averaging kernel. Figure 7 assumes that our result for a given observation is expressed as a column-averaged volume mixing ratio and shows the cumulative weights of the vertical averaging. We can see from the plots that, for example, the mixing ratio in the bottom 5 km above the local surface has a 40% − 55% weight in the reported result, depending on the details of the opacity profile and observing geometry, and that the bottom 1 km has a weight of about 10%. Although the shape of these weighting functions expressed as cumulative volume-weighted kernels is dominated by the overall decrease in atmospheric density with height, when expressed as mass-weighted averaging kernels (Figure 8) there are indeed substantial vertical variations – in all cases the mass response falls off to near zero between 30 and 35 km above the surface, and peaks between 3 and 5 km above the surface.

# 4   Results

## *4.1   Data selection*

Out of a data set of 136 ChemCam passive sky observations acquired prior to sol 1293 (the cutoff for MSL's August 2016 PDS data release), 113 pass our quality controls for inclusion in the results reported here. These quality controls eliminate observations with saturated pixels at wavelengths greater than 690 nm, observations where the continuum-fit for aerosol properties failed to converge to $\chi^2 < 4$, and observations for which no solution for $H'$ (or $H'_{ice}$) was found. The maximum continuum-fit $\chi^2$ of 4 is chosen based on the observation that successful fits have an average $\chi^2$ of 0.20 with only two instances larger than 0.5. (Note that the uncertainty assigned to the continuum-data points was arbitrary so the small values of $\chi^2$ are not cause for concern.) We also eliminate observations for which the combination of noise and detector background signal is too large, based on the standard deviation of the continuum-removed ratio in an absorption-line free portion of the spectrum between 800 nm and 810 nm. We choose 0.002 as the threshold standard deviation above which we reject the observation. In addition, we reject observations when the reduced-$\chi^2$ of any of the gas absorption line fits exceeds 20. These last two thresholds are chosen empirically to filter out observations whose error bars would be so large that they don't contribute useful information. All of the observations that do not meet these two thresholds occurred prior to sol 790, before we started placing a high priority on matching the sky brightness at the low- and high-elevation-angle pointings.

We accept reduced-$\chi^2$ as large as 20 because we believe that we understand the process that causes the excess data variance – spatially variable detector background – and because our uncertainty estimates (section 2.11) take that excess variance into account. The largest reduced-$\chi^2$ values only occur when a large amount of time, more than ~100 sols, separates the sky observation from a calibration observation, meaning that changes in the detector background pattern over time are an important contributor to the excess variance.

Note that in this paper we use the now-conventional definition of numbering Mars Years (MY) proposed by Clancy et al. (2000), in which MY 1 starts at Mars $L_s = 0°$ on April 11, 1955. MSL landed in Gale crater just after MY 31, $L_s=150°$. The ChemCam data set that we present here extends from MY 31 $L_s = 291°$ to MY 33 $L_s=127°$.

### 4.2    Water vapor

Figure 9 shows the complete data set for water vapor column abundance with corrections supplied by smoothed scaling to $CO_2$ measurements (Eq. 4). (See Appendix B section B.8 for a discussion of smoothed scaling vs. non-smoothed scaling.) Also included in this figure for comparison is a representation of the CRISM-derived Gale-localized water vapor retrievals from Toigo et al. (2013). This figure shows precipitable microns scaled to a 6.1 mb surface pressure, as is traditionally done in the Mars water vapor literature. Note that the line from Toigo et al. (2013) is their Fourier-analysis-based fit shown in their figure 15 (but re-scaled to 6.1 mb reference pressure) and contains only annual and semi-annual components. It was derived from data points spanning MY 28 through early MY31 and is based on the statistics of individual ~20-meter-scale CRISM pixels (cf. Toigo et al. 2013 section 4.2.2) within a 136° – 140° East and 4° – 7° South box (A. Toigo, personal communication).

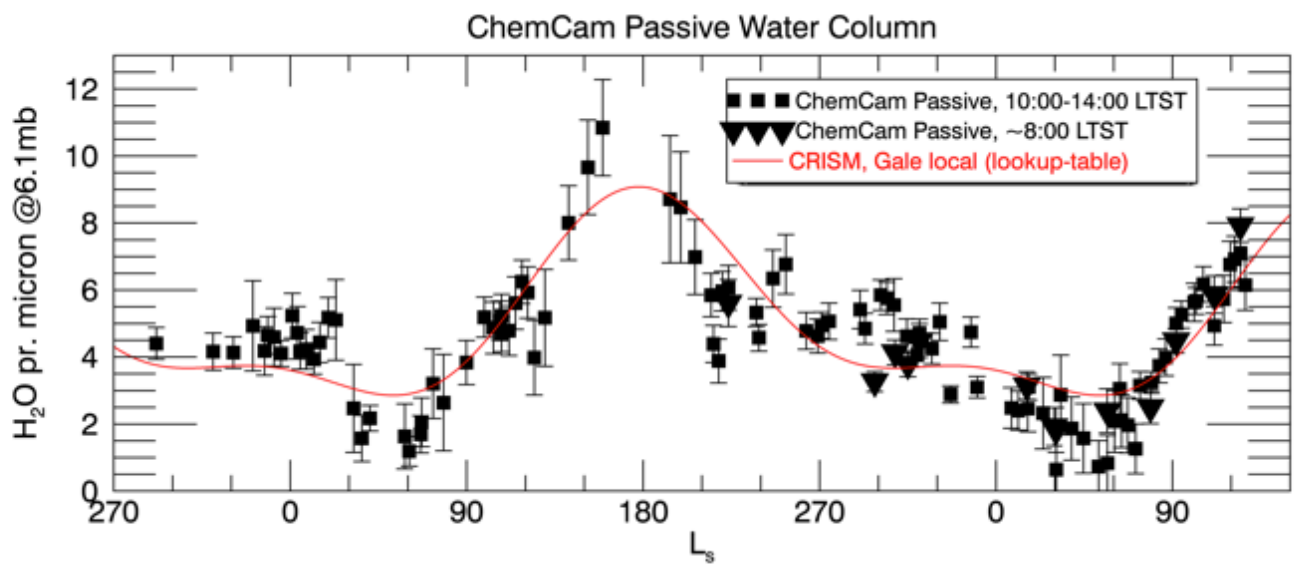

**Figure 9:** Water column abundance from: ChemCam passive sky with smoothed $CO_2$ scaling, (black square and triangles); Toigo et al.'s (2013) two-component fit to Gale-crater-localized CRISM lookup-table-based retrievals. ChemCam results from local-true solar times earlier than 10:00 are plotted with downward-facing triangles, and those with local-true solar times between 10:00 and 14:00 are plotted with squares. Water column is scaled to a 6.1 mb surface pressure. The x-axis begins in Mars Year 31 and continues into Mars Year 33.

### 4.2.1   Comparison to Gale-Crater-localized, look-up-table-based CRISM retrievals

The median uncertainty of the ChemCam passive sky $H_2O$ data set shown in Figure 9, which has smoothed $CO_2$ scaling and the column abundance scaled to a 6.1mb pressure surface, is +/- 0.60 precipitable microns. (This would increase only slightly to +/- 0.65 if non-smoothed $CO_2$ scaling was used.) So far the ChemCam $H_2O$ data set shows no significant interannual changes, except for the period near $L_s$=15°. It also closely matches the Toigo et al. (2013) two-Fourier-component fit to CRISM look-up-table-based water vapor results, although with two interesting exceptions. (Note that Toigo et al.'s (2013) two-component shows only negligible differences from their 60° of Ls wide boxcar moving average of Gale-Crater-localized lookup-table-based water vapor results.) One interesting exception is the $L_s$=30° to $L_s$=70° period in both MY 32 and MY 33. Other exceptions occur near $L_s$=220° and near $L_s$=300° in MY 32. The interannually repeating difference between $L_s$=30° and $L_s$=70° represents a factor of almost two less water vapor than implied by the Toigo et al. (2013) look-up-table-based CRISM data, and suggests a significant climatological minimum that is resolved only by ChemCam. The differences at $L_s$=220° and $L_s$=300°, could be connected with interannual changes in circulation during these variably dusty seasons, or with differences in the observability of water vapor at different wavelengths (720 and 825 nm vs. 2600 nm) and from different viewpoints (surface vs. orbit) during high opacity periods.

### 4.2.2   Comparison to other orbital data sets

To better understand the significance of these differences, we plot in Figure 10 all Mars Years of the ChemCam water vapor on a single 0° – 360° x-axis and then add multiple versions of CRISM and Mars Global Surveyor (MGS) Thermal Emission Spectrometer (TES) water vapor column abundance data sets. The TES data are the retrievals presented by Smith (2002, 2004) except that they are the reprocessed version of those retrievals, which have updated $H_2O$-in-$CO_2$ line-broadening coefficients and which have been used in more recent inter-dataset comparisons such as Maltagliati et al. (2011). The CRISM data include the previously discussed Toigo et al. (2013) lookup-table-retrieval two-Fourier-component fit and results from the original CRISM "full" retrieval of Smith et al. (2009). For both TES and the CRISM full retrievals we have smoothed the data sets with a 20° of $L_s$ FWHM Gaussian-weighted moving average. For the CRISM retrievals we use only the first three Martian years of data because data points become relatively sparse after that. For purposes of comparison we define the "local" area of Gale crater as a latitude-longitude box ranging from 136° – 140° East and 4° to 7° South, matching the definition (A. Toigo, personal communication) used for the Toigo et al. (2013) results. We also define a Gale crater "region" as 120° – 160° East, 0° to 10° South, and an "all longitudes" zonal-average as 0° – 360° East, 0° to 10° South. We have experimented with alternative definitions of the Gale crater "region", but we find that as we do so the resulting curves simply vary smoothly between those of the "local" and "all longitudes" extremes. It should be noted that each individual TES observation and each individual "full retrieval" CRISM observation samples a surface area – ~10 km x ~10 km for TES (Smith, 2004) and 2 km x 2 km for CRISM (Smith et al., 2009) – that is quite large compared to the individual CRISM pixels used by the CRISM lookup-table retrieval. However these individual-observation footprints are still small compared to the size of our "local" and "regional" area definitions, so even though we filter TES and CRISM data points based on their center lat.-lon. the individual-observation footprint sizes don't significantly alter the size of the area effectively sampled by the averages.

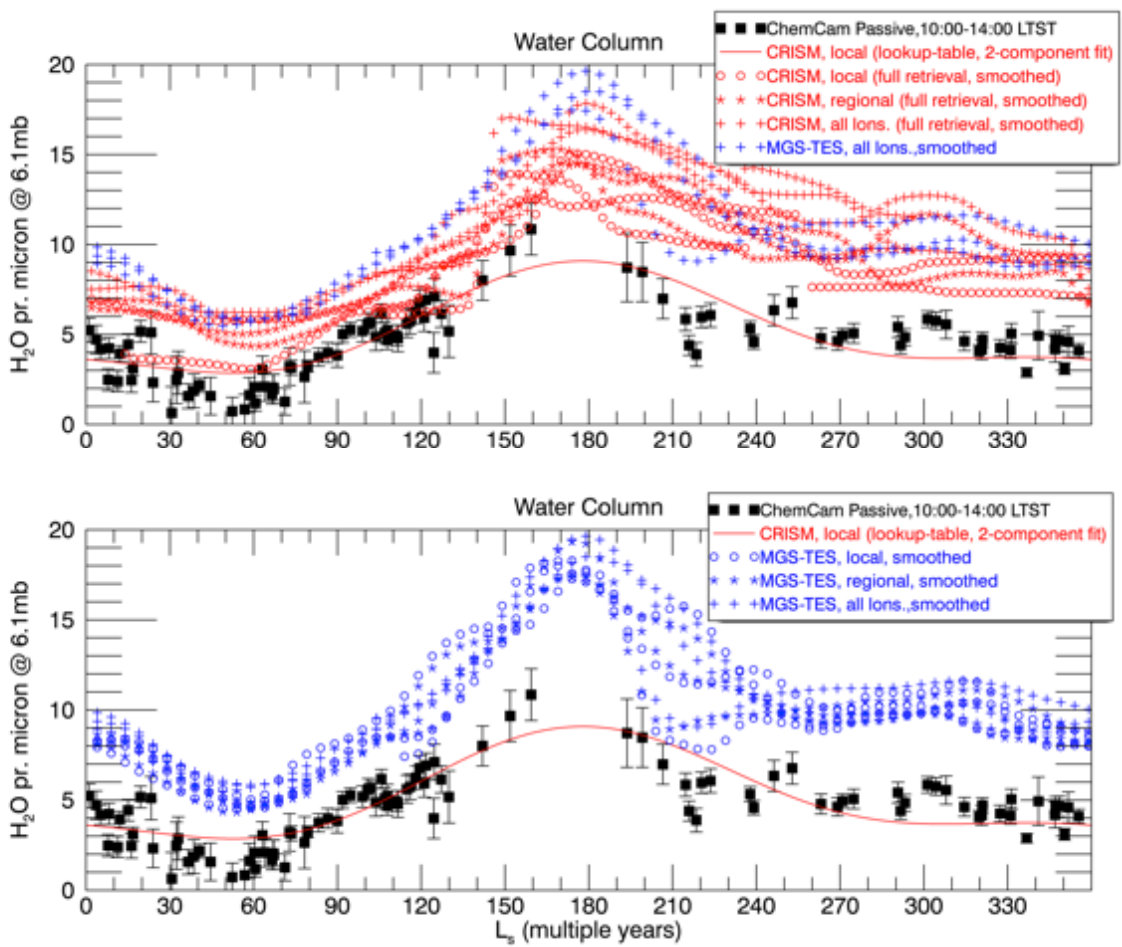

**Figure 10:** Water column from ChemCam passive sky (black squares) and from Toigo et al.'s (2013) two-component fit to Gale-crater-localized CRISM lookup-table-based retrievals (red line) compared with various representations of CRISM and TES data sets. The details of these data sets are described in the text. All Mars years are plotted on the same x-axis. The ChemCam data plotted here is limited to results from 10:00 to 14:00 LTST. Water column is scaled to a 6.1 mb surface pressure. The two panels are identical except that some data sets are omitted from each panel for clarity.

The most striking feature of Figure 10 is that ChemCam consistently shows less water vapor than the orbital data sets, except for the Gale-localized lookup-table-derived CRISM and perhaps the Gale-localized full-retrieval CRISM data sets    in some seasons. It is also apparent that, despite the differences in magnitude, and with one notable ChemCam exception in the previously mentioned $L_s$=30° to $L_s$=70° period, the seasonal pattern and periods of significant inter-annual variability are essentially identical in all of the data sets. This is true at zonal, regional, and local scales and the observed patterns are broadly consistent with the now well-known global scale features of the Martian water cycle. (These well-known features include the aphelion minimum, the maximum at northern fall equinox when the pulse of vapor from the northern polar cap reaches the equator, and the intermediate but variable levels during the dusty perihelion season. See, e.g., Smith (2004), Smith et al. (2009), Maltagliati et al. (2011).) Also note that there is a modest trend in both the TES and CRISM data sets towards less water vapor in Gale regional data as compared to the zonal-average data. The significant differences among data sets in the magnitude of water vapor abundance appear to occur only at the local scale.

We would like to know whether and to what extent the persistent relatively low water vapor abundances are a feature of Gale Crater rather than a feature of the ChemCam data set. However the differences among the various local-scale data sets present some difficulties: the lookup-table-derived CRISM matches the ChemCam data set but isn't well corroborated by the Gale-localized full-retrieval CRISM time series, and meanwhile the Gale-localized TES data set is no different from regional-scale TES data set.

There are two issues with the two local-scale CRISM time series that may explain why they differ. The Gale-localized full-retrieval CRISM time series is based on only a small number of individual retrievals (82 spanning three Mars years) and thus even the smoothed data is heavily influenced by some combination of random spatial sampling biases or simply random retrieval or instrumental uncertainties. The harmonic fit to CRISM lookup-table-based data that we have taken from Toigo et al. (2013) is more statistically robust because of the small number of harmonics and large number of individual CRISM pixels that it uses. The other local scale CRISM issue is a limitation of the lookup-table retrievals: according to Toigo et al. (2013, c.f. their Fig. 4) there may be some modest look-up-table-induced negative biases relative to the full retrievals, but these look-up-table biases are very likely (based on the Toigo et al. [2013] figure) no more than $1-2$ precipitable microns. At the low end where the look-up table water vapor values are near ~3 precipitable microns the biases are very likely less than 1 precipitable micron. Overall we can conclude that the water vapor depletion is most likely a real Gale Crater feature because the upper limit of likely CRISM look-up table biases won't substantially change the agreement with ChemCam. However further work to reconcile the CRISM data sets will be needed to verify this conclusion.

The disagreement between ChemCam water vapor and the TES Gale-localized water vapor also needs to be considered. Despite being essentially the same as the CRISM data at zonal and regional scale, TES data at local scale does not follow CRISM data in showing lower water vapor and hence is not similar to ChemCam data. A similar unexplained phenomenon in local-scale TES data occurs at the Opportunity rover site where essentially the entire data set of upward-looking water vapor retrievals from the Mars Exploration Rover's (MER) mini-TES instrument show substantially less water vapor than is seen by TES over that location (Smith et al., 2006). (This occurs to at the Spirit rover site as well but to a lesser extent and only during $L_s=30°-180°$.) Even without explanation this observation indicates that local-scale depletion can occur without being observable by TES and thus it is reasonable for us to conclude that that is likely what is happening at Gale. Obviously an explanation for why TES is unable to observe the depletion phenomenon would strengthen the case that it is real, but that is beyond the scope of this paper as it will require a new research effort to evaluate TES and mini-TES averaging kernels and reanalyze TES and mini-TES data sets.

The previously described ChemCam minimum from $L_s=30°$ to $L_s=70°$ continues to stand out in that none of the other data sets in Figure 10 show anything approaching the $> 50\%$ drop in water abundance that ChemCam sees repeatedly near $L_s=30°$, and none of the other data sets in Figure 10 have absolute water column abundances as low as $1-2$ precipitable microns. However the upward looking mini-TES (Smith et al., 2006) observations, observing in the vicinity of the Gusev crater and Meridiani Planum landing sites, do see water column abundances in this very low $1-2$ precipitable microns range in the same $L_s=30°$ to $L_s=70°$ season. Smith et al. (2006) do not offer any explanation for these very low column abundances periods but they are important because they indicate that the phenomenon is not unique to ChemCam or to Gale Crater.

Meanwhile the $L_s$=220° and $L_s$=300° periods where ChemCam diverges from the CRISM harmonic fit are now clearly seen to be periods of significant interannual variability, making these differences unsurprising even though we still don't know whether they represent interannual variations in actual water vapor mass or instead variations in the visibility of the water vapor mass through obscuring aerosol. A statistically significant sample of concurrent ChemCam and CRISM measurements could potentially resolve this question due to their very different viewing geometry, but no Gale-local CRISM observations exist in the relevant season within the span of the ChemCam time series. Detailed analysis of TES and CRISM vertical averaging kernels for water vapor including the response of those kernels to dust loading may also be helpful in resolving this question.

### 4.2.3   *Error due to non-uniform vertical distributions of water vapor*

A complication in comparing our ChemCam results with TES and CRISM is that in our ChemCam retrievals we treat water vapor as being uniformly mixed while the Smith (2002, 2004), Smith et al. (2009) and Toigo et al. (2013) TES and CRISM retrievals assume a water vapor distribution capped by a saturation level. Our averaging kernels (Figures 7 and 8) let us estimate the significance of this difference in distribution. Based on Maltagliati et al. (2011) the saturation level can be as low as 7 km above the geoid or roughly 10 km above the local surface at Gale. If all of a given column mass of water vapor was confined to but uniformly mixed within the bottom 10 km, then reading the cumulative weight of 0.77 for 10km altitude off of Figure 7 and approximating the scale height as 10km, we find that our retrieval would return 1.22 times as much water as it does for the uniformly mixed case. (For the sake of simple calculations we are making the approximation that the mixing ratio drops to zero at the saturation level – this is an acceptable approximation because of the combination of steep lapse rates and the roughly exponential saturation vapor pressure curve.) In other words if the true column-averaged volume mixing ratio is 100 ppm we would retrieve a column-averaged mixing ratio of 100 ppm if the water vapor was in fact uniformly mixed, but we would retrieve a column-averaged volume mixing ratio of 122 ppm if the water vapor was actually confined to the bottom 10 km in accordance with the saturation level approach. Thus if the saturation level approach accurately reflects the real vertical distribution, our ChemCam results are biased high by *at most* 22%, and only in the cold aphelion season. Such a hypothetical bias could be significant for comparison with other data sets. For example if the potential for ChemCam results to be biased high by vertical distribution assumptions and the potential for CRISM results to be biased low by the look-up-table approximations were both realized, CRISM vs. ChemCam comparisons would look significantly different although still reasonably well matched. It is worth noting here that a worst case bias in ChemCam results from HITRAN line intensity parameter uncertainties (see section 3.5) would be almost as important as a worst case vertical distribution bias, and so it could similarly become significant but only if combined with look-up-table errors or other CRISM errors that act in the opposite direction. Even worst-case line parameter bias combined with worst case vertical distribution bias would leave ChemCam water vapor results substantially lower than TES and regional-to-global scale CRISM.

Note that potential biases from non-uniform water vapor distributions do not get substantially larger when water is concentrated into even thinner near surface layers. The worst case positive bias would occur for a purely hypothetical case of water vapor confined in the bottom ~7 km above the local surface, in which case there would be a 30% positive bias. Water layers even thinner than that give less bias, for example the $1-3$ km thick layers that Titov et al. (1999) argue for at the Pathfinder landing site would be given only a $5-15\%$ positive bias in ChemCam retrievals. Of course these estimates of bias are for the *column averaged* mixing ratio or equivalently for the total water column. The true mixing ratio within such a hypothetical thin layer would have to be larger by a substantial factor to yield a given retrieved water column.

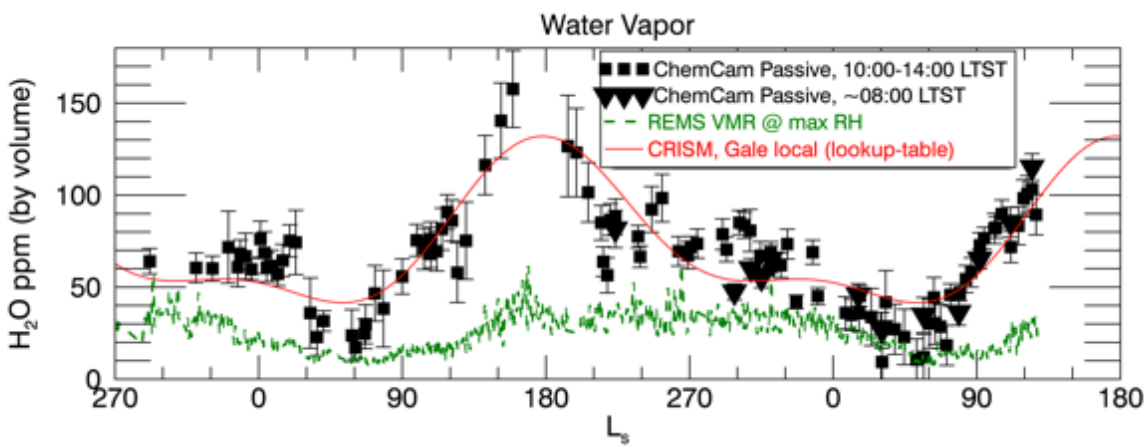

**Figure 11:** Column-averaged volume mixing ratios from: ChemCam passive sky (black squares and triangles); Toigo et al.'s (2013) two-component fit to Gale-crater-localized CRISM lookup-table-based retrievals (red line). The x-axis begins in Mars Year 31 and continues into Mars Year 33. Also plotted is the REMS-H-derived *in-situ* volume mixing ratio at the time of maximum relative humidity on each sol (dashed green line), which always occurs in the early morning before sunrise. Note that this REMS-H value cannot be directly compared to the other quantities plotted here because it has not been measured during daylight hours.

### 4.2.4 Comparison to REMS humidity sensor measurements

In Figure 11 we convert the ChemCam water column abundances to column-averaged volume mixing ratios in order to compare them with volume mixing ratios inferred from MSL's REMS-H relative humidity measurements (Harri et al., 2014b). Note that these in-situ REMS-H volume mixing ratios cannot be directly compared to the ChemCam column-averaged volume mixing ratios, because they are not available during daylight hours. Thus these comparisons are not intended as a simple cross-validation exercise – they are intended to address the physics controlling the water vapor vertical profile and its diurnal evolution (cf. Savijärvi et al. 2015, 2016).

The column-averaged volume mixing ratio is found by simply dividing the mass of water vapor in the column (based on the retrieved value of precipitable microns) by the total atmospheric column mass (based on the REMS-measured surface pressure) to find the column mass mixing ratio, and then multiplying by the ratio of molecular weights (=2.4) to give the volume mixing ratio. The column-averaged volume mixing ratio, in units of ppm, is thus related to the scaled-to-6.1-mb water column abundance, in units precipitable microns, by a constant factor of 14.5 (For mass mixing ratio the corresponding factor is 6.0).

For Figure 11, the REMS humidity sensor (REMS-H) relative humidity (RH) values are converted to mass mixing ratio using the saturation vapor pressure over ice at the REMS-H inlet temperature as described by Savijärvi et al. (2015) and Martínez et al. (2015), and then to volume mixing ratio using the ratio of molecular weights. The REMS-derived mixing ratios are not available during daytime hours because the RH is too low for reliable measurements, but during the evening and nighttime hours REMS-H mixing ratios are typically observed to decrease as the night progresses until a mixing ratio minimum and relative humidity maximum is reached in the pre-dawn period (Savijärvi et al. 2015, 2016). For this paper we take the approach of Martínez et al. (2016a,b) and use the mixing ratio at the time of maximum relative humidity, which is the most precise and reliable value and which corresponds to the pre-dawn mixing ratio minimum. In fact the REMS data that we show in Figure 11 is identical to figure 8 of Martínez et al. (2016b) except for REMS data set updates and the conversion from mass to volume mixing ratio. As described in Martínez et al. (2016b), among the full set of REMS-H measurements, only those taken during the first four seconds of measurements after the relative-humidity sensor has been turned on after ~5 min or more of inactivity are reliable. This is because heating of the sensor by the REMS control electronics causes an artificial decrease in the relative humidity values after the first four seconds of operations.

Figure 11 shows that the pre-dawn (maximum RH) in-situ mixing ratios are in all seasons consistently lower that the column-averaged mixing ratio by a large factor, which ranges from ~1.4 to slightly larger than 5. The ratio of the pre-dawn in-situ mixing ratio to the column-averaged mixing ratio is equal to the ratio that Jakosky et al. (1997) call the ratio of "lander-derived abundance" to "orbiter-derived abundance". They define that ratio to be the precipitable water column that would be inferred by assuming constant mass mixing of a nighttime near-surface in-situ lander measurement divided by the actual precipitable water column derived from orbital measurements. Savijärvi et al. (2016) define the same ratio in their models as $R$, and so we will follow this convention, which means

$$R = (\text{REMS VMR @ max-RH}) / (\text{ChemCam column-averaged VMR}). \qquad (5)$$

Thus we find, looking at Figure 12, that $R$ based on ChemCam and pre-dawn REMS ranges from 0.19 to 0.73 for Gale crater, which compares to a range of 0.3 to 0.7 for Viking Lander 1 and 0.3 to 1.0 for Viking Lander 2 as reported by Jakosky et al. (1997). (An updated analysis of Viking Orbiter water vapor data [Fedorova et al., 2010] would raise the Jakosky et al. $R$ values slightly.) Both Jakosky et al. (1997) and Savijärvi et al. (2016) present models that include adsorption of water on near-surface soil particles and argue, for Viking Lander sites and for Gale Crater respectively, that the depletion of nighttime in-situ water vapor relative to the column can indeed be explained by temperature-dependent soil adsorption. The Viking in-situ water vapor measurements are however indirect, being based on a nighttime temperature inflections (Ryan and Sharman, 1981). The REMS-H measurements meanwhile, even combined with the Savijärvi et al. (2016) models which can relate pre-dawn mixing ratios to mixing ratios within the daytime convective boundary layer, don't constrain the precipitable water column and hence don't constrain $R$. We must therefore rely on ChemCam passive sky measurements or orbit based measurements to provide the denominator of $R$, and as previously discussed ChemCam resolves at least one feature that the orbital data sets do not while providing much more frequent coverage.

While low values of $R$ are certainly consistent with diurnal water vapor interactions with the surface, there is also the potential for large-scale seasonal circulation patterns to create a background vertical gradient in water vapor. $R$ is essentially a combination of these effects with any nighttime surface interaction, and the diurnal mesoscale crater circulation (e.g. Rafkin et al. 2016) could play a role as well. It seems unlikely however that large-scale circulation can account for $R$ values as low as 0.2. GCMs such as Richardson et al. (2002) and Navarro et al. (2014) (which like most GCMs don't include a regolith adsorption scheme) show mixing ratios one scale height above the equatorial surface being as much as 2 or 3 times larger than at the surface during the northern mid-summer season when such an effect is likely to be most pronounced. After vertical averaging (using the ChemCam vertical averaging kernel, which is biased towards low altitudes – see sections 3.7 and 4.2.3) accounting for the actual shapes of the Navarro et al. and Richardson et al. vertical distributions, which peak near one scale height and decrease above and below, that factor of 2 or 3 implies an $R$ no lower than ~0.55. Nevertheless it will be important to account for the potentially considerable uncertainties of the REMS mixing ratios – see Martínez et al. (2016a) for an initial analysis of these – and to conduct detailed modeling before drawing any conclusions about surface interactions based on $R$.

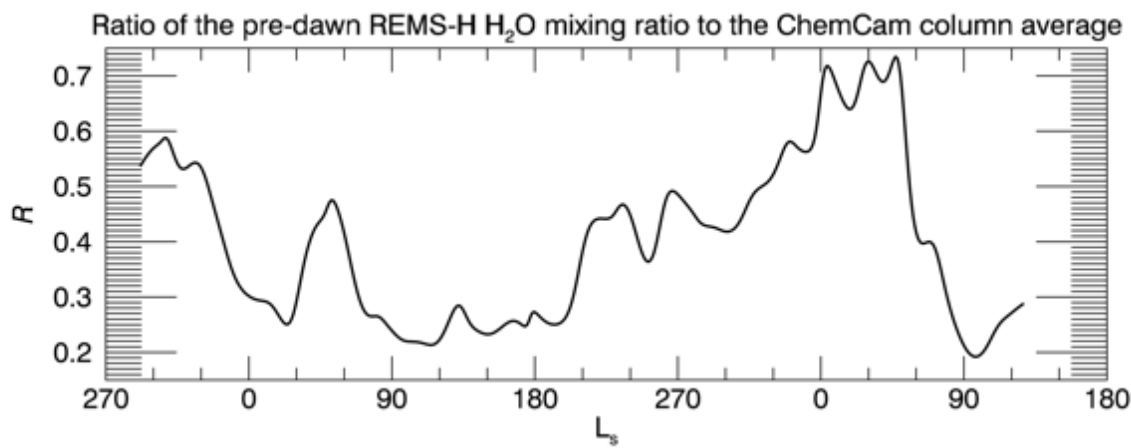

**Figure 12:** The ratio $R$ as defined in the text. The x-axis begins in Mars Year 31 and continues into Mars Year 33.

The non-uniform vertical distribution of water vapor implied by the observed $R$ in Fig. 12 naturally leads to the question of whether the uniform vertical distribution that we assumed in retrieving the ChemCam column-averaged VMR is biasing that VMR value in certain periods and hence biasing $R$ in certain periods. It turns out however that this effect is small. The worst case is the one addressed in section 4.2.3 where the mixing ratio is uniform up to a low saturation level 10km above the local surface. (Such a profile could be compatible with the low observed $R$ if the near-surface water vapor depletion was extremely shallow and/or completely diurnal as Jakosky et al. [1997] and Savijärvi et al. [2016] proposed.) This worst case has the ChemCam-retrieved column-average mixing ratio being biased high by 22% which means $R$ could be biased low by 22%. Correcting that hypothetical bias would raise the lowest $R$ values from 0.2 to 0.25. Having the water vapor mixing ratio instead increasing linearly with height up to some level can lead to a bias for $R$ of up to 10% in the opposite direction, so the largest possible impacts on the $R$ patterns that we observe would be to decrease the amplitude of any particular feature by 30%. Vertical distributions such as those in the Richardson et al. (2002) and Navarro et al. (2014) GCMs that we have previously discussed lead to near-zero (less than 5%) bias because in those cases the mixing ratio decreases both above *and* below a near-one-scale-height peak.

The HITRAN line parameter uncertainty effects (section 3.5) are in the worst case similar in magnitude to the potential biases from the vertical profile, but of course they would change the entire $R$ time series by a fixed percentage and not affect the magnitude of any trends. That percentage change could be positive or negative but supposing a hypothetical correction were to increase $R$ it would raise the lowest $R$ values from 0.2 to 0.24.

Overall, despite their large difference in magnitudes the ChemCam mixing ratios and the REMS-H pre-dawn mixing ratio appear to be significantly coupled in that their minima and maxima and rises and falls occur at essentially the same time. Both mixing ratios also share a long period of nearly constant or very slowly declining values from $L_s$=220° to $L_s$ =330° in MY 32, while all of the orbital data sets show a wide variety of differing trends reflecting the different Martian years in which they were measured and the significant interannual variability in that season. (As can be seen in Figure 12, the slow decline of ChemCam mixing ratios during this period is slightly larger than that of REMS-H.)

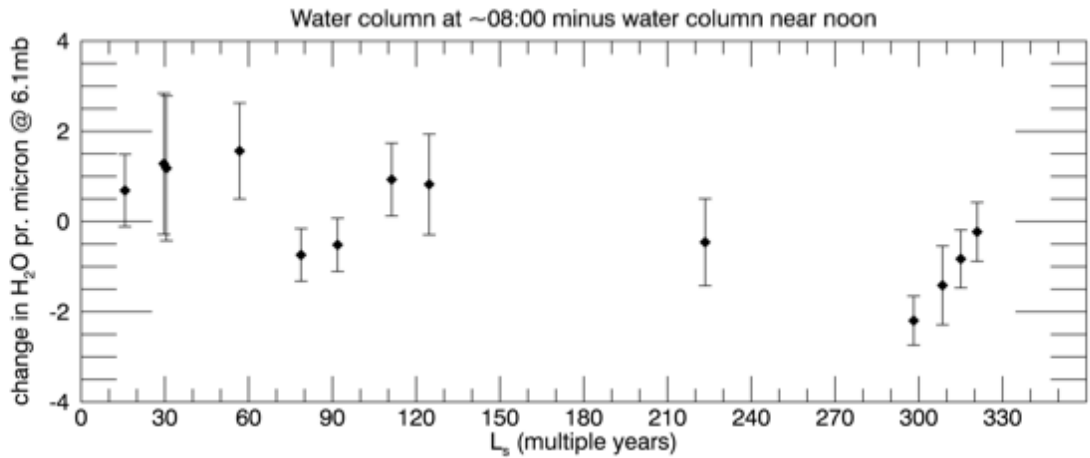

**Figure 13:** Diurnal changes in water vapor measured by ChemCam passive sky, calculated as described in the text by differencing early morning and near-noon observations.

Even though the REMS-H and ChemCam mixing ratios appear to be coupled, their ratio $R$ may have a seasonal pattern with a minimum just after $L_s=90°$ and a rising trend from that season until at least $L_s = 300°$. However the behavior of $R$ during the $L_s=300° – 90°$ period is different in the two Martian years that we have observed, so if there really is an inter-annually repeating component we will need a longer data set to clearly distinguish it. Most of the inter-annual difference in $R$ is caused by the previously noted inter-annual change in the water vapor column around $L_s = 15°$, which appeared minor in the context of the orbital data sets but now seems quite significant relative to the REMS-H data. Meanwhile the same $L_s = 30°$ to $L_s=70°$ period that stands out as anomalous in ChemCam when compared to orbital data sets also stands out in comparison to REMS-H, at least in the first Martian year, as evidenced by a distinct peak in $R$. In the 2nd Martian year the $L_s = 30°$ to $L_s=70°$ phenomenon has a less obvious expression in the $R$ ratio because of the $L_s = 15°$ inter-annual change, but it does show up in the form of a sharp decline in $R$ occurring near $L_s=70°$.

### 4.2.5   Absence of diurnal changes in precipitable water column

To calculate diurnal changes we define observations before 10:00 LTST as "early AM" and those between 11:00 and 14:00 LTST as "near noon". In practice the early AM observations are always close to 08:00 LTST and the "near noon" observations typically range from 11:00 to 13:00 LTST. For each early AM observation we calculated the diurnal difference using the nearest-neighbor of the near-noon observations occurring prior to the early-AM observation and the nearest-neighbor of the near-noon observations occurring after the early-AM observation. If the time differences between these two nearest neighbors and the early-AM observation are within a factor of two of each other, linear interpolation is used to find a "near noon" value for the diurnal change calculation. Otherwise, the nearest of the two nearest neighbors is used.

In Figure 13 we plot time-of-day changes in the ChemCam precipitable water column to assess whether any detectable amount of mass is involved in the large day-night differences implied by $R$. There are no detectable changes in the water column. Only one out of 13 early morning measurements shows a difference of greater than $2\sigma$ from nearby near-noon measurements. We would need to see several instances of changes greater than a threshold of about 1.5 microns to conclude that such changes are occurring. The lack of diurnal signature in precipitable water column is as predicted by the Savijärvi et al. (2016) model, which shows only a very shallow layer affected by surface interactions and essentially no change in the water column once it is scaled by the surface pressure (which has a 10% diurnal amplitude due to tides).

### *4.3    Aerosols*

Four aspects of the aerosol retrieval results that we have obtained to-date make them not fully satisfactory: 1) poor initial guesses and occasional undefined results for the *H'* vertical profile parameter (see section 2); 2) the large sensitivity of water ice aerosol particle size to retrieval assumptions (see section 3); 3) unknown sensitivity to uncertainties in aerosol optical constants, phase functions, and potential 3-dimensional geometric effects. However, important aspects of the ChemCam passive sky aerosol results to-date are likely to be robust, and even the more questionable results provide valuable information albeit with some ambiguity. We therefore present our current understanding of the aerosol information provided by ChemCam passive sky observations below, keeping in mind that it should be considered preliminary.

#### *4.3.1    Dust and ice opacity*

Figure 14 shows the ChemCam-derived opacity of dust and ice at a reference wavelength of 880 nm, with the sum of dust and ice 880 nm opacity constrained by the time-interpolated opacity determined from Mastcam direct sun imaging as previously discussed. TES climatology-based estimates of 880nm opacities are shown for comparison, with 880 nm extinction opacities from TES being estimated based on the extinction-vs.-absorption correction factors – 1.3 and 1.5 for dust and ice aerosol respectively – of Clancy et al. (2003) and on the aerosol particles size assumed in the original TES retrievals (Smith et al., 2000, Smith 2004). The sum of ChemCam dust and ice opacity (which as previously discussed is constrained to equal Mastcam opacity measurements) is typically larger than the corresponding TES climatology sum (50% larger on average) because the average elevation of the climatology grid point is about 4 km (~0.4 scale heights) higher than MSL. The most notable result in Figure 14 is that dust opacity observed so far in MY 33 is essentially identically to that of MY 32, but the ice opacity in MY33 has been larger than MY 32 by a factor of ~2. Elevated ice opacities relative to the preceding year are actually apparent as early as $L_s$=290° in MY 32, and factor-of-two changes are much larger than the uncertainties implied by our sensitivity analyses. It is unlikely that an un-accounted-for systematic issue, for example 3-dimenional topography changes as the rover moves, would consistently change the retrieved dust/ice fraction by exactly the amount needed to alter the absolute magnitude of ice while keeping dust unchanged from the previous year. Thus the interannual change in ice extinction should be taken seriously even though previous interannual comparisons (e.g. Smith 2004) have not reported anything comparable.

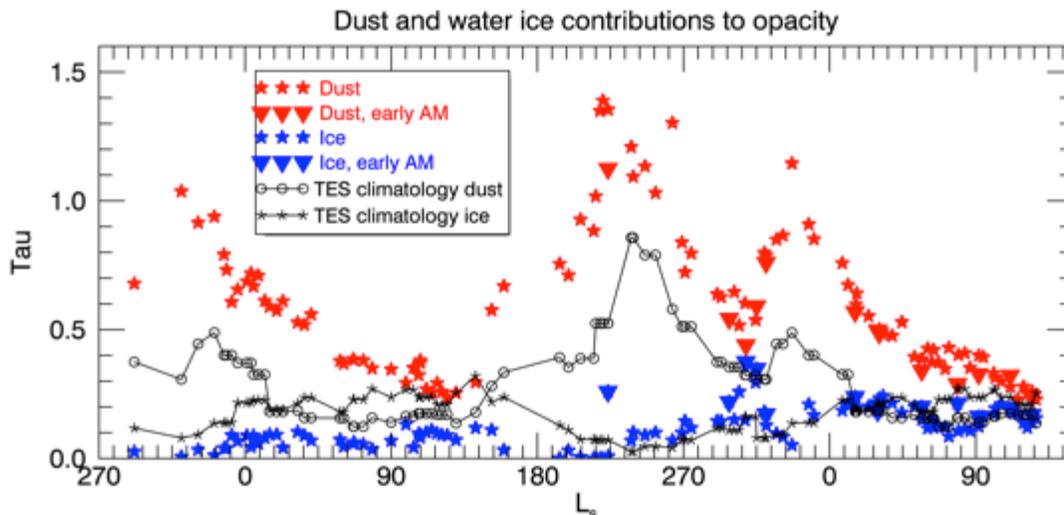

**Figure 14:** Dust and water ice contribution to column opacity at a reference wavelength of 880 nm. ChemCam observations are shown with red (dust) and blue (ice) points. TES climatology is shown with black lines and open circles (dust) or black lines with asterisks (ice). ChemCam points plotted with a triangle are for observations before 10:00 LTST. The x-axis begins in Mars Year 31 and continues into Mars Year 33.

Figure 14 also shows an interesting diurnal pattern in dust and ice opacity, and this is shown more clearly in Figure 15 where the diurnal changes are plotted. There is a clear pattern of higher ice opacity in the morning (positive changes on Fig. 15) and lower dust opacity in the morning (negative changes on Fig. 15); this persists in both the cold season ($L_s = 60° – 90°$) and the dusty season ($L_s = 210° – 310°$). This pattern is larger than the known uncertainties, but in this case this result should be deemed questionable because the morning gains in water ice are nearly identical to the losses in dust. In other words morning gains in water ice are typically accompanied by only very small changes in total opacity. Although water ice may well be nucleating on dust particles and perhaps coincidentally causing a one-for-one exchange of opacity contribution at 880 nm, an error in the aerosol optical constants or shape assumptions could cause a phase function error and hence systematics that depend on the sun angle. Further investigation will be required to rule out the latter possibility before we can be confident that we have detected a diurnal pattern.

### 4.3.2   *Effective particle size*

Figure 16 shows the retrieved effective particle radius for dust. Ice particle size as previously discussed is quite sensitive to input uncertainties, so some or perhaps even all of the ice particle size variability is controlled by systematic errors. Therefore, although retrieved ice effective particle radius values are available in the supplementary material, they are not included in Fig. 16. The dust particle size in contrast has negligible sensitivity, in the vast majority of cases, to the systematic uncertainties that we have quantified.

It is important to consider that we have not quantified the influences of uncertainties and approximations in the optical constants and particle shapes. The potential effects of three-dimensional local topography and spatially variable surface albedo are a concern as well. However, all of these poorly quantified factors would be expected to cause apparent particle size changes whenever the pointing direction and lighting geometry change, while most such direction and lighting changes, including changes in time of day and 180° changes in viewing direction, produce no changes in the retrieved dust particle size. Consequently the dust particle size is likely robust to these pointing-dependent systematics in most seasons.

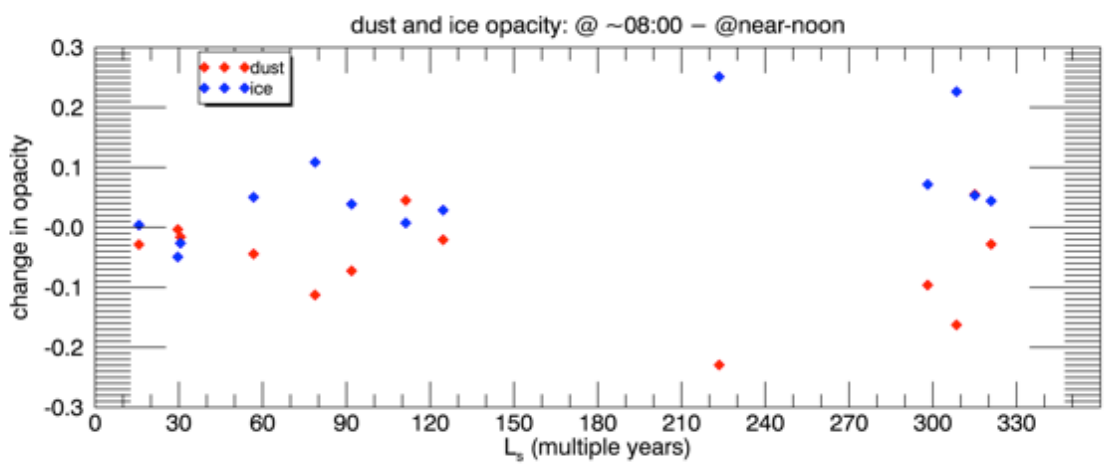

**Figure 15:** The same as Figure 13 but for dust (red points) and ice (blue points) opacity.

A brief period around $L_s$=110° in MY 32 stands out as the only exception to the pointing-direction independence of retrieved particle sizes. During this period we experimented with pointing both the high-elevation-angle and low-elevation-angle pointings at either 270° or 90° azimuth, while also acquiring some observations with both pointings at the nominal (180° for this season) azimuth. All of the non-nominal pointings during this particular period show larger dust particle sizes than those of the nominal pointings. This phenomena is likely caused by the loss of measurement sensitivity to dust particle size that occurs with this particular geometry, i.e. when both the low elevation angle and high elevation angle pointings are viewing nearly perpendicular to the sun azimuth. This is likely due to the dust scattering phase function being largely independent of particle size at a particular scattering angle. (There are a few east- or west-pointed cases after MY 32 $L_s$ = ~220° (sol 790) that don't experience the low-sensitivity problem because the high elevation angle and low elevation angle pointings have different azimuths instead of identical azimuths.) This loss of sensitivity is evident in Figure 4 panels C & D. Essentially, the spectrum becomes insensitive to dust aerosol size in these cases, and therefore the results for dust can be dominated by systematics, just as is usually the case for ice aerosol particle size. Meanwhile the spectrum becomes *more* sensitive to ice particle size in this unusual geometry (likely due to the ice aerosol scattering phase function having more particle size sensitivity at that particular angle), which suggests that the retrieved ice particle size may speculatively be more accurate in these unusual cases. This suggestion will remain speculative until we conduct sensitivity experiments for these rare cases.

Another puzzling feature of the ChemCam dust particle size results that *might* be attributable to systematic errors is the smaller sizes for $L_s$ > 60° in MY 33, relative to MY 32. The MY 33 $L_s$ $> 60°$ dust $r_{eff}$ values are $0.2 - 0.5$ microns smaller, a change two to five times larger than the most severe systematic effect identified in our sensitivity analyses. Poorly understood pointing- or geometry-dependent systematic effects are unlikely to be the explanation, because the interannual differences emerges after $L_s = 60°$ with no changes in observing geometry and no diurnal variability. Similarly, any error in dust optical constants or assumed particle shapes would not likely be time dependent. Thus our preferred interpretation is that the inter-annual particle size change is real, and perhaps related to the previously-discussed inter-annual increase in ice aerosol abundance, although that ice aerosol increase appears much earlier.

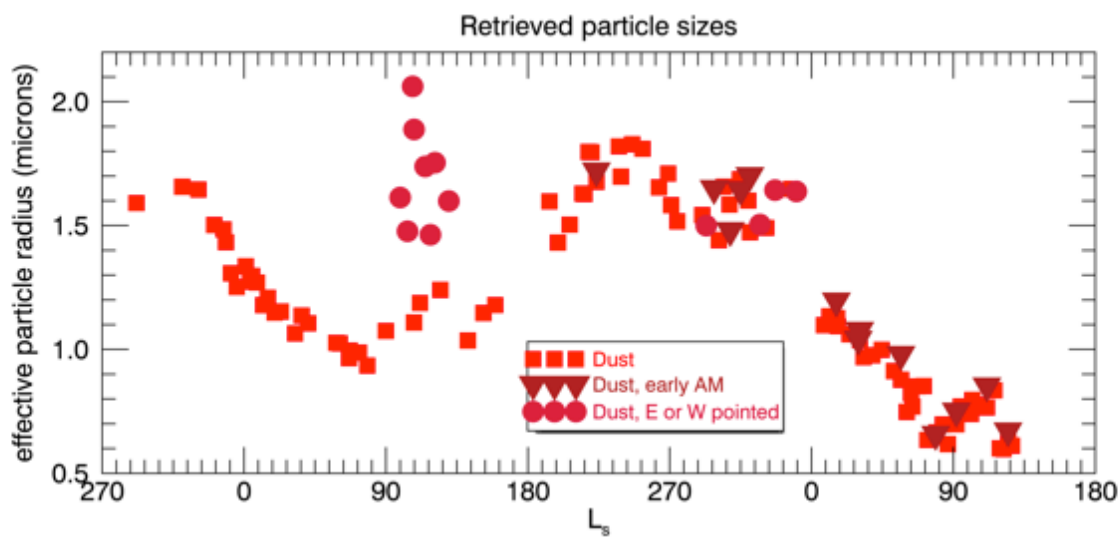

**Figure 16:** Dust effective particle radius as retrieved from ChemCam passive sky. Triangles are for observations before 10:00 LTST, circles are for observations where the low-elevation angle pointing was pointed at either 90° or 270° azimuth, and squares are for all other observations. The x-axis begins in Mars Year 31 and continues into Mars Year 33.

If there *was* a systematic error from some unknown source, our set of sensitivity analyses in Table 3 (section 3) suggests that errors in aerosol ice opacity should be closely coupled to errors in dust particle size, with erroneously small dust particle sizes being correlated with erroneously high dust opacity fractions and hence erroneously *small* ice opacities. The similar (but not identical) effects of dust particle size and dust opacity fraction on the modeled spectrum (Figure 4) suggest the same thing: as shown in panels A and B of Fig. 4 increasing the dust particle size and increasing the dust opacity fraction both typically increase the modeled continuum ratio, so if the dust particle size becomes systematically higher that will tend to move the dust opacity fraction in the opposite direction making it systematically lower in order to continue to approximately match a given observed ratio spectrum. This is true regardless of whether we are introducing a new systematic error or eliminating a hypothetical existing systematic error. Thus if we "corrected" a hypothetical dust particle size error to increase MY 33 dust particle sizes toward those of MY 32, we would also be substantially decreasing the dust opacity fraction and *increasing the ice aerosol opacities* making the interannual change in ice abundance even more extreme.

One clear and robust pattern that does emerge for aerosol particle size is a strong, smooth, seasonal behavior in dust $r_{eff}$, one which shows evidence of interannual repeatability in the timing of its trends and maxima and minima. This repeatability in the timing of trends and extreme values does not always apply to the magnitude of dust particle size, which as previously discussed is substantially smaller after $L_s = 60°$ in MY 33 relative to MY 32. The pattern that we observe includes a minimum in dust particle size at or near aphelion ($L_s = \sim 70°$) and a maximum at or near perihelion ($L_s = 250°$). The aphelion minimum repeats in both Martian years, but we have repeat coverage only for $L_s = 290°$ through $L_s = 130°$ and thus we have observed the perihelion maximum only once so far. There is a secondary minimum in the vicinity of $L_s = 140°$ that appears to be repeatable although we don't fully cover this period in MY 33.

The dust particle size pattern as a whole shows a strong correlation with dust opacity (Fig. 14) although there are some noteworthy differences in the timing of maxima and minima. In particular the minimum dust particle size precedes the minimum dust opacity by 60° of $L_s$. Meanwhile the maximum dust particle size period lags the dust opacity maximum by about 20° of $L_s$ in the one Mars year (MY 32) that we have observed it so far. This particle size maximum is also relatively smooth and broad compared to the sharp spike of maximum opacity.

The seasonal dust particle size pattern observed by ChemCam is entirely consistent with the dust particle size pattern derived by Clancy et al. (2003) from MGS-TES emission-phase-function (EPF) observations, although the TES EPF results cover a much wider range of latitudes, show much greater scatter, and are limited to low-ice-opacity cases.

### 4.3.3   Vertical structure

Figure 17 plots the quantity $H'$, which was defined in section 2.5 as the ratio of gas density scale height to aerosol extinction coefficient scale height. Table 4 summarizes the physical interpretation of $H'$, but we must keep in mind that the true vertical profile is not necessarily well described by a single scale-height-based parameter. We must also keep in mind that we generally are solving only for $H'_{dust}$, but that in reality the vertical profile of $H'_{ice}$ could deviate from our MGS-TES-based assumption.

As we have previously discussed, whenever varying $H'_{dust}$ alone does not yield a solution, we produce a useable water-vapor-abundance solution by varying $H'_{ice}$. Since we can't simultaneously constrain both parameters, when this happens we essentially have no solution for $H'$, and in these cases we set $H'_{dust}=5$ for the sake of displaying them on Figure 17. We have chosen a large positive number to represent these no-solution cases because they are all cases where more near-surface aerosols are needed to reach a solution, but this does not mean that the true vertical distribution has $H'_{dust}>5$. In fact increasing $H'_{dust}$ beyond 5 never has any significant effect, which is why a value of 5 is beyond our search grid and so is a convenient indicator of undefined results. Cases of $H'_{dust}=5$ are observations for which our vertical profile parameterizations fails because there appears to be more scattering in the bottom two scale heights of the atmosphere than can be provided by dust alone. Not surprisingly this happens only for the smallest values of dust opacity fraction.

**Table 4**: Physical Interpretation of $H'$

| | | |
|---|---|---|
| $H' > 1$ | Low altitude aerosol maximum | The aerosol extinction coefficient decreases more rapidly than the gas density. |
| $H' = 1$ | "Well mixed" aerosol | The aerosol extinction coefficient divided by the gas density is a constant |
| $0 < H' < 1$ | High altitude aerosol maximum | The gas density decreases more rapidly than the aerosol extinction coefficient. |
| $H' = 0$ | Uniform mixing by volume. | The aerosol extinction coefficient is constant with height. |
| $H' < 0$ | Detached aerosol layer. | The aerosol extinction coefficient increases with altitude |

In addition to the ambiguity inherent in our simplified parameterization of the aerosol vertical profile, there are three types of errors that we must consider that are potentially significant and not well represented by our calculated (and displayed in Fig. 17) errors bars. The first of these is an unexplained error in $CO_2$ band depth fitting that is discussed further in Appendix B section B.8. This appears to be a purely random error that can cause outliers and excess scatter relative to our estimated error bars, but shows no evidence of a systematic bias or trend and is only prominent after $L_s=110°$ in Mars Year 33. A more serious concern is the possibility that the diurnal pattern (Fig. 15) in ice vs. dust opacity is in fact a systematic error. If this was the case, then such an error *might* also influence the modeled $CO_2$ band depth and cause a compensating error in $H'_{dust}$. The third potentially significant source of error is the $CO_2$ line intensity uncertainty discussed in section 3.5 – this error tends to raise or lower the entire time series leaving the temporal pattern intact, but it could lower the entire time series by up to 1 or raise it by up to 2 if the extremes of the potential error were realized. The aerosol vertical profile parameter is also subject to additional errors including errors in the determination of other aerosol parameters, but the effects of these are captured in our other sensitivity experiments (section 3) and turn out to be only marginally significant compared to the calculated statistical uncertainties. Since the statistical uncertainties are themselves often quite small compared to the key features of Fig. 17, the additional sensitivity-experiment-derived errors are similarly inconsequential.

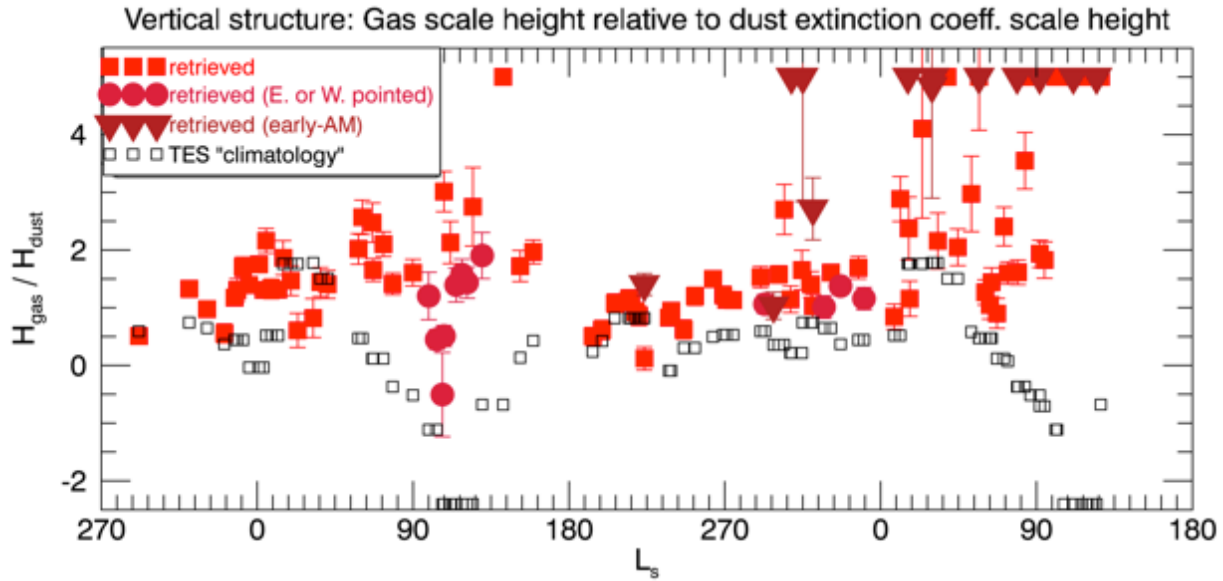

**Figure 17:** Vertical structure parameter $H'_{dust}$ retrieved from ChemCam passive sky compared with the initial guess $H'_{dust}$ from TES climatology (small black open squares). The shapes of the ChemCam data points represent the type of observation: triangles are for observations before 10:00 LTST, circles are for observations where the low-elevation angle pointing was pointed at either 90° or 270° azimuth, and squares are for all other observations. The x-axis begins in Mars Year 31 and continues into Mars Year 33.

The possibility of an unknown systematic error, the $CO_2$ line intensity uncertainty issue, and the limitations of our parameterization all indicate a need for further investigation and confirmation of our aerosol profile results, but these results are significantly different from expectations in a several ways and thus worth noting here. The expectations, based on TES climatology (Guzewich et al. 2013) and corroborated by Mars Climate Sounder (MCS) data sets (e.g. Heavens et al., 2011) are represented by open squares on Fig. 17. Another important part of our expectations about the vertical profile of aerosols is that the ice aerosols form a capping layer above the dust, and that expectation is built into the TES climatology that initializes our retrievals and supported by MCS (e.g. Kleinböhl et al., 2009) and CRISM (Smith et al., 2013). (By "capping layer", we mean that the ice layer altitude coincides with the approximate top of the dust layer, a configuration which suggests that ice formation limits or "caps" the vertical extent of the dust.)

The failure of our vertical profile retrievals ($H'_{dust}$ set to 5 in Fig. 17) in most of the cases where the ice opacity fraction is high suggests that there is more low altitude ice than expected from the TES representation of the capping layer phenomenon. (This phenomenon in our data set is robust even at the extremes the potential $CO_2$ line intensity error.) Since TES observations, and indeed all from-orbit aerosol observations, have limited sensitivity to the vertical profile in the lowest scale height of the atmosphere, the ChemCam results for $H'$ are arguably just as credible as those of in-orbit limb sounders. More extensive inter-instrument comparisons are needed to assess whether the disagreement is limited to TES and whether alternative profiles exist that can explain all of the observations.

The other deviations from expectations shown in Fig. 17 point toward an excess of low-altitude dust. $H'_{dust}$ is persistently larger than the TES climatology and consistently larger than one. It also appears to reach a maximum near $L_s = 90°$ at a time when TES climatology shows a minimum. This period around $L_s = 90°$ would have $H'_{dust}$ significantly larger than the TES climatology even at the positive extreme of the potential $CO_2$ line intensity error (which gives the most negative possible shift in $H'_{dust}$).

In both cases it is, so far, unclear which instrument is most likely to be affected by systematics, or whether the differences are simply local phenomena specific to Gale Crater. Note that the TES tendency to have $H'_{dust} < 1$ in most seasons at the equator is consistent with the "high-altitude tropical dust maximum" described in MCS data (Heavens et al., 2011), but both TES and MCS may simply be failing to accurately resolve the bottom 10 km. The ChemCam results *are* generally consistent with the vertical distribution parameterization retrieved by Wolff et al. (2006) from joint, multiple phase angle, MGS TES & MER mini-TES observations. These shows the aerosol confined near the surface, i.e. with Conrath parameter (Conrath, 1975) approaching 1.0, between $L_s = 30°$ and $L_s = 150°$ at both Gusev Crater and Meridiani Planum.

An additional complication for understanding the ChemCam vertical profile results is the observation by Moore et al. (2016) that aerosol opacity in the air below the crater rim (i.e. the lowest 2 km of atmosphere) is *depleted* relative to the column opacity in most seasons, which seemingly points toward $H'_{dust} < 1$ in apparent contradiction to the ChemCam results. This can be reconciled by recalling that the $H'_{dust}$ parameter addresses aerosol extinction from the surface up to ~20 km (two scale heights up), and so the mixing ratio could be generally increasing toward the surface in the bottom 20 km of atmosphere but then drop off within the bottom ~2 km. The bottom ~2 km are below the crater rim and so most influenced by the isolating (e.g. Rafkin et al. 2016) local crater circulation and the associated very shallow boundary layer. Thus it appears that the simplest aerosol profile that fits the available observations is actually rather complicated, with processes that are enhancing the aerosol concentration near the surface on regional and global scales and competing processes that are depleting it on a local scale.

# 5    Conclusions

We have presented a procedure to retrieve water vapor column abundance and aerosol properties from ChemCam passive-mode sky observations. This paper includes the results of 113 successful retrievals spanning one-and-a-half Martian years from $L_s = 291°$ in MY 31 to $L_s = 127°$ in MY 33.

## *5.1    Water Vapor*

Our initial results give water vapor column abundances with a precision of +/- 0.6 precipitable microns. Sensitivity tests and analysis of water vapor results indicates that systematic errors are no larger than +/- 0.3 precipitable microns, except for the effects of deviations from the assumed uniformly-mixed vertical profile and for uncertainties in the HITRAN 2008 (Rothman et al., 2009) line intensity parameters. For extreme cases, which are only plausible for the northern summer season, a condensation-level-limited vertical profile could create a positive bias of up to 22%. Correcting for hypothetical line intensity parameter errors could make the entire water vapor time series lower or higher by between −16% – +20%.

The ChemCam water vapor abundances show one peculiar seasonal feature that is not present in any of the orbital data sets, that being a distinct minimum period between $L_s = 30°$ and $L_s = 70°$ where the scaled-to-6.1 mb column abundance is in the $1 – 2$ precipitable micron range.

A similar minimum *is* however present in upward-looking MER mini-TES water vapor retrievals (Smith et al. 2006).

Otherwise, the ChemCam-retrieved water abundances show the same seasonal behavior and the same timing of seasonal minima and maxima as the TES, CRISM, and REMS-H data sets. In addition, CRISM water vapor retrievals that are localized to Gale crater appear to match the magnitude of ChemCam column abundances quite closely in most seasons, although further CRISM data analysis will be needed to verify this. However, ChemCam water vapor column abundances are substantially smaller than regional- and zonal-scale TES and CRISM averages while at the same time the column-averaged ChemCam water volume mixing ratios are larger than pre-dawn REMS-H in-situ values by a factor of ~1.4 to 5. Pending further analysis of REMS-H volume mixing ratio uncertainties, the differences between ChemCam and REMS-H pre-dawn mixing ratios appear to be much too large to be explained by large scale circulations (based on existing GCMs that don't consider regolith adsorption), which supports the hypothesis of substantial diurnal interactions of water vapor with the surface as proposed by Jakosky et al. (1997) and Savijärvi et al. (2016). We don't see these diurnal changes in comparisons of early-morning ChemCam water vapor data with near-noon ChemCam water vapor data, but this is consistent with the Savijärvi et al. (2016) prediction that only a very shallow layer of the atmosphere participates in the diurnal response.

If the surface interactions favored by ChemCam vs. REMS-H mixing ratio comparisons are in fact occurring, one likely mechanism is the temperature dependent adsorption favored by Jakosky et al. (1997) and Savijärvi et al. (2016), i.e. nighttime adsorption of water vapor followed by daytime desorption. Nighttime frost formation, which Martínez et al. (2016b) found indirect evidence for on a few sols, may also contribute.

Regardless of the mechanism of surface interaction, the failure, to date, of ChemCam LIBS to detect diurnal change in the hydrogen content of surface soils places an upper limit on the amount of mass exchanged. However, that upper limit (~1 wt. % water in the upper millimeter of soil, corresponding to ~20 precipitable microns water uniformly distributed over that thickness of soil) (Meslin et al. 2013) is larger than the total atmospheric column mass of water (Fig. 10). Thus, the ChemCam LIBS analyses performed so far are not sensitive enough to detect the amounts of water exchanged in the adsorption/desorption process.

### 5.2    Aerosols

Our aerosol retrievals should be considered preliminary but they yield mostly-reliable results for the dust and water ice aerosol contributions to column opacity and partially reliable results for dust aerosol particle size and for parameterized aerosol vertical profiles.

The most striking and apparently robust aerosol result is a two-fold increase in water ice aerosol opacity in the second year of ChemCam passive sky observations. We see no evidence for systematic errors that could produce such a change, but we are also not aware of any prior observations or models that suggest the possibility of interannual water ice aerosol changes of this magnitude. Clearly it will be important to seek confirmation of this result in other data sets.

For dust aerosol particle size, we see a strong and smoothly varying seasonal pattern that is consistent with Clancy et al.'s (2003) MGS-TES emission-phase-function (EPF) results. The ChemCam particle sizes, unlike those derived from TES EPFs, show very little scatter about the smooth seasonal trend, perhaps because they cover only a single location. The seasonal pattern of dust particle sizes is correlated with dust opacity, but with some important exceptions.

For the aerosol vertical profiles we consistently find more low-level aerosol than expected from TES climatology and from Mars Climate Sounder vertical profiles. In fact mixing ratios are indicated to be consistently larger in the bottom scale height than at ~20km, which seems to contradict the Heavens et al. (2011) observations of a "high-altitude tropical dust maximum". ChemCam is most likely more sensitive to the bottom scale height of aerosols than the MCS and TES limb-sounding retrievals, but given the preliminary nature of ChemCam aerosol results it is too early to say which instrument is providing the most accurate profile.

# Appendix A. External data sets

## A.1 Surface pressure

The surface pressure required for our modeling comes from MSL's Rover Environmental Monitoring Station (REMS) pressure sensor (Harri et al., 2014a). It is read from REMS ENVRDR data products (which are archived by the PDS). Note that the PDS-stored MODRDR REMS data products provide more accurate surface pressure information and we intend to make use of them for future work. We used the REMS ENVRDR only for legacy reasons, having verified that the difference between ENVRDR and MODRDR pressure value is negligible for our applications. (The ENVRDR pressure product is systematically lower than the MODRDR product due to expected sensor drift by an amount which grows to 0.5% by sol 1293, the last sol used in this paper. This produces a comparable or smaller percentage error in absolute water vapor column abundances but tends to cancel itself out in pressure-scaled column abundances or mixing ratios.)

We normally interpolate the pressure linearly in time to the mid-point local mean solar time of the passive sky observations. If the passive sky observation's time falls in a REMS data gap longer than 90 minutes in duration, we reject the REMS data for that sol unless the passive sky observation is within 15 minutes of the end or beginning of the REMS data gap, in which case we apply nearest-neighbor interpolation. When we reject the REMS data for given sol, we repeat the search for valid REMS data on adjacent sols within a range of nine sols from the passive sky observation, selecting the nearest valid sol.

## A.2 Total column opacity

Total column opacity is measured by MSL's Mastcam using direct solar imaging in narrow-band solar filters center at 440 and 880nm. This data set is described by Lemmon (2014). We use a simple linear interpolation in time from the Mastcam opacity measurements to the midpoint time of the passive sky observation. The interpolated 880 nm opacity is then used as a fixed constraint for all passive sky retrievals. Although we do not explicitly consider the local true solar time of the Mastcam measurements in the interpolation process, all ChemCam passive sky observations that occur at unusual times of day are paired with Mastcam opacity measurements performed immediately or nearly immediately before or after (i.e. within five minutes before the start or after then end of the passive sky observation).

## A.3 Surface albedo

For the lambert surface albedo that our radiative transfer model (section 2.6 & Appendix B section B.2) requires, we have adopted a weighted sum of endmember spectra from Mustard and Bell (1994), using a CRISM atmospherically corrected lambert albedo data cube (Arvidson et al., 2005) to guide our choice of weights. Using the PDS-archived CRISM mosaic with product-id

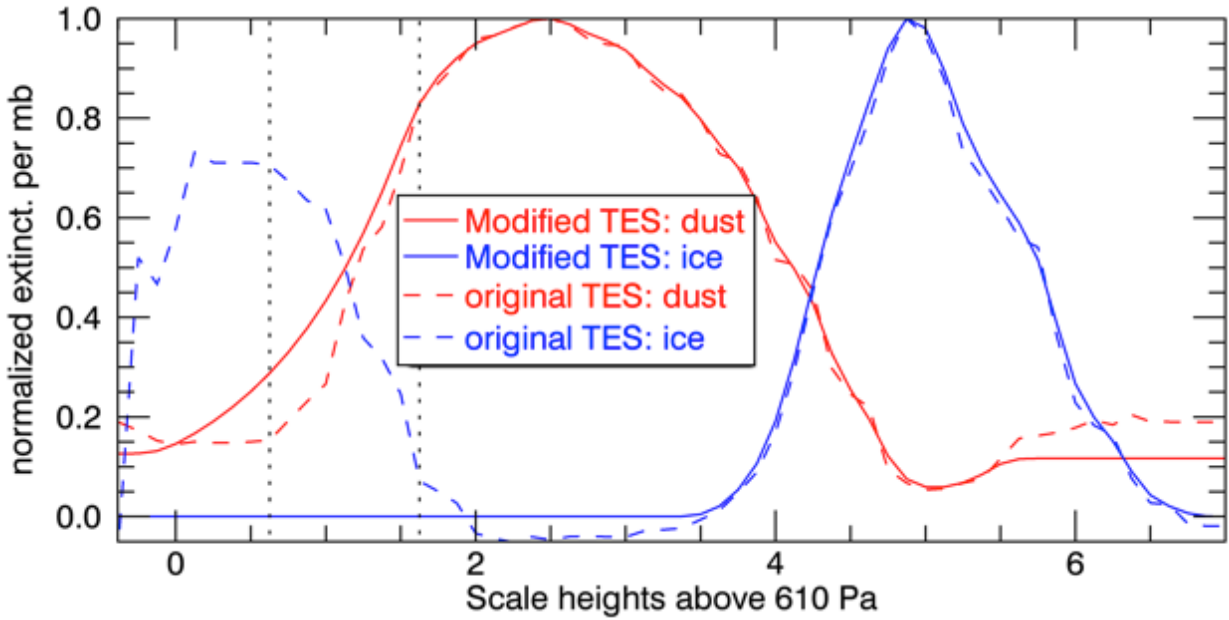

**Figure A1:** The original aerosol extinction profiles (dash lines) compared with the modified profiles (solid lines) used in initial retrievals. The profiles are in units of km$^{-1}$ per mb and normalized to a maximum of 1.0. These profiles are sourced from TES climatology as described in section 2.4.2 and Appendix A section A.4 and come from the Mars Year 24, $L_s$=238° time bin. The modified dust profile below two scale heights has H$^'_{dust}$ = −0.1 in this case.

"T0902_MRRAL_05S138_0256_1", we selected all mosaic points within 40 km of the MSL landing site and found that the average atmospherically corrected lambert albedo spectra was well matched by a combination of 80% – 90% Mustard and Bell (1994) bright region endmember with 20–10% Mustard and Bell (1994) dark region endmember. Since the CRISM mosaic coverage was not complete in the vicinity of the MSL operating area and since our subjective assessment was that the coverage was modestly biased towards darker regions, we adopted a 90% bright region 10% dark region mixture. Since our 1-d radiative transfer can at best only approximate the heterogeneous surface albedos influencing the passive sky observing area, this adopted albedo represents an adequate approximation, but of course we must consider the surface albedo uncertainties in our sensitivity analyses, which we did in section 3. We do not use the CRISM hyperspectral data directly because of its relatively course spectral sampling. For the Mustard and Bell (1994) data it is still necessary to interpolate the spectral data to a fine 0.2 nm sampling grid and then smooth it with at 20 nm FWHM Gaussian kernel. This interpolation/smoothing procedure eliminates spurious high frequency spectral features that would otherwise propagate to the continuum-removed ratio and interfere with gas retrievals.

### A.4 Aerosol vertical profile.

#### A.4.1      Modifications to the TES climatological aerosol profiles

The aerosol extinction profiles, taken from the TES data set as described in section 2.4.2, are modified before being used in our retrievals. This section describes these modifications. These modifications are necessary to remove near-surface ice signatures that are likely spurious, to remove anomalously high high-altitude mixing ratios, and to introduce the idealized parameterization (section 2.5) of the dust profile in bottom two scale heights. The profile modifications consist of the following steps:

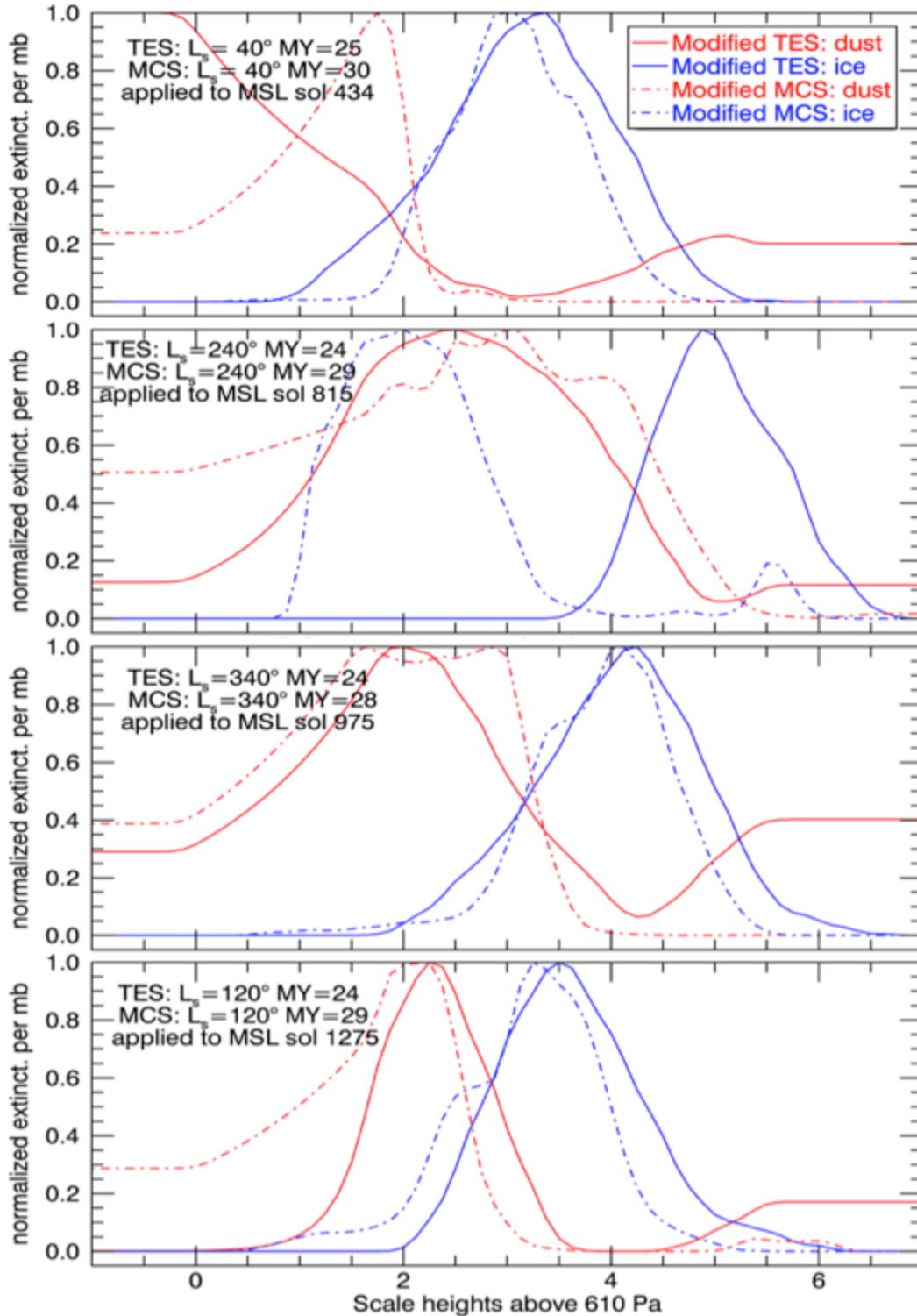

**Figure A2**: Profiles of dust (red) and ice (blue) aerosol extinction per unit millibar normalized to a maximum value of 1.0. Each panel compares the TES-derived (solid lines) and MCS-derived (dash-dot line) modified aerosol profiles for a particular observation sequence – these are the profiles used for our initial guess vertical profile. (TES is used for nominal retrievals and MCS for sensitivity testing.) The four cases shown here correspond to the four cases in table A1.

1.  Negative extinction coefficient values (which can occur due to TES retrieval uncertainties) are set equal to zero.

2.  Questionable near surface ice signatures are eliminated by finding local minima in ice aerosol mixing ratio below the altitude at which total column opacity reaches 0.01. Ice aerosol extinction below the highest of these local minima is set to zero and the extinction thus removed from ice is assigned to dust.

3.  Any questionable high-altitude mixing ratios are eliminated by forcing all mixing ratio values more than 5.5 scale heights above the local surface to not exceed the mixing ratio value at 5.5 scale heights. Although the extinction from physically plausible aerosols above 5.5 scale heights is negligible for ChemCam observations, if this step were not performed there would be a small risk that anomalously high values introduced by TES retrieval uncertainties could influence our retrievals.

4.  In the lowest two scale heights above the local surface, the dust aerosol profile is replaced with the idealized, parameterized profile described in section 2.5

5.  After *all other* modifications to the aerosol profiles are complete, we smooth the profiles with a 0.25 scale height FWHM Gaussian kernel to decrease the magnitude of discontinuities and noise in the profiles.

Figure A1 shows an example of the net effect of our aerosol profile modifications.

### *A.4.2        MCS vs. TES vertical profiles*

We conducted several sensitivity experiments with MCS-derived vertical profiles replacing our standard vertical profiles in order to assess whether our results are affected by our choice of the source for our initial guess vertical profiles. As discussed (section 3.6) in the main text we found no significant sensitivity to this choice, apart from a small effect in the low-opacity icy (sol 1275) case that was essentially the same as the sensitivity to the *H'* vertical profile parameter. Here we provide more information about the MCS-derived profiles, a detailed comparison (Fig. A2) of the TES- and MCS-derived profiles, and more sensitivity experiment results (Table A1).

Our MCS vertical profiles for both temperature and aerosols are MCS Level 2 PDS products (pipeline version 4.3.11) that are binned spatially and temporarily in the same manner as our TES vertical profiles. When multiple years are available for a given seasonal bin, we choose the first year that gives a useable profile. After binning the MCS profiles are modified, parameterized, and smoothed according to the procedure in the previous section, except that steps #1, #2, and #3 are not necessary and not performed. In some cases the MCS dust profile does not extend down to one scale height about the surface – in those cases the parameters for the parameterized portion of the profile are calculated based on whatever portion of the dust profile is available between one and two scale heights. Similarly, the MCS ice profile normally does not extend all the way to the surface – we substitute zeros for the unavailable ice extinction coefficients prior to the smoothing in step #5.

The resulting MCS-derived profiles are compared to the corresponding TES profiles in Fig. A2. Recall that the overall opacities for dust and ice *are not* determined in any way by the MCS or TES data sets, and for that reason all profiles are normalized to a maximum value of 1 in Fig. A2. Considering Fig. A2 with respect to those features most likely to affect our retrievals, we see that MCS and TES differ substantially in the bottom two scale heights. However that uncertainty is already accounted for by our solved-for *H'* parameter. We also see that the location of water ice extinction is similar in MCS and TES, except for the sol 815 case where the dust opacity is large and the ice opacity contribution is relatively very small.

As previously discussed and shown once again in Table A1, none of the MCS-vs.-TES perturbation tests shown any effect on retrieved water vapor. Table A1 also gives further evidence that MCS-vs.-TES effects on aerosol parameters are quite small, with the main notable difference occurring in the sol 1275 case that was discussed in section 3.5. Looking again at Fig. A2 we can see that for the sol 1275 case the ice extinction in MCS extends slightly lower than TES – this is the initial-guess feature responsible for the seemingly large but not physically significant difference in solved-for $H'_{ice}$.

**Table A1**: Additional perturbation experiments for MCS vs. TES vertical profile comparison

| Perturbation experiments for sol 434, LTST 11:10 | Inputs | | | Results | | | | | |
|---|---|---|---|---|---|---|---|---|---|
| | Initial $H'$ | Surface lambert albedo at 800 nm | Mastcam opacity ($\tau$) at 880 nm | Dust fraction | Dust $r_{eff}$ ($\mu$m) | Ice $r_{eff}$ ($\mu$m) | $\chi^2$ for Aerosol | $H', H'_{ice}$[2] | Water Vapor pr. $\mu$m[1] |
| Nominal | 1.40 | 0.288 | 0.63 | 0.89 | 1.1 | 1.5 | 0.18 | 1.4 ±0.2 | 3.2 ±0.6 |
| MCS profile (MY 30) | 0.20 | 0.288 | 0.63 | 0.90 | 1.1 | 1.4 | 0.17 | 1.3 | 3.2 |

| Perturbation experiments for sol 815, LTST 11:36 | Inputs | | | Results | | | | | |
|---|---|---|---|---|---|---|---|---|---|
| | Initial $H'$ | Surface lambert albedo at 800 nm | Mastcam opacity ($\tau$) at 880 nm | Dust fraction | Dust $r_{eff}$ ($\mu$m) | Ice $r_{eff}$ ($\mu$m) | $\chi^2$ for Aerosol | $H', H'_{ice}$[2] | Water Vapor pr. $\mu$m[1] |
| Nominal | -0.1 | 0.288 | 1.28 | 0.95 | 1.8 | 1.5 | 0.28 | 0.8 ±0.1 | 8.2 ±0.7 |
| MCS profile (MY 29) | 0.8 | 0.288 | 1.28 | 0.93 | 1.8 | 1.7 | 0.30 | 0.9 | 8.2 |

| Perturbation experiments for sol 975, LTST 10:10 | Inputs | | | Results | | | | | |
|---|---|---|---|---|---|---|---|---|---|
| | Initial $H'$ | Surface lambert albedo at 800 nm | Mastcam opacity ($\tau$) at 880 nm | Dust fraction | Dust $r_{eff}$ ($\mu$m) | Ice $r_{eff}$ ($\mu$m) | $\chi^2$ for Aerosol | $H', H'_{ice}$[2] | Water Vapor pr. $\mu$m[1] |
| Nominal | 0.37 | 0.288 | 1.21 | 0.96 | 1.6 | 2.1 | 0.19 | 1.4 ±0.1 | 4.2 ±0.4 |
| MCS profile (MY 28) | 0.44 | 0.288 | 1.21 | 0.95 | 1.7 | 2.0 | 0.19 | 1.4 | 4.2 |

| Perturbation experiments for sol 1275, LTST 12:46 | Inputs | | | Results | | | | | |
|---|---|---|---|---|---|---|---|---|---|
| | Initial $H'$ | Surface lambert albedo at 800 nm | Mastcam opacity ($\tau$) at 880 nm | Dust fraction | Dust $r_{eff}$ ($\mu$m) | Ice $r_{eff}$ ($\mu$m) | $\chi^2$ for Aerosol | $H', H'_{ice}$[2] | Water Vapor pr. $\mu$m[1] |
| Nominal | -2.40 | 0.288 | 0.42 | 0.67 | 0.60 | 2.5 | 0.14 | 5.0, 5.0 ±0.5 | 9.3 ±1.1 |
| MCS profile (MY 29) | 0.43 | 0.288 | 0.42 | 0.70 | 0.56 | 3.7 | 0.19 | 5.0, −0.7 | 9.3 |

# Appendix B. Additional methodology details

### B.1 Wavelength calibration

Each ChemCam pixel was carefully calibrated to wavelength pre-flight, and these wavelengths are documented in the "ccam_default_wave" file within the "document" directory of the ChemCam PDS archive. The true wavelength of each pixel however is a function of spectrometer temperature and a weak function of time. This function is determined from regular LIBS measurements of the titanium calibration target (e.g. Wiens et al. 2013) and is documented in the "wave_cal_coeffs" and "wave_cal_coeffs_500" files in the "document" directory of the ChemCam PDS archive. We use this wave calibration function to calculate the wavelength of each EDR pixel, using the average wavelength over the course of a passive sky observation sequence because variations during a single sequence are negligible. Unlike standard ChemCam data processing, for passive sky observations we do not interpolate the pixel data to a common wavelength, since doing so would smear out the pixel-to-pixel varying component of the detector background. Also unlike standard ChemCam data processing, which use "wave_cal_coeffs" function coefficients for sols prior to sol 500 and "wave_cal_coeffs_500" for the rest, we use the coefficients in "wave_cal_coeffs_500" for the entire data set. The RMS error of the resulting wavelength calibration is < 0.2 pixels (< 0.05 nm), with a maximum error of 0.4 pixels.

### B.2 Radiative transfer model details

This section describes the assumptions and model parameter choices that we have made for our radiative transfer models. As described in the main text of this paper, we use a discrete ordinates (c.f. Thomas and Stamnes, 1999) model with a correlated-k approximation (Lacis and Oinas, 1991) for gas absorption, and the underlying model is identical to that of Smith et al. (2009) except for the addition of the delta-approximation (e.g. Goody and Yung, 1989). However most of the assumed aerosol property details, model parameters, and other assumptions are specific to this work, and these are detailed below.

For this work, we have used spherical water ice particles (optical constants from Warren, 1984) with log-normal size distributions described by $v_{eff} = 0.1$ and $r_{eff}$ ranging from 1.0 to 4.0 microns. For dust we have used cylindrical dust particles with an axial ratio of one as suggested by Wolff et al. (2009), as well as the Wolff et al. (2009) optical constants. Our dust particle size distributions are log-normal with $v_{eff} = 0.3$ and $r_{eff}$ ranging from 0.5 to 2.5 microns. We supply the aerosol properties to our radiative transfer code in the form of pre-computed look-up tables containing extinction cross sections, single-scattering albedos, and the Legendre expansion terms for the phase function. These look-up tables sample $r_{eff}$ at 0.1 micron intervals and interpolate linearly in $r_{eff}$. In wavelength, they are interpolated and smoothed to 0.2 nm intervals with a 20 nm FWHM Gaussian kernel in the same manner as the surface albedo (Appendix A section A.3). The discrete ordinates code itself uses 97 equally-spaced log-pressure vertical levels from the local surface to 10 scale heights above the local surface, plus, for numerical accuracy, 3 widely spaced levels extending the grid to 20 scale heights. When fitting the continuum for aerosol properties we use 64 streams and 80 phase function moments, which we found by numerical experimentation to be the minimum necessary for accurate results at all scattering angles. For fitting gas absorptions we use only 24 streams and 24 phase function moments, which we have found to be sufficient for continuum-removed spectra.

We take gas absorption line parameters from HITRAN 2008 (Rothman et al., 2009). Gas absorption cross sections are supplied, with interpolation, to our discrete ordinates code from pre-computed lookup tables with 9 temperatures from 280 to 120 K, 10 pressures from 0.1 to 3000 Pa, and wavelength sampling at intervals equal to one-tenth the nominal width ($\Delta_{res}$) of the applied instrument line-spread function. We have found that with very weak VNIR absorptions that only a small number of correlated-k quadrature points are needed to model the gas absorptions with negligible loss of accuracy – we use 8 quadrature points. We account for both air-broadening and self-broadening for all molecules. Although in principle the Earth-specific air-broadening coefficients supplied by HITRAN should be corrected for the different composition of the Martian atmosphere, we have found that our model has no sensitivity to such corrections due once again to the fact that the absorptions in question are very weak. We have therefore accepted the terrestrial coefficients for $O_2$ and $CO_2$. For water vapor we have applied a correction factor of 1.5 for consistency with Smith (2002) and Smith et al. (2009).

In computing the gas absorption look up table, we use an instrument line spread function of:

$$\phi(x) = \frac{\sin(\pi|x|)}{\pi|x|} + \frac{23\,|x|\sin(\pi|x|)}{28\,\pi(1-x^2)} \quad \text{for} \quad x = -2 \rightarrow 2 \tag{B1}$$

$$= \; 0 \quad \text{for} \quad |x| > 2,$$

$$x \; \equiv \; 2\frac{\lambda - \lambda_0}{\Delta_{res}} \tag{B2}$$

with central wavelength $\lambda_0$ and nominal wavelength resolution $\Delta_{res}$. The FWHM of this function is 0.89 $\Delta_{res}$. We use $\Delta_{res}$. = 0.47 nm and thus FWHM = 0.42 nm. Wiens et al. (2012) (see their Figure 8) implies a FWHM of 0.47 nm for the typical operating temperatures of the VNIR spectrometer on Mars (2 ± 8 °C), but their plots describe the width of a real emission line and so represents an upper limit. This upper limit was considered in our sensitivity analysis (section 3). The other line-spread-function related uncertainty that we must consider is a small mismatch between our adopted analytical line-spread function and the observed line shape (see Wiens et al., 2012, figure 8c). This mismatch is confined to the wings and represents a difference of 3.5% in the area under the curve for a given FWHM and thus a potential error of that same magnitude in the peak line intensity after normalization.

The major radiative transfer model contribution to our uncertainties turns out to be the HITRAN line parameters. Based on an intensity-weighted average of the line parameter uncertainty indices in HITRAN 2008 (Rothman et al., 2009), the uncertainty in the intensity line parameters within the absorption bands used in this paper is 5% – 10% for water vapor and 10% – 20% for $CO_2$. For the sensitivity analysis in this paper we adopt the midpoint of these estimated uncertainty ranges – i.e. 7.5% for water vapor and 15% for $CO_2$. We then assume the worst case for correlation between line intensity errors among the hundreds (for $CO_2$) to thousands (for $H_2O$) of individual lines in our absorption bands, i.e. we assume that they are 100% correlated and so the intensities of the absorption bands are just as uncertain as the individual line parameters. Since we find no sensitivity to the line broadening parameters, only the intensity parameter uncertainties contribute to the uncertainty in our results.

A given intensity line parameter error contributes the reciprocal of that error to the error in the retrieved abundances. Correcting such a line intensity error would change the retrieved abundance by the same percentage as that of the line intensity error. (For example if we treat the $CO_2$ line strength as in error by -15%, the $CO_2$ abundance is in error by a factor of 1/0.85 relative to the corrected value and correcting for this error would change the reported value by -15%).

$H_2O$ line parameter errors affect retrieved $H_2O$ directly. Thus water vapor results corrected for hypothetical water vapor line parameter errors could be globally higher or lower by 7.5%.

$CO_2$ line intensity errors affect the $H'$ parameter (see section 3.5) and also contribute to water vapor uncertainties because $CO_2$ is used to correct the water vapor abundance (section 2.10). The nature of a hypothetical correction for $CO_2$ line intensity errors can best be understood graphically. The hypothetical correction be read off of a graph like that in Fig. 6: we know how much the $CO_2$ abundance is in error due to a given line intensity error and so we choose a location on the $CO_2$ abundance axis that is higher or lower by the amount needed to compensate for the error and read off the $H_2O$ column or $H'$ result at that point. For example in Fig. 6 if $CO_2$ line intensity is hypothesized to be too small by 15%, we compensate for that by using the 96% / 0.85 = 113% location on the $CO_2$ axis and thereby reading off a corrected $H_2O$ column of 8.2 precipitable microns. (For comparison if we assume that the line intensities are correct we use 96% on the $CO_2$ axis and we get are standard result 7.0 precipitable microns for this observation.) We could make the similar hypothetical correction to $H'$ by plotting $H'$ versus $CO_2$ abundance and applying the same procedure. Note that in actual practice such as for the sensitivity tests in section 3.5 we do not use the previously described graphical approach for either $H'$ or $H_2O$. Instead we apply the procedures of section 2.10 with an alternative target value for $CO_2$ abundance to create our hypothetical corrections, but this procedure is essentially equivalent to the graphical approach in concept and produces essentially identical results.

### B.3  Adaptive model look-up table

The adaptive model look-up table is generated and maintained as follows:
1. The adaptive look-up table starts with model calculation points at 3 mixing ratio values spanning the range of plausible values.
2. Additional model calculation points are iteratively added in the gaps between the existing model calculation points until all sets of three adjacent points in the table pass a linearity test at all wavelength samples. The linearity test is that the value of the middle point must be within $1 \times 10^{-5}$ of a line between the other points. Since we will be using the model to match data that is the ratio of high- and low-elevation-angle pointings, we perform the linearity test on the ratio of high and low elevation-angle pointings in the model.
3. When a query to the look-up table is made, the look-up table returns a result by interpolating linearly between the model calculation points.
4. If a query is made that is beyond the bounds of the initial range, the range is expanded and step #2 is repeated until all calculation points again meet the linearity criterion.

### B.4  Calculating the spatially uniform component of the detector background

As was discussed in section 2.7.1, our calculation of the spatially uniform component of the detector background is slightly more complicated than simply finding the minimum DN level in the minimum-photon-signal region of the non-illuminated detector rows. This section of the appendix describes the spatially uniform detector background calibration in detail and shows that its uncertainties are negligible.

The calculation begins by finding the minimum of a parabola fitted to the minimum photon signal region between 859 nm and 869 nm nominal wavelength. We do this for both illuminated row collects and non-illuminated row collects. Letting $d$ be the spatially uniform detector background component, $s_0$ and $s_1$ be the non-illuminated and illuminated rows fitted minimums, respectively, and letting $\acute{f}$ be the ratio of total photon response (including readout photons) in illuminated rows relative to the non-illuminated rows, we estimate the spatially uniform dark current as

$$d = s_0 - \frac{s_1 - s_0}{\acute{f} - 1}. \tag{B3}$$

We perform this calculation for each non-illuminated row collect in the observation sequence, using the nearest illuminated row collect as the source of $s_1$. All passive sky observations include a VNIR non-illuminated row collect at the beginning and again at the end of the sequence. (Type 3 sequences include additional VNIR non-illuminated row collects.) To determine the value of $d$ to apply to any given illuminated-rows collect, the $d$ value calculated at each non-illuminated row collect is interpolated linearly in time. Then a single scalar value for $d$ is assigned to each collect, and it is subtracted from all pixels in the spectrum before further processing.

With the correct value of $\acute{f}$, $d$ is independent of the photon signal and so independent of wavelength in the vicinity of $859 - 869$ nm. Thus if we take $s_0$ and $s_1$ from individual pixels rather than from fitted minimums, the resulting plot of $d$ will be a straight line (plus noise). Using this fact we estimate $\acute{f}$ to be inversely proportional to integration time and equal to 28 for 100 millisecond integration time. We observe that the term $\frac{s_1 - s_0}{\acute{f} - 1}$ is always small, 5% of $d$ at most and never more than 0.1% of the signal at wavelengths of interest. The magnitude of $\frac{s_1 - s_0}{\acute{f} - 1}$ is an upper limit on the uncertainty of the spatially uniform background calculations, and numerical experiments show that perturbations of this magnitude have no effect on passive sky ratio spectra or continuum removed ratio spectra.

### B.5 Formal definition of model fitting and spatially-variable background correction in continuum-removed ratio spectra

To describe our model fitting procedure for the continuum and our correction for the spatially variable component of detector background, we will need some formal definitions. We use $i$ for the pixel position of any given spectrum pixel. Denoting the continuum removed ratio spectrum by $\boldsymbol{C}$, and denoting the low-elevation-angle and high-elevation-angle average DN values after spatially-uniform background subtraction by $\boldsymbol{A}$ and $\boldsymbol{B}$, respectively, our continuum removal procedure gives

$$C_i = \frac{A_i / B_i}{S_i \left( A_{i-(w-1)} / B_{i-(w-1)} \cdots A_{i+(w-1)} / B_{i+(w-1)} \right)} \tag{B4}$$

with the function $S_i$ defined by

$$S_i \left( x_{i-(w-1)} \cdots x_{i+(w-1)} \right) = \frac{1}{w} \sum_{j=i-\frac{w-1}{2}}^{j=i+\frac{w-1}{2}} (x_j) \left/ \frac{1}{w} \sum_{j=i-\frac{w-1}{2}}^{j=i+\frac{w-1}{2}} \left( \frac{x_j}{\frac{1}{w} \sum_{k=j-\frac{w-1}{2}}^{k=j+\frac{w-1}{2}} (x_k)} \right) \right. . \tag{B5}$$

Hereafter we will use the notation

$$\langle x_j \rangle_i \equiv \frac{1}{w} \sum_{j=i-\frac{w-1}{2}}^{j=i+\frac{w-1}{2}} (x_j) . \tag{B6}$$

The parameter $w$ is the boxcar smoothing kernel width at each iteration and is required to be odd. As previously mentioned, we use $w = 75$ pixels.

The $A_i$ and $B_i$ spectra from sky observations, which we will call $(A_{sky})_i$ and $(B_{sky})_i$ are composed of random noise $\varepsilon$, spatially variability detector background $e_i$, and signal $(A^*_{sky})_i$ and $(B^*_{sky})_i$:

$$\left(A_{sky}\right)_i = \left(A^*_{sky}\right)_i + e_i + (\varepsilon_A)_i$$
$$\left(B_{sky}\right)_i = \left(B^*_{sky}\right)_i + e_i + (\varepsilon_B)_i.$$

(B7)

The spatially variable detector background $e_i$ is a function of temperature and integration time.

The signal can be decomposed as the product of solar irradiance $\pi F$, instrument response $g$, the broadband surface & atmosphere scattering and absorption $H$, and the narrow-band surface & atmosphere scattering and absorption $(1 - h)$. So:

$$\left(A^*_{sky}\right)_i = \left(H_{A_{sky}}\right)_i \left(1 - \left(h_{A_{sky}}\right)_i\right) F_i g_i$$
$$\left(B^*_{sky}\right)_i = \left(H_{B_{sky}}\right)_i \left(1 - \left(h_{B_{sky}}\right)_i\right) F_i g_i.$$

(B8)

Our radiative transfer model returns I/F rather than DN and of course lacks a detector background pattern or random noise, thus:

$$(A_{model})_i = (A^*_{model})_i = \left(H_{A_{model}}\right)_i \left(1 - \left(h_{A_{model}}\right)_i\right)$$
$$(B_{model})_i = (B^*_{model})_i = \left(H_{B_{model}}\right)_i \left(1 - \left(h_{B_{model}}\right)_i\right).$$

(B9)

The $h$ term is observed to be of order $10^{-3}$ and is only non-zero because of the presence of gas absorption lines in the Martian atmosphere (although scattering is an important factor in determining the magnitude of $h$).

Define $(e_2)_i$ as the effect of the spatially variable detector background on the ratio spectrum such that

$$\frac{\left(A_{sky}\right)_i}{\left(B_{sky}\right)_i} = \frac{\left(A^*_{sky}\right)_i + e_i + (\varepsilon_A)_i}{\left(B^*_{sky}\right)_i + e_i + (\varepsilon_B)_i} = \left(A^*_{sky}\right)_i / \left(B^*_{sky}\right)_i + \left(e_{2_{sky}}\right)_i + (\varepsilon_{AB})_i,$$

(B10)

where $\varepsilon_{AB}$ is yet another purely random error. To first order in $e/(B^*_{sky})_i$:

$$\left(e_{2_{sky}}\right)_i \approx \frac{e_i}{\left(B^*_{sky}\right)_i}\left(1 - \frac{\left(A^*_{sky}\right)_i}{\left(B^*_{sky}\right)_i}\right),$$

(B11)

which is an acceptable approximation because $e_i/(B^*_{sky})_i$ is typically observed to be < 5% for 99% of pixels and < 9% for 100% of pixels. $e_i/(B^*_{sky})_i$ will however depend on integration time and detector temperature – these typical values are for 200 milliseconds at -5.75°C. Putting (B10) and (B11) together

$$\frac{\left(A_{sky}\right)_i}{\left(B_{sky}\right)_i} = \frac{\left(A^*_{sky}\right)_i + e_i + (\varepsilon_A)_i}{\left(B^*_{sky}\right)_i + e_i + (\varepsilon_B)_i} \approx \frac{\left(A^*_{sky}\right)_i}{\left(B^*_{sky}\right)_i}\left(1 + e_i\left(\frac{1}{\left(A^*_{sky}\right)_i} - \frac{1}{\left(B^*_{sky}\right)_i}\right)\right) + (\varepsilon_{AB2})_i.$$

(B12)

To simplify (B12), define $e_3$ and make an approximation good to 1st order in $e_i/(B^*_{sky})_i$:

$$\left(e_{3_{sky}}\right)_i \equiv e_i\left(\frac{1}{\left(A^*_{sky}\right)_i} - \frac{1}{\left(B^*_{sky}\right)_i}\right) \approx e_i\left(\frac{1}{\left(A_{sky}\right)_i} - \frac{1}{\left(B_{sky}\right)_i}\right),$$

(B13)

which is in practice always smaller than $e_i/(B^*_{sky})_i$ since both $(A^*_{sky})_i$ and $(B^*_{sky})_i$ are typically around 6000. In fact as previously discussed we design our observations so that the low-elevation angle and high-elevation angle sky brightnesses, i.e. $(A^*_{sky})_i$ and $(B^*_{sky})_i$, are as nearly equal as possible, so in the ideal case $e_3$ approaches zero. In the typical case it is smaller than $e_i/(B^*_{sky})_i$ by a factor of ~10, meaning $e_3$ is typically around $5\times10^{-3}$. Had we kept second order terms in $e_i/(B^*_{sky})_i$ they would give an additional $e_3$ term of around $2.\,5\times10^{-4}$. Since this is of the same magnitude as the typical (for 200 milliseconds exposures at -5.75°C) $\varepsilon_{AB2}$ random noise level, there could be some modest benefit to keeping higher order terms. We will consider this in future work, but given that temporal changes in $e_i$ inherently limit the accuracy of our spatially variable dark current correction we will take the simpler 1st order approach in this paper.

Now making use of (B13) we can insert (B12) into the definition of a continuum-removed ratio spectrum in (B4) and (B5) to decompose the continuum removed ratio into signal and detector background terms.

$C_{sky_i}$

$$= \frac{\left(A^*_{sky}\right)_i / \left(B^*_{sky}\right)_i}{S_i\left(A^*_{i-(w-1)}/B^*_{i-(w-1)} \cdots A^*_{i+(w-1)}/B^*_{i+(w-1)}\right)} \left( 1 \; + \; \left(e_{3\,sky}\right)_i + \langle\langle\left(e_{3\,sky}\right)_k\rangle_j\rangle_i \right.$$

$$\left. - \frac{\text{cov}_i\left(\left(e_{3\,sky}\right)_j, \left(H_{A\,sky}\right)_j / \left(H_{B\,sky}\right)_j\right)}{\langle\left(H_{A\,sky}\right)_j / \left(H_{B\,sky}\right)_j\rangle_i} + \langle\frac{\text{cov}_j\left(\left(e_{3\,sky}\right)_k, \left(H_{A\,sky}\right)_k / \left(H_{B\,sky}\right)_k\right)}{\langle\left(H_{A\,sky}\right)_k / \left(H_{B\,sky}\right)_k\rangle_j}\rangle_i \right)$$

$$\quad\quad\text{(B14)}$$

$+ (\varepsilon_{ABC})_i$ .

As before, to arrive at (B14) we have performed a series expansion on $1/(1 + e_3)$ terms and dropped terms of 2nd order in $e_3$ and smaller. We have also performed a series expansion on the $1/(1-h)$ terms imbedded in the $A^*$ and $B^*$ terms and made use of the fact that $h$ is even smaller than $e_3$. The function $\text{cov}_e(x, y)$ is the covariance of the given variables within the same width $w$ moving average box implied by the $<>$ brackets. The last two terms involving the covariance are less than $10^{-5}$ based on the magnitude of deviations of $H_a/H_b$ from its mean (less than 0.25% within the smoothing width) and the magnitude of smoothed deviations in the variable detector background when it is directly observed at night. We therefore drop these last two terms. The $<<e_3>>$ term turns out to be a factor of 20 smaller than $e_3$ (based on directly observed spatially variable background), but we will retain an approximation of this term as:

$$\langle\langle\left(e_{3\,sky}\right)_k\rangle_j\rangle_i \approx \langle\langle e_k\rangle_j\rangle_i \left(\frac{1}{\left(A_{sky}\right)_i} - \frac{1}{\left(B_{sky}\right)_i}\right). \quad\quad\text{(B15)}$$

This approximation works because $e$ and $h$ terms that are part of $A$ and $B$ can be eliminated by a series expansion that drops terms smaller than $e_i/B_i$, and that leaves only $H_a$ and $H_b$ which have no narrow band features by definition and so can be treated as approximately constant within the smoothing width. In practice the $(1/A - 1/B)$ term differs from its smoothed version by 5% in the most extreme cases, which means the error from the approximation in (B15) is smaller than $e_3$ by at least a factor of 400 and so less than $10^{-5}$ in absolute terms.

To apply our detector background correction we need to know how to scale the detector background terms so we define:

$$\left(e_{4\,\text{sky}}\right)_i \equiv \left(e_{3\,\text{sky}}\right)_i + \langle\langle\left(e_{3\,\text{sky}}\right)_k\rangle_j\rangle_i \approx \left(\langle\langle e_k\rangle_j\rangle_i + e_i\right)\left(\frac{1}{\left(A_{\text{sky}}\right)_i} - \frac{1}{\left(B_{\text{sky}}\right)_i}\right). \tag{B16}$$

This leads to:

$$C_{\text{sky}_i} = \frac{\left(A^*_{sky}\right)_i/\left(B^*_{sky}\right)_i}{S_i\left(A^*_{i-(w-1)}/B^*_{i-(w-1)}\cdots A^*_{i+(w-1)}/B^*_{i+(w-1)}\right)}\left(1 + \left(e_{4\,\text{sky}}\right)_i\right) + (\varepsilon_{ABC})_i\,. \tag{B17}$$

The last manipulation needed to define our background calibration procedure is to expand the continuum-removed ratio of $A^*$ and $B^*$ to fully separate it from the detector background terms. So we substitute (B8) in for $A^*$ and $B^*$ and carry out the series expansion as usual, dropping second order terms in $h$ ($h^2$ is of order $10^{-6}$) and terms where $h$ is multiplied by $e_4$ ($e_4$ is essentially identical in magnitude to $e_3$ so these dropped terms are of order $5\times10^{-6}$ or smaller). We also drop the covariances of $h$ terms with $H_A/H_B$ terms that arise from the smoothing, because $H_a$ and $H_b$ lack narrow band feature by definition and in practice we can see in our models that these covariances are less than $10^{-6}$. The result is:

$C_{\text{sky}_i}$

$$= 1 - \left(h_{A\,\text{sky}}\right)_i + 2\langle\left(h_{A\,\text{sky}}\right)_j\rangle_i - \langle\langle\left(h_{A\,\text{sky}}\right)_k\rangle_j\rangle_i + \left(h_{B\,\text{sky}}\right)_i - 2\langle\left(h_{B\,\text{sky}}\right)_j\rangle_i - \langle\langle\left(h_{B\,\text{sky}}\right)_k\rangle_j\rangle_i$$

$$+ \left(e_{4\,\text{sky}}\right)_i + (\varepsilon_{ABC})_i\,. \tag{B18}$$

For our radiative transfer model, $e_4$ and $\varepsilon$ are of course absent so:

$C_{\text{model}_i}$

$$= 1 - \left(h_{A\,\text{model}}\right)_i + 2\langle\left(h_{A\,\text{model}}\right)_j\rangle_i - \langle\langle\left(h_{A\,\text{model}}\right)_k\rangle_j\rangle_i + \left(h_{B\,\text{model}}\right)_i - 2\langle\left(h_{B\,\text{model}}\right)_j\rangle_i$$

$$- \langle\langle\left(h_{B\,\text{model}}\right)_k\rangle_j\rangle_i \tag{B19}$$

For our calibration target observations $h_A = h_B$ because both calibration targets are receiving the exact same illumination from the sky and the direct solar beam. Also, the same approximations work for the calibration targets because $H_a$ / $H_b$ for the calibration targets depends only on the target reflectances that as expected are no more variable within the smoothing width than is scattered sky light (based on pre-flight measurements, e.g. Johnson et al., 2015). Therefore:

$$C_{\text{cal}_i} = 1 + \left(e_{4\,\text{cal}}\right)_i + (\varepsilon_{\text{cal}})_i\,. \tag{B20}$$

The spatially variable detector background is a function of detector temperature $T$ (in °C) and is proportional to the integration time $t$. Relative to $e_{\text{ref}}$, its value at some reference temperature and integration time,

$$\left(\langle\langle e_k\rangle_j\rangle_i + e_i\right) \approx \frac{t}{t_{\text{ref}}}\left(\frac{T + 15}{T_{\text{ref}} + 15}\right)\left(\langle\langle(e_{\text{ref}})_k\rangle_j\rangle_i + (e_{\text{ref}})_i\right). \tag{B21}$$

Since the equation (B16) definition for $e_4$ holds for the calibration observations as well as for sky observations:

$$\left(e_{4\text{sky}}\right)_i \approx \gamma \left(e_{4\text{cal}}\right)_i \frac{t_{\text{sky}}}{t_{\text{cal}}} \left(\frac{T_{\text{sky}} + 15}{T_{\text{cal}} + 15}\right) \frac{\left(\dfrac{1}{\left(A_{\text{sky}}\right)_i} - \dfrac{1}{\left(B_{\text{sky}}\right)_i}\right)}{\left(\dfrac{1}{\left(A_{\text{cal}}\right)_i} - \dfrac{1}{\left(B_{\text{cal}}\right)_i}\right)}. \tag{B22}$$

We introduce the $\gamma$ detector background scale factor as an additional parameter – one we will fit for – to approximately account for any inaccuracies in the scaling of $e_4$ and to approximately account for changes in the variable detector background over time.

Recall that $A_{\text{sky}}$ and $B_{\text{sky}}$ and $A_{\text{cal}}$ and $B_{\text{cal}}$ are all things that we directly observe, and that $C_{\text{sky}}$ and $C_{\text{cal}}$ are calculated from them according to (B4). So we use the calibration target observations to calculate $e_{4\text{cal}}$ via (B20), then $e_{4\text{sky}}$ via (B22). Comparing (B18) and (B19) we see that the model parameters that we seek will produce a $C_{\text{model}}$ that differs from the observation-derived $C_{\text{sky}}$ only by $e_{4\text{sky}}$ plus random noise. Thus we vary gas abundances to minimize:

$$\chi^2 = \sum_i \left(\frac{C_{\text{model}_i} - C_{\text{sky}_i} - \left(e_{4\text{sky}}\right)_i}{\left(\varepsilon_{\text{total}}\right)_i}\right)^2. \tag{B23}$$

The data standard errors used for weighting are:

$$\left(\varepsilon_{\text{total}}\right)_i{}^2 = \left(\varepsilon_{\text{ABC}}\right)_i{}^2 + \left(\left(\varepsilon_{\text{cal}}\right)_i \gamma \frac{t_{\text{sky}}}{t_{\text{cal}}} \left(\frac{T_{\text{sky}} + 15}{T_{\text{cal}} + 15}\right) \frac{\left(\dfrac{1}{\left(A_{\text{sky}}\right)_i} - \dfrac{1}{\left(B_{\text{sky}}\right)_i}\right)}{\left(\dfrac{1}{\left(A_{\text{cal}}\right)_i} - \dfrac{1}{\left(B_{\text{cal}}\right)_i}\right)}\right)^2. \tag{B24}$$

$\varepsilon_{\text{ABC}}$ and $\varepsilon_{\text{cal}}$ are the bootstrap standard errors for the continuum-removed ratio spectra of the sky observation and the calibration target observation, respectively.

### B.6 Lack of significant cross-talk between absorption fitting regions

The $w-1$ pixel distance ($w$ is defined above in Appendix B section B.5) over which any one spectral sample can influence others due to smoothing in our (equation B5) definition of continuum-removed spectra is equal to 15.2 nm, so in theory the $O_2$ fitting range could be influenced by the $CO_2$ absorption line which has not yet been calculated at the time of $O_2$ fitting, but we can and do safely neglect this effect because after passing through two stages of boxcar smoothing that influence is $< 5 \times 10^{-6}$ in magnitude (based on our models) and affects only 8 out of 100 pixels on the extreme long wavelength side of the $O_2$ fitting range. Similarly, the $CO_2$ fitting is influenced by water vapor absorption lines that have not yet been calculated at the time of the $CO_2$ fitting. In this case the water features in question are too minor to be useful in our $H_2O$ fitting range, but they are significant enough that neglecting them as we do produces an absolute error of $\sim 4 \times 10^{-6}$ averaged over the $CO_2$ fitting range (it gradually decreases from $\sim 1.5 \times 10^{-5}$ on the long wavelength end to zero on the short wavelength end). Although this is significantly smaller in magnitude than the random noise its slowly varying nature led to a concern that it might significantly bias $CO_2$ fitting, so we conducted sensitivity testing on the significance of neglecting these features during $CO_2$ fitting and found that they in fact have no detectable effect. In general, we have found no evidence that expanding the model calculation range for any given species or changing the order in which species are fit has any effect on the results. However we chose to fit $O_2$ first because it has the strongest and broadest absorption lines, which makes it most useful for identifying the calibration observation that offers the best match to the detector background pattern by the process we describe below.

### *B.7 Calculating the spatially variable detector background*

To calculate the $e_{4cal}$ in (B22) from calibration-target calibration observations, we simply perform the same processing steps used to create the continuum-removed ratio spectra from sky observations, with the bright calibration target collects and dark calibration target collects taking the place of the low elevation angle collects and high elevation angle collects, respectively. Equation (B20) then gives $e_{4cal}$. Calculating $e_{4cal}$ from nighttime calibration observations presents a small complication in that, although at night we observe $e$ directly, $A^*$ and $B^*$ are zero. ($e$ is spatially variable detector background prior to any ratioing, smoothing, or continuum removal and $A^*$ and $B^*$ are detector-background -free spectra– see Eq. (B7).) In order to apply the same procedures with these nighttime calibration observations, we assign arbitrary non-zero values to $A^*$ and $B^*$, then calculate $A_{night}$ and $B_{night}$ from Eq. (B7), and from there proceed exactly as with the calibration-target calibration observations.

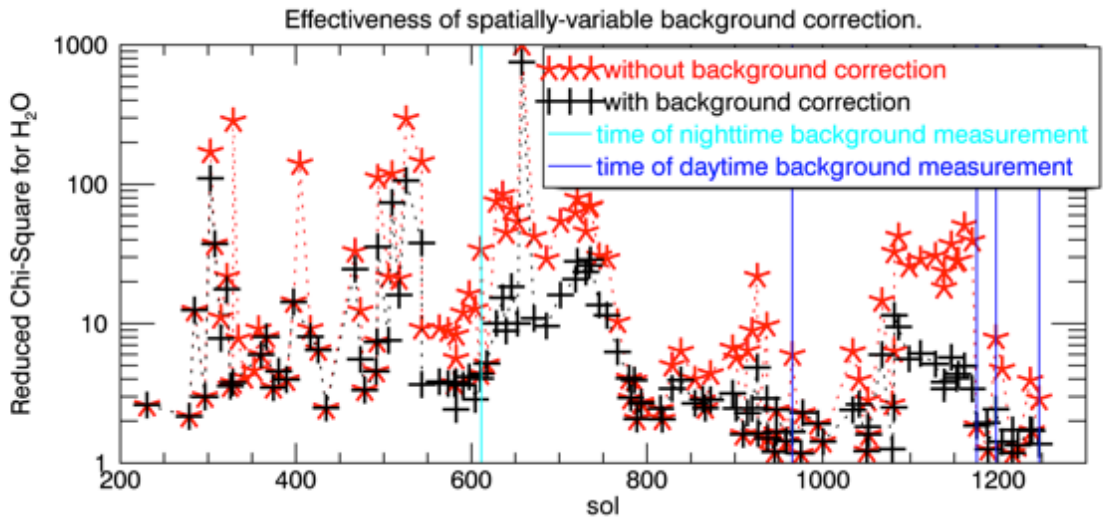

**Figure B1:** Comparison goodness-of-fit obtained with (black crosses) and without (red asterisks) the spatially variable background correction. The vertical lines show the timing of the calibration observations used in the background correction.

Figure B1 shows the reduced-chi-square of model fits to the $H_2O$ absorption line with and without the correction for spatially variable detector background. It also shows the timing of calibration observations acquired to-date. The correction produces a very substantial improvement unless the original reduced-chi-square was already very small due to spatially variable detector background being small in that particular observation. In all cases where we apply the detector background, the improvement in $\chi^2$ passes an F-test with greater than 99% confidence. We do see less improvement in reduced-chi-square as the time between the calibration and the sky observation increases, although this effect is not detectable when the time difference is less than 30 sols.

### B.8 Smoothed vs. non-smoothed CO₂ scaling

Smoothed and non-smoothed $CO_2$ scaling were discussed in section 2.10, and we adopted the smoothed scaling approach for all of the results presented in the main body of this paper. We noted that, in principle, sky conditions could change more rapidly than the averaging window of our smoothed scaling and thereby affect the accuracy of our results. Here we show that this does not seem to be the case, i.e. we show that smoothing does not appear to negatively affect accuracy. We also discuss the implications of the differences that we do observe between smoothed scaling and non-smoothed scaling results.

Except for some of the most recent data points in Mars Year 33, smoothed $CO_2$ scaling results show no significant differences from the single-point $CO_2$ scaling results (Fig. B2). The transition from high to low column abundance at $L_s$=~30° in Mars Year 32 appears smoother in the single-point results, which could be consistent with changes in atmospheric conditions interfering with the accuracy of the smoothed-scaling approach, but the difference is confined to a single data point and is not statistically significant, so it does not meaningfully influence our interpretations.

After $L_s$=110° in Mars Year 33, the differences between the two scaling approaches becomes significant, with single-point scaling showing a large statistically significant scatter in the retrieved column and the smoothed scaling showing very little scatter around a smooth trend in column abundance. This discrepancy indicates that all of the scatter in the single-point-scaled results for this time period comes from abrupt changes in the apparent $CO_2$ abundance. Although in principle abrupt real changes in the atmosphere could cause abrupt changes in both the real water column and apparent $CO_2$ abundance (via changes in the vertical distribution of aerosols), in order for those abrupt changes to disappear entirely in the smoothed-scaling version those abrupt changes in aerosol distribution and water column would have to coincidentally balance each other so as to produce no changes in $H_2O$ absorption band depth. Since such a coincidence seems unlikely, our preferred explanation is that an unknown but apparently random source of error is affecting the $CO_2$ band after $L_s$=110° in Mars Year 33. This error is not associated with increased fit residuals and so is not captured by our uncertainty estimation procedures or error bars. One reasonably likely potential cause of this error appearing after MY = 33 $L_s$=110° is the development of anomalous behavior in one or two pixels that happen to line up with one of the $CO_2$ absorption peaks. Since these peaks (see Fig. 5) are very narrow (unlike for $O_2$) and few in number (unlike for $H_2O$), a single unluckily-placed anomalous pixel can heavily influence the fitted result and still have a small contribution to $\chi^2$. In future work, it may be possible to mitigate this error by filtering out certain pixels, or by including higher order terms in our spatially-variable background correction. Regardless of the nature of the error, using smoothed $CO_2$ scaling substantially diminishes it, and so we will use smoothed $CO_2$ scaling for $H_2O$ results in the remainder of this paper. The fact that we cannot detect an improvement in $H_2O$ results from smoothed $CO_2$ scaling before MY=33 $L_s$=110° does not mean that the source of error in $CO_2$ fitting was absent, it simply means that its effect was smaller than uncertainties in $H_2O$ itself. In fact when solving for $O_2$, the absorption lines of which can be fit with substantially higher relative precision than that of $H_2O$, using smoothed $CO_2$ scaling substantially reduces data point scatter which suggest that the unknown source of error is in fact present and simply not large enough to be significant for $H_2O$ with its larger intrinsic uncertainties and larger natural variability. It may however be significant for $O_2$ in much of the data set, which is why we are not ready to present $O_2$ results in the present paper.

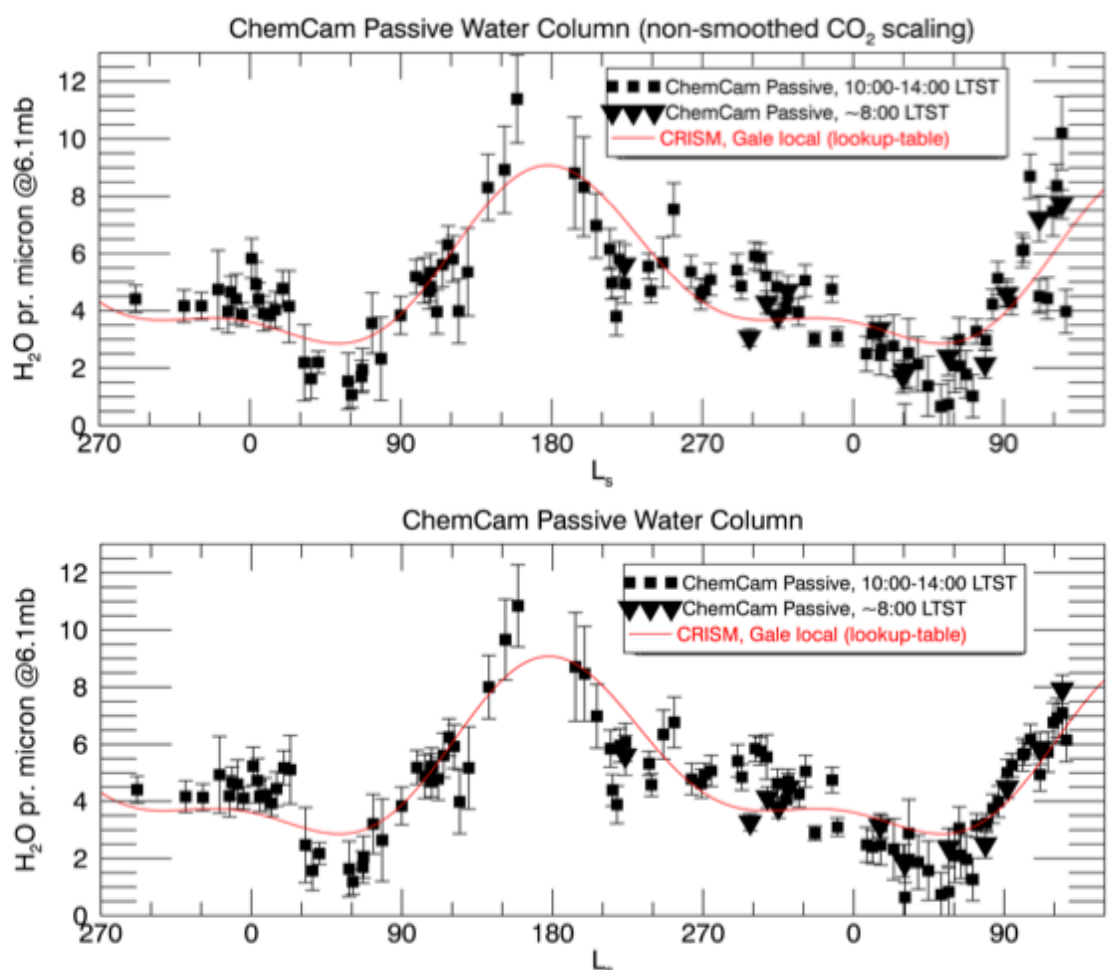

**Figure B2**: ChemCam passive sky water vapor column results with non-smoothed (top) and smoothed (bottom) CO2 scaling. These plots are identical to Fig. 9 in the main text except for the non-smoothed $CO_2$ scaling in the top panel. The bottom panel is completely identical to Fig. 9 and is presented again here only for ease of comparison.

## Appendix C. ChemCam Passive Sky Data in the PDS

### C.1 Sky data

The ChemCam Engineering Data Records (EDRs) archived by NASA's Planetary Data System (PDS) are the starting point for all ChemCam passive sky data analysis. Passive observations can be identified from the "INSTRUMENT_ELEVATION" data element within the "SITE_DERIVED_GEOMETRY_PARMS" group of the PDS EDR headers. This data element will be a value greater than 0 for all passive sky observations. All passive sky observations will also have the "LASER_MODE" data element equal to "NO". A group of EDRs that make up an individual ChemCam passive sky observation sequence will be a set of EDRs that meet the INSTRUMENT_ELEVATION and LASER_MODE criteria and which all have the same "SEQUENCE_ID" value and the same or nearly the same "PLANET_DAY_NUMBER" value. Each EDR includes data from one of the "collects" that make up a passive sky observing sequence. A very small number of ChemCam passive observations pointed above the horizon have been acquired for other purposes, and so to confirm that a group of observations is a passive sky observation sequence check that the pattern of observations matches or is very similar to one of the three passive sky types in Table 1. The role of a particular collect within the sequence can be determined by examining both the previously mentioned "INSTRUMENT_ELEVATION" data element and the START_ROW_XXXX and STOP_ROW_XXXX data elements.

### C.2 Calibration target data

Since the ChemCam passive sky calibration target sequences are a small minority of ChemCam calibration target measurements and even a minority of passive-mode calibration target measurements, they must be identified by selecting groups of EDRs with the same SEQUENCE_ID and same or nearly the same "PLANET_DAY_NUMBER" that have the a passive-sky-like pattern of illuminated and non-illuminated collects *as well as* appropriate LASER_MODE and ARTICULATION_DEVICE_ANGLE values. The chosen calibration targets can be identified from the ARTICULATION_DEVICE_ANGLE data element of the RSM_ARTICULATION_STATE group of the PDS EDR headers. The requested azimuth and elevation will be within 5 milliradians of the following values (given in radians) for each target – target #5: 0.1632 az., 1.1019 el.; target #9: 0.1561az., 1.0884 el.; target #10: 0.1741 az., 1.0884 el.. The positions targeted by the RSM for the bright targets are offset from the center of the targets (by less than 5 milliradians) in order to avoid the region of the target darkened by LIBS laser shots. Although we try to avoid shadowed calibration targets in the interest of consistent illumination conditions, we have found that shadowed calibration targets produce results that are indistinguishable from non-shadowed targets after continuum removal. If only one of the two targets were in shadow, that calibration sequence would likely be unusable due to residual atmospheric absorption lines, but this has not yet occurred.

## Acknowledgements

We would like to thank the MSL science and operations teams, and the MSL "ENV" theme group, for their extensive support with planning and executing the observations described here. We thank Ray Arvidson and Valerie Fox for their assistance with CRISM-derived surface albedos. We also thank Geronimo Villanueva for his valuable advice on HITRAN line parameter uncertainties. In addition, we acknowledge the thorough and thoughtful contributions of two reviewers: Alexey Pankine and Armin Kleinböhl.

Authors McConnochie and Smith acknowledge funding from the MSL project's Participating Scientist program. ChemCam operations were supported by the NASA Mars Exploration Program.

# A. REFERENCES